\documentclass[11pt,onecolumn,final]{IEEEtran} 
\usepackage[final]{graphicx}
\usepackage{graphicx}
\usepackage{amsfonts,amssymb}
\usepackage{latexsym}
\usepackage{amsmath}
\newtheorem{theorem}{Theorem}[section]
\newtheorem{lemma}{Lemma}[section]
\newtheorem{definition}{Definition}[section]
\newtheorem{corollary}[theorem]{Corollary}

\newtheorem{remark}{Remark}[section]
\newtheorem{example}{Example}[section]
\newtheorem{claim}{Claim}[section]

\newcommand{\bi}{\begin{itemize}}
\newcommand{\ei}{\end{itemize}}
\newcommand{\ben}{\begin{enumerate}}
\newcommand{\een}{\end{enumerate}}
\newcommand{\beq}{\begin{equation}}
\newcommand{\eeq}{\end{equation}}
\newcommand{\beqa}{\begin{eqnarray}}
\newcommand{\eeqa}{\end{eqnarray}}

\def\BibTeX{{\rm B\kern-.05em{\sc i\kern-.025em b}\kern-.08em
    T\kern-.1667em\lower.7ex\hbox{E}\kern-.125emX}}

\begin{document}

\title{Pseudocodewords of Tanner Graphs}

\author{\begin{tabular}{cc}
Christine A. Kelley& Deepak Sridhara\\
Department of Mathematics& Seagate Technology\\
The Ohio State University& 1251 Waterfront Place \\
Columbus, OH 43210, USA.& Pittsburgh, PA 15222, USA.\\
ckelley@math.ohio-state.edu& deepak.sridhara@seagate.com\vspace{0.1in}
\end{tabular} \vspace{0.1in}
\thanks{This work was done when the first author was at the University of
Notre Dame and the Institute of Mathematics at the University of
Zurich and when the second author was at the Indian Institute of
Science. The first author was supported by a Center for Applied
Mathematics fellowship from the University of Notre Dame and the
second author was supported by an Institute Mathematics Initiative
fellowship from the Indian Institute of Science and a DRDO-India
grant. This work was also supported in part by NSF Grant No.
CCR-ITR-02-05310. Some of the material in this paper was previously
presented at ISIT 2004 (Chicago, USA) and ISITA 2004 (Parma, Italy).
The first author is now with the Department of Mathematics at The
Ohio State University, Columbus, OH 43210, USA and the second author
is now with Seagate Technology, 1251 Waterfront Place, Pittsburgh,
PA 15222, USA.} 
\\ \centerline{To appear in Nov. 2007 issue of IEEE 
Transactions on Information Theory}}

\markboth{Submitted to IEEE Transactions on Information Theory}
{Kelley and Sridhara}

\maketitle

\vspace{0in}

\begin{abstract}
This papers presents a detailed analysis of pseudocodewords of
Tanner graphs. Pseudocodewords arising on the iterative decoder's
computation tree are distinguished from pseudocodewords arising
on finite degree lifts. Lower bounds on the minimum
pseudocodeword weight are presented for the BEC, BSC, and AWGN
channel. Some structural properties of pseudocodewords are examined, and pseudocodewords and graph
properties that are potentially problematic with min-sum iterative decoding are identified.
An upper bound on the minimum degree lift needed to realize a
particular irreducible lift-realizable pseudocodeword
is given in terms of its maximal component, and it is shown that all irreducible lift-realizable
pseudocodewords have components upper bounded by a finite value $t$ that is dependent on the graph structure.
Examples and different Tanner graph representations of individual
codes are examined and the resulting pseudocodeword
distributions and iterative decoding performances are analyzed.
The results obtained provide some insights in relating the structure of the Tanner graph to the pseudocodeword
distribution and suggest ways of designing Tanner graphs with good minimum pseudocodeword weight.
\end{abstract}


\section{Introduction}
Iterative decoders have gained widespread attention due to their
remarkable performance in decoding LDPC codes. However, analyzing their
performance on finite length LDPC constraint graphs has nevertheless
remained a formidable task.  Wiberg's dissertation \cite{wiberg} was among the earliest
works in characterizing iterative decoder convergence on finite-length
LDPC constraint graphs or Tanner graphs. Both \cite{wiberg} and
\cite{horn} examine the convergence behavior of the {\em min-sum}
iterative decoder \cite{forney_forward_backward} on cycle codes, a special
class of LDPC codes having only degree two variable nodes, and they
provide some necessary and sufficient conditions for the decoder to converge.
Analogous works in \cite{forney} and \cite{pascal} explain the behavior of
iterative decoders using the lifts of
the base Tanner graph. The common underlying idea in all these works is the
role of {\em pseudocodewords} in determining decoder
convergence.

Pseudocodewords of a Tanner graph play an analogous role in
determining convergence of an iterative decoder as {\em codewords}
do for a maximum likelihood decoder. The error performance of a
decoder can be computed analytically using the distance distribution
of the codewords in the code. Similarly, an iterative decoder's
performance may be characterized by the pseudocodeword distance.
Distance reduces to {\em weight} with respect to the all-zero
codeword. Thus, in the context of iterative decoding, a minimum
weight pseudocodeword \cite{forney} is more fundamental than a
minimum weight codeword. In this paper we present lower bounds on
the minimum pseudocodeword weight $w_{\min}$ for the BSC and AWGN
channel, and further, we bound the minimum weight of {\em good} and
{\em bad} pseudocodewords separately.

Paper \cite{pascal} characterizes the set of pseudocodewords in terms of a polytope that includes
pseudocodewords that are realizable on finite degree graph covers of the base Tanner graph,
but does not include all pseudocodewords that can arise on the
decoder's {\em computation tree} \cite{wiberg,frey}. In this paper, we investigate the usefulness of the
graph-covers-polytope definition
of \cite{pascal}, with respect to the min-sum iterative decoder, in characterizing the set of pseudocodewords of a
Tanner graph. In particular, we give examples of computation trees that have several pseudocodeword configurations that
may be bad for iterative decoding whereas the corresponding polytopes of these graph do not contain these bad pseudocodewords.
We note however that this does not mean the polytope definition of pseudocodewords is inaccurate; rather, it is exact
for the case of linear programming decoding \cite{feldman}, but incomplete for min-sum iterative decoding.

As any pseudocodeword is a convex linear combination of a finite
number of {\em irreducible} pseudocodewords, characterizing
irreducible pseudocodewords is sufficient to describe the set of all
pseudocodewords that can arise. It can be shown that the weight of
any pseudocodeword is lower bounded by the minimum weight of its
constituent irreducible pseudocodewords \cite{pascal}, implying that
the irreducible pseudocodewords are the ones that are more likely to
cause the decoder to fail to converge. We therefore examine the
smallest lift degree needed to realize irreducible lift-realizable
pseudocodewords. A bound on the minimum lift degree needed to
realize a given pseudocodeword is given in terms of its maximal
component. We show that all lift-realizable irreducible
pseudocodewords cannot have any component larger than some finite
number $t$ which depends on the structure of the graph. Examples of
graphs with known $t$-values are presented.

The results presented
in the paper are highlighted through several examples. These include
 an LDPC constraint graph  having all
pseudocodewords with weight at least $d_{\min}$ (the minimum
distance of the code), an LDPC constraint graph with both good and
low-weight (strictly less than $d_{\min}$) bad pseudocodewords, and
an LDPC constraint graph with all bad non-codeword-pseudocodewords.
Furthermore, different graph representations of individual codes
such as the $[7,4,3]$ and $[15,11,3]$ Hamming codes are examined in
order to understand what structural properties in the Tanner graph
are important for the design of good LDPC codes. We observe that
despite a very small girth, redundancy in the Tanner graph
representations of these examples can improve the distribution of
pseudocodewords in the graph and hence, iterative decoding
performance.

This paper is organized as follows. Definitions and terminology are
introduced in Section 2. Lower bounds on the pseudocodeword weight
of lift-realizable pseudocodewords are derived in Section 3, and a
bound on the minimum lift degree needed to realize a particular
lift-realizable irreducible pseudocodeword is given. Section 4
presents examples of codes to illustrate the different types of
pseudocodewords that can arise depending on the graph structure.
Section 5 analyzes  the structure of pseudocodewords realizable in
lifts of general Tanner graphs. Finally, the importance of the graph
representation chosen to represent a code is highlighted in Section
6, where the [7,4,3] and [15,11,3] Hamming codes are used as case
studies. Section 7 summarizes the results and concludes the paper.
For readability, the proofs are given in the appendix.

\section{Background} In this section we establish the necessary
terminology and notation that will be used in this paper, including
an overview of pseudocodeword interpretations, iterative decoding
algorithms, and pseudocodeword weights.  Let $G=(V,U;E)$ be a
bipartite graph comprising of vertex sets $V$ and $U$, of sizes $n$
and $m$, respectively, and edges  $E\subset \{(v,u)|\ v\in V, u\in
U\}$. Let $G$ represent a binary LDPC code $\mathcal{C}$ with
minimum distance $d_{\min}$. Then $G$ is called a Tanner graph (or,
LDPC constraint graph) of $\mathcal{C}$. The vertices in $V$ are
called {\em variable} nodes and represent the codebits of the LDPC
code and the vertices in $U$ are called {\em constraint} nodes and
represent the constraints imposed on the codebits of the LDPC code.
\vspace{0.15in}

\begin{definition}
A codeword ${\bf c}$ in an LDPC code $\mathcal{C}$ represented by
a Tanner graph $G$ is a binary assignment to the variable nodes of
$G$ such that every constraint node is connected to an even
number of variable nodes having value 1, i.e., all the parity
check constraints are satisfied.
\label{tannergraph_defn}
\end{definition}
\vspace{0.15in}

\subsection{Pseudocodewords}

\subsubsection{Computation Tree Interpretation}

\begin{figure}
\centering{\resizebox{3.8in}{2.5in}{\includegraphics{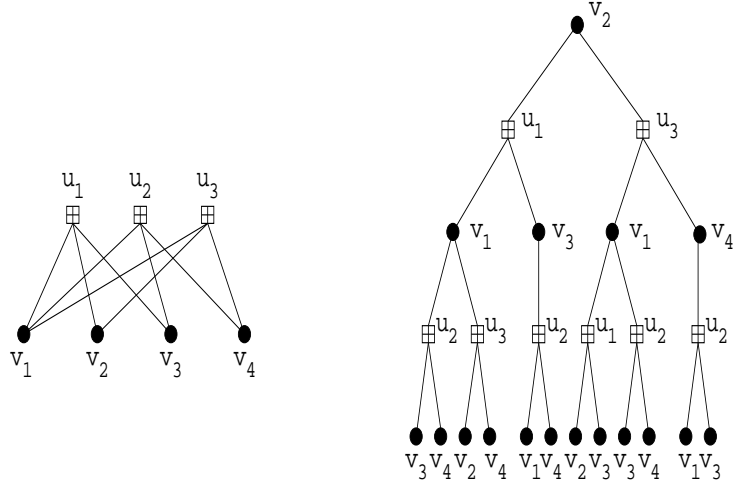}}}
\caption{A graph and its computation tree after two iterations of message
passing.}
\label{comp_tree_fig}
\end{figure}

Wiberg originally formulated pseudocodewords in terms of the computation
tree, as described in \cite{wiberg}, and this work was extended by Frey et al
in \cite{frey}. Let $C(G)$ be the computation tree,
corresponding to the {\em min-sum} iterative decoder, of the base LDPC
constraint graph $G$ \cite{wiberg}. The tree is formed by enumerating the Tanner graph from an
arbitrary variable node, called the {\em root} of the tree, down through the desired number of
layers corresponding to decoding iterations. A computation tree enumerated for ${\ell}$ iterations and having
 variable node $v_i$ acting as the root node of the
tree is denoted by $C_i(G)_{\ell}$. The shape of the computation
tree is dependent on the scheduling of message passing used by the
iterative decoder on the Tanner graph $G$. In
Figure~\ref{comp_tree_fig}, the computation tree $C_2(G)_{2}$ is
shown for the flooding schedule \cite{gallager}. (The variable nodes
are the shaded circles and the constraint nodes are the square
boxes.) Since iterative decoding is exact on cycle-free graphs, the
computation tree is a valuable tool in the exact analysis of
iterative decoding on finite-length LDPC codes with cycles.

A binary assignment to all the variable nodes in the computation
tree is said to be {\em valid} if every constraint node in the
tree is connected to an even number of variable nodes having
value 1.
A codeword ${\bf c}$ in the original Tanner graph $G$ corresponds
to a valid assignment on the computation tree, where for each
$i$, all nodes representing $v_i$ in the computation tree are assigned the same value.
A pseudocodeword ${\bf p}$, on the other hand, is a valid
assignment on the computation tree, where for each $i$, the nodes
representing $v_i$ in
the computation tree need not be assigned the same value.

For a computation tree $C(G)$, we define a {\em local configuration}
at a check $u_j$ in the original Tanner graph $G$ as the average of
the local codeword configurations at all copies of $u_j$ on $C(G)$.
A valid binary assignment on $C(G)$ is said to be {\em consistent}
if all the local configurations are consistent. That is, if a
variable node $v_i$ participates in constraint nodes $u_a$ and
$u_b$, then the coordinates that correspond to $v_i$ in the local
configurations at $u_a$ and $u_b$, respectively, are the same. It
can be shown that a consistent valid binary assignment on the
computation tree is a pseudocodeword that also lies in the polytope
of \cite{pascal} (see equation \ref{polytope}), and therefore
realizable on a lift-graph of $G$. A consistent valid binary
assignment on $C(G)$ also has a compact length $n$ vector
representation ${\bf p}=(p_1,p_2,\dots,p_n)$ such that if a check
node $u_a$ has variable nodes $v_{a_1},v_{a_2},\dots,v_{a_r}$ as its
neighbors, then projecting ${\bf p}$ onto the coordinates $a_1,
a_2,\dots, a_r$ yields the local configuration of $u_a$ on the
computation tree $C(G)$. An inconsistent but valid binary assignment
on $C(G)$ on the other hand has no such compact vector
representation. The computation tree therefore contains
pseudocodewords some of which that lie in the polytope of
\cite{pascal} and some of which that do not.

\subsubsection{Graph Covers Definition}

A degree $\ell$ cover (or, lift) $\hat{G}$ of $G$ is defined in the
following manner: \vspace{0.15in}

\begin{definition}{\rm A finite degree {\em $\ell$ cover} of $G=(V,U;E)$ is
a bipartite graph $\hat{G}$ where for each vertex $x_i\in V \cup U$, there is a {\em
cloud}
$\hat{X}_i=\{\hat{x}_{i_1},\hat{x}_{i_2},\dots,\hat{x}_{i_{\ell}}\}$
of vertices in $\hat{G}$, with $deg(x_{i_j})=deg(x_i)$ for all
$1\le j \le \ell$,
and for every $(x_i, x_j)\in E$, there are $\ell$ edges from $\hat{X}_i$
to $\hat{X}_j$ in $\hat{G}$ connected in a $1-1$ manner.
}\label{graphcover}
\end{definition}
\vspace{0.15in}

\begin{figure}
\centering{\resizebox{3.1in}{1.3in}{\includegraphics{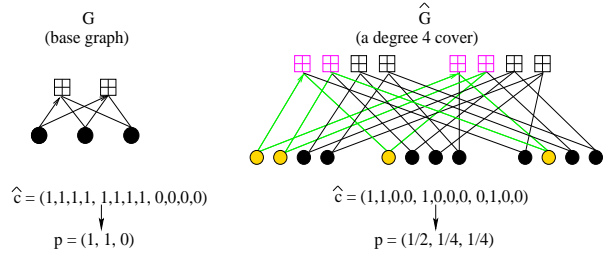}}}
\caption{A pseudocodeword in the base graph (or a valid codeword in a lift).}
\label{pscw_fig}
\end{figure}

Figure~\ref{pscw_fig} shows a base graph $G$ and a degree four cover
of $G$.

A codeword ${\bf \hat{c}}$ in a lift graph $\hat{G}$ of a Tanner
graph $G$ is defined analogously as in Definition
\ref{tannergraph_defn}. \vspace{0.15in}

\begin{definition}{\rm Suppose that ${\bf \hat{\bf c}} =
({\hat{c}}_{1,1},{\hat{c}}_{1,2},\dots,{\hat{c}}_{1,\ell},{\hat{c}}_{2,1},\dots,{\hat{c}}_{2,\ell},\dots)$ is a codeword in
the Tanner graph $\hat{G}$ representing a degree $\ell$ lift of $G$. A {\em
pseudocodeword ${\bf p}$ of $G$} is a vector
$(p_1,p_2,\dots,p_n)$ obtained by reducing a codeword ${\bf \hat{\bf c}}$, of the
code in the lift graph $\hat{G}$, in the following way:
\begin{center}{ ${\bf
\hat{c}}=({\hat{c}}_{1,1},\dots,{\hat{c}}_{1,\ell},{\hat{c}}_{2,1},\dots,{\hat{c}}_{2,\ell},\dots)
\rightarrow  ({\hat{c}}_{1,1}+{\hat{c}}_{1,2}+
\dots+{\hat{c}}_{1,\ell},{\hat{c}}_{2,1}+{\hat{c}}_{2,2}+
\dots+{\hat{c}}_{2,\ell},\dots) = (p_1,p_2,\dots,p_n)$=${\bf p}$, }
\end{center}
\vspace{-0in}$\mbox{ where }p_i = ({\hat{c}}_{i,1}+{\hat{c}}_{i,2}+
\dots+{\hat{c}}_{i,\ell}). $
}
\label{pscw_defn}
\end{definition}
\vspace{0.15in}

Note that each component of the pseudocodeword is merely the number
of 1-valued variable nodes in the corresponding variable cloud of
$\hat{G}$, and that any codeword ${\bf c}$ is trivially a
pseudocodeword as ${\bf c}$ is a valid codeword configuration in a
degree-one lift.  Pseudocodewords as in this definition are called
{\em lift-realizable} pseudocodewords and also as {\em unscaled
pseudocodewords} in \cite{koetter2}. We will use this definition of
pseudocodewords throughout the paper unless mentioned otherwise.
\vspace{0.15in}

\begin{remark}
It can be shown that that the components of a lift-realizable pseudocodeword satisfy the following set of inequalities.
At every constraint node $u_i$, that is connected to variable nodes $v_{i_1},v_{i_2},\dots, v_{i_r}$, the pseudocodeword components satisfy
\begin{eqnarray}
p_{i_1}\le p_{i_2}+p_{i_3}+\dots+p_{i_r}\nonumber \\
p_{i_2}\le p_{i_1}+p_{i_3}+\dots+p_{i_r}\nonumber \\
\vdots \hspace{1.0in} \nonumber\\
p_{i_r}\le p_{i_1}+p_{i_2}+\dots+p_{i_{r-1}}
\label{pscw_ineq}
\end{eqnarray}
\end{remark}
\vspace{0.15in}

\begin{definition}
A pseudocodeword that does not correspond to a codeword in the
base Tanner graph is called a non-codeword pseudocodeword, or
{\em nc-pseudocodeword}, for short.
\label{nc_pscw_def}
\end{definition}
\vspace{0.15in}

\subsubsection{Polytope Representation}
The set of all pseudocodewords associated with a given Tanner graph $G$ has an elegant
geometric description \cite{pascal,feldman}.
In \cite{pascal}, Koetter and Vontobel characterize the set of pseudocodewords via the
{\em fundamental cone}. For each parity check $j$ of degree $\delta_j$, let $C_j$
denote the $(\delta_j,
\delta_j -1,2)$ simple parity check code, and let $P_{\delta_j}$ be a $2^{\delta_j - 1}\times
\delta_j$ matrix with the rows being the codewords of $C_j$.
The {\em fundamental polytope at check $j$} of a Tanner
graph $G$ is then defined as:
\begin{equation}P^{GC}(C_j) =\{\omega
\in \mathbb{R}^{\delta_j}: \omega = xP_{\delta_j}, x \in \mathbb{R}^{2^{\delta_j - 1}}, 0\le x_i
\le 1,
\sum_{i}x_i = 1\}, \end{equation}

and the {\em fundamental polytope of $G$} is defined as:  \begin{equation} P^{GC}(G)
= \{ \omega
\in \mathbb{R}^n: \omega_{N(j)} \in P^{GC}(C_j), j = 1,\ldots,m \},
\label{polytope}
\end{equation}
\vspace{0.15in}

We use the superscript GC to refer to pseudocodewords arising from
graph covers and the notation $\omega_{N(j)}$ to denote the vector
$\omega$ restricted to the coordinates of the neighbors of check
$c_j$. The fundamental polytope gives a compact characterization of
all possible lift-realizable pseudocodewords of a given Tanner graph
$G$. Removing multiplicities of vectors,
 the {\em fundamental cone
$F(G)$} associated with
$G$ is obtained as:
\[ F(G) =
\{\mu \omega \in \mathbb{R}^n:\omega \in P^{GC}(G), \mu \ge 0\}.\] A
lift-realizable pseudocodeword ${\bf p}$ as in  Definition
\ref{pscw_defn} corresponds to a point in the graph-covers polytope
$P^{GC}(G)$.

In \cite{feldman}, Feldman also uses a polytope to characterize the pseudocodewords in
linear programming (LP) decoding and this polytope has striking similarities with the polytope
of \cite{pascal}. Let $E(C_j)$ denote the set of all configurations that satisfy the code $C_j$ (as
defined above). Then the {\em feasible set} of the LP decoder is given by:
\[P^{LP}(G) = \{c \in \mathbb{R}^{n}: x_j \in \mathbb{R}^{2^{\delta_j-1}}, \sum_{S \in E(C_j)}
x_{j,S} = 1, c_i = \sum_{S \in E(C_j),\  i \in S}{ x_{j,S} } \  \forall i
\in N(j), \]\[  0 \le x_{j,S} \le 1,\   \forall S \in E(C_j), j\in\{1,\ldots,m\}\}
\]

\vspace{0.15in}
\begin{remark} {\rm It can be shown that the polytopes of \cite{pascal} and \cite{feldman} are
identical, i.e., $P^{GC}(G) =P^{LP}(G)$ \cite{coleman}.}
\end{remark}
\vspace{0.15in}

\begin{definition} {\rm The {\em support} of a vector ${\bf
x}=(x_1,\ldots,x_n)$, denoted $\operatorname{supp}({\bf x})$, is the set of indices $i$ where $x_i \ne 0$. }
\end{definition}
\vspace{0.15in}

\begin{definition}{\rm \cite{Di} A {\em stopping set} in $G$ is a subset
$S$ of $V$ where for each $s \in S$, every neighbor of $s$ is
connected to $S$ at least twice.  } \label{s_set} \end{definition}
\vspace{0.15in}

\begin{figure}
\centering{\resizebox{4in}{1.15in}{\includegraphics{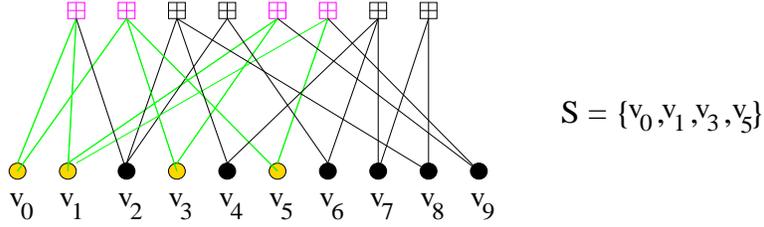}}}
\caption{A stopping set $S = \{v_0,v_1,v_3,v_5\}$ in $G$.}
\label{sset_fig1}
\end{figure}

The size of a stopping set $S$ is equal to the number of elements in $S$.
A stopping set is said to be {\em minimal} if there is no smaller
sized nonempty stopping set contained within it. The smallest minimal stopping set is
 called a {\em minimum} stopping set, and its size is denoted by
$s_{\min}$. Note that a minimum stopping set is not necessarily unique.
Figure~\ref{sset_fig1} shows a stopping set in the graph. Observe that
$\{v_4,v_7,v_8\}$ and $ \{v_3,v_5,v_9\}$ are two minimum stopping sets of
size $s_{\min} = 3$, whereas $\{v_0,v_1,v_3,v_5\}$ is a minimal stopping
set of size 4.

On the erasure channel, pseudocodewords of $G$ are essentially stopping
sets in $G$ \cite{forney, pascal, feldman} and thus, the non-convergence of
the iterative decoder is attributed to the presence of stopping sets. Moreover, any
stopping set can potentially prevent the iterative decoder from converging.

One useful observation is that that the support of a lift-realizable
pseudocodeword as in Definition \ref{pscw_defn} forms a stopping set
in $G$.  This is also implied in \cite{pascal} and \cite{feldman}.\\
\vspace{0.15in} \vspace{0.15in}

\begin{lemma}
The support of a lift-realizable pseudocodeword ${\bf p }$ of $G$ is the incidence vector of a
stopping set in $G$.
\label{supportpscw_sset_lemma}
\end{lemma}
\vspace{0.15in}

\begin{definition} {\rm A pseudocodeword ${\bf p} = (p_1,\ldots,p_n)$ is
{\em irreducible}  if it cannot be written as a sum of two or more codewords
or pseudocodewords. }
\label{pscw_irred_defn}
\end{definition}
\vspace{0.15in}

Note that irreducible pseudocodewords are called {\em minimal} pseudocodewords in \cite{pascal}
 as they correspond to vertices of the polytope $P^{GC}(G)$, and
in the scaled definition of pseudocodewords in \cite{pascal}, any pseudocodeword is a convex linear combination of these irreducible pseudocodewords.
We will see in subsequent sections that the irreducible pseudocodewords,
as defined above, are the ones that can potentially cause the min-sum
decoder to fail to converge.

\subsection{Pseudocodewords and Iterative Decoding Behavior}

The feature that makes LDPC codes attractive is the existence of
computationally simple decoding algorithms. These algorithms either converge
iteratively to a sub-optimal solution that may or may not be the maximum
likelihood solution, or do not converge at all.  The most common of these algorithms
are the min-sum (MS) and the
sum-product (SP) algorithms \cite{forney_forward_backward, factor_graphs}.
These two algorithms are graph-based message-passing algorithms applied on
the LDPC constraint graph. More recently, linear programming (LP) decoding has been applied to decode
LDPC codes. Although LP decoding is more complex, it has the advantage that when it decodes to a codeword, the
codeword is guaranteed to be the maximum-likelihood codeword (see \cite{feldman}).

A message-passing decoder exchanges messages along the edges of the code's
constraint graph. For binary LDPC codes, the variable nodes assume the values one or
zero; hence, a message can be represented either as the probability vector
$[p_0,p_1]$, where $p_0$ is the probability that the variable node assumes
a value of $0$, and $p_1$ is the probability that the variable node
assumes a value of $1$, or as a log-likelihood ratio (LLR) $\log(\frac{p_0}{p_1})$, in which case the
domain of the message is the entire real line $\mathbb{R}$.

Let ${\bf c}=(c_1,\dots,c_n)$ be a codeword and let ${\bf w}=(w_1,\dots,w_n)$ be the input to the decoder from the channel.
That is, the log-likelihood ratios (LLR's) from the channel for the codebits
$v_1,\dots,v_n$ are $w_1,\dots,w_n$, respectively. Then the optimal
maximum likelihood (ML) decoder estimates the codeword \[ c^{*} = arg\min_{{\bf c}\in \mathcal{C}}(
c_1w_1+c_2w_2+\dots+c_nw_n) = arg\min_{{\bf c}\in \mathcal{C}}{\bf c}{\bf w}^T.\]

Let $\mathcal{P}$ be the set of all pseudocodewords (including all codewords) of the graph $G$.
Then the graph-based min-sum (MS) decoder essentially estimates \cite{pascal}
\[ {\bf x}^{*} =arg\min_{{\bf x}\in \mathcal{P}}{\bf x}{\bf w}^T.\]
We will refer to the dot product ${\bf x}{\bf w}^T$ as the {\em cost-function} of the vector ${\bf x}$ with
respect to the channel input vector ${\bf w}$. Thus, the ML decoder estimates the codeword with the lowest cost whereas
the sub-optimal graph-based iterative MS decoder estimates the pseudocodeword with the lowest cost.

The SP decoder, like the MS decoder, is also a message passing decoder that operates
on the constraint graph of the LDPC code. It is more accurate than the MS decoder as it takes into account all
pseudocodewords of the given graph in its estimate. However, it is still sub-optimal compared to the ML decoder.
Thus, its estimate may not not always correspond to a single codeword (as the ML decoder), or a single
pseudocodeword (as the MS decoder). A complete description, along with the update rules, of the MS and SP decoders may be found
in \cite{forney_forward_backward}.

In this paper we will focus our attention on the graph-based min-sum
(MS) iterative decoder, since it is easier to analyze than the
sum-product (SP) decoder. The following definition characterizes the
iterative decoder behavior, providing conditions when the MS decoder
may fail to converge to a valid codeword. \vspace{0.15in}

\begin{definition}{\rm \cite{horn} A pseudocodeword ${\bf
p}$=$(p_1,p_2,\dots,p_n)$ is {\em good} if for all input weight vectors
${\bf w}=(w_1,w_2,\dots,w_n)$ to the min-sum iterative decoder, there is a
codeword ${\bf c}$ that has lower overall cost than ${\bf p}$, i.e., ${\bf
c}{\bf w}^T < {\bf p}{\bf w}^T$.}
\label{good_pscw_defn}
\end{definition}
\vspace{0.15in}

\begin{definition}{\rm A pseudocodeword ${\bf p}$ is {\em bad} if there is a
weight vector ${\bf w}$ such that for all codewords ${\bf c}$,
${\bf c}{\bf w}^T >{\bf p}{\bf w}^T $.}
\label{bad_pscw_defn}
\end{definition}
\vspace{0.15in} Note that a pseudocodeword that is bad on one
channel is not necessarily bad on other channels since the set of
weight vectors ${\bf w}$ that are possible depends on the channel.

Suppose the all-zeros codeword is the maximum-likelihood (ML) codeword for
an input weight vector ${\bf w}$, then all non-zero codewords ${\bf c}$
have a positive cost, i.e., ${\bf c}{\bf w}^T >0$.
In the case where the all-zeros codeword is the ML codeword, it is
equivalent to say that a pseudocodeword ${\bf p}$ is bad if there is a
weight vector ${\bf w}$ such that for all codewords ${\bf c}$,
${\bf c}{\bf w}^T \ge 0$ but ${\bf p}{\bf w}^T < 0$.

As in classical coding where the distance between codewords affects error
correction capabilities, the distance between pseudocodewords affects
iterative decoding capabilities. Analogous to the classical case, the
distance between a pseudocodeword and the all-zeros codeword is captured
by weight. The weight of a pseudocodeword depends on the channel, as
noted in the following definition.

\vspace{0.15in}

\begin{definition}{\rm \cite{forney}
Let ${\bf p}= (p_1,p_2,\dots,p_n)$
be a pseudocodeword of the code $C$ represented by the Tanner graph $G$, and let $e$ be
the smallest number such that the sum of the $e$ largest $p_i$'s is at
least $\frac{\sum_{i=1}^{n}p_i}{2}$.
Then the {\em weight} of ${\bf p}$ is:
\bi
\item  $w_{BEC}({\bf p}) = |\operatorname{supp}({\bf p})|$ for the binary erasure
channel (BEC);

\item $w_{BSC}({\bf p})$ for the binary symmetric channel (BSC)
is:\vspace{-0.05in}
\[ w_{BSC}({\bf p})= \left\{ \begin{array}{cc}
           2e, & \mbox{if } \sum_{e}p_i = \frac{\sum_{i=1}^{n}p_i}{2}\\
2e-1, & \mbox{if } \sum_{e}p_i >\frac{\sum_{i=1}^{n}p_i}{2}
\end{array} \right. ,  \]
where $\sum_e p_i$ is the sum of the $e$ largest $p_i$'s.
\item $w_{AWGN}({\bf p})= \frac{\ (p_1+p_2+\cdots+p_n)^2}{(p_1^2+p_2^2+\cdots+
p_n^2)}$ for the additive white
Gaussian noise (AWGN) channel.
\ei
}
\label{pscw_wt_defn}
\end{definition}
\vspace{0.15in}

Note that the weight of a pseudocodeword of $G$ reduces to the
traditional Hamming weight when the pseudocodeword is a codeword of
$G$, and that the weight is invariant under scaling of a
pseudocodeword. The {\em  minimum} pseudocodeword weight of $G$ is
the minimum weight over all pseudocodewords of $G$ and is denoted by
$w_{\min}^{BEC}$ for the BEC (and likewise, for other channels).
\vspace{0.15in}

\begin{remark}{\rm The definition of pseudocodeword and pseudocodeword
weights are the same for generalized Tanner graphs, wherein the constraint nodes
represent subcodes instead of simple parity-check nodes. The difference is
that as the constraints impose more conditions to be satisfied, there are
fewer possible nc-pseudocodewords. Therefore, a code
represented by an LDPC constraint graph having stronger subcode
constraints will have a larger minimum pseudocodeword weight
than a code represented by the same LDPC constraint graph having
weaker subcode constraints.}
\end{remark}
\vspace{0.15in}

\subsection{Graph-Covers-Polytope Approximation}
\begin{figure}[h]
\centering{\resizebox{3.5in}{2in}{\includegraphics{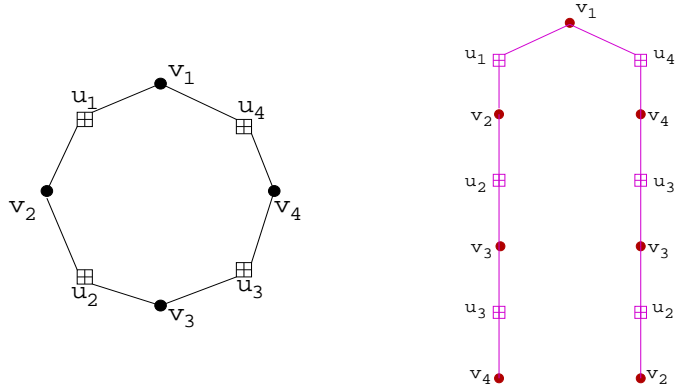}}}
\caption{A Tanner graph and computation tree (CT) for the [4,1,4] repetition code.}
\label{rep4_1_4orig}
\end{figure}

In this section, we examine the graph-covers-polytope definition of \cite{pascal}
in characterizing the set of pseudocodewords of a Tanner graph with respect to min-sum iterative decoding. Consider the $[4,1,4]$-repetition
code which has a Tanner graph representation as shown in Figure~\ref{rep4_1_4orig}.
The corresponding computation tree for three iterations of message passing is also shown in the figure.
The only lift-realizable pseudocodewords for this graph are $(0,0,0,0)$
and $(k,k,k,k)$, for some positive
integer $k$; thus, this graph has no nc-pseudocodewords. Even on the computation tree,
the only valid assignment assigns the same value for all the nodes on the computation tree. Therefore,
there are no nc-pseudocodewords on the graph's computation tree as well.

\begin{figure}[h]
\centering{\resizebox{3.5in}{1.6in}{\includegraphics{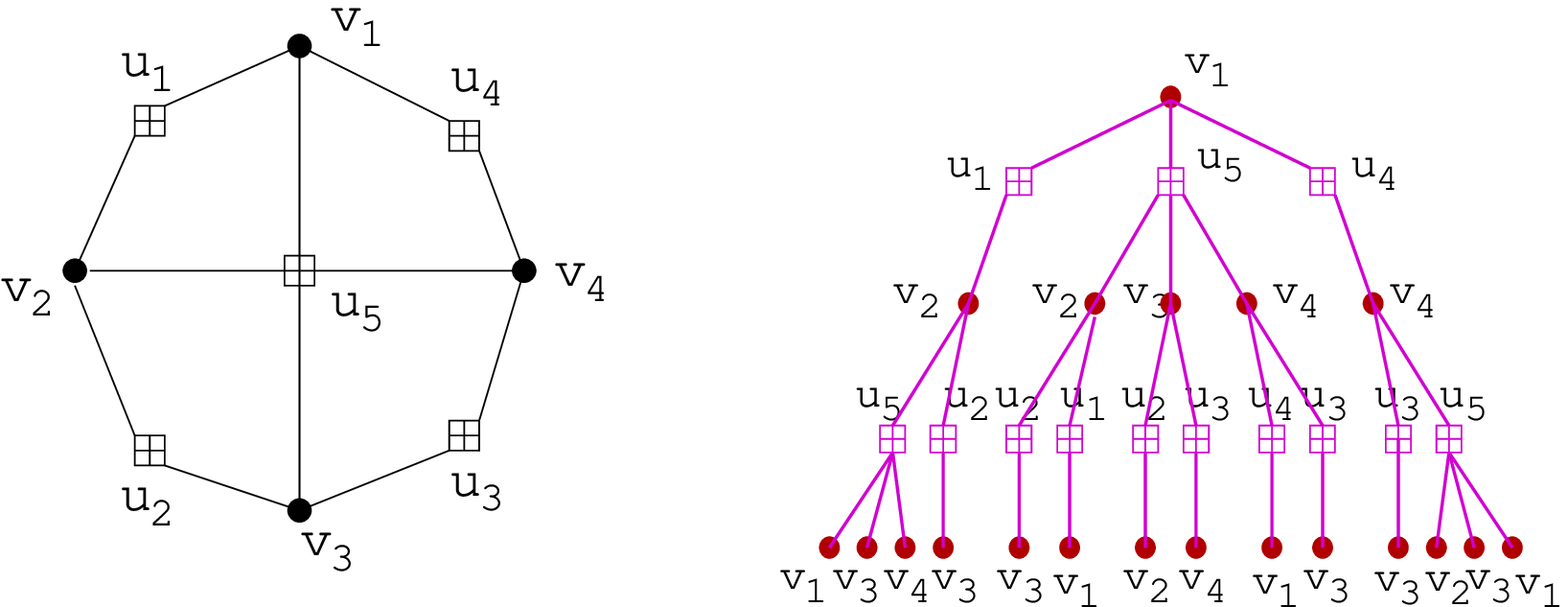}}}
\caption{Modified Tanner graph and CT for the [4,1,4] repetition
code.} \label{rep4_1_4new1}
\centering{\resizebox{3.5in}{1.6in}{\includegraphics{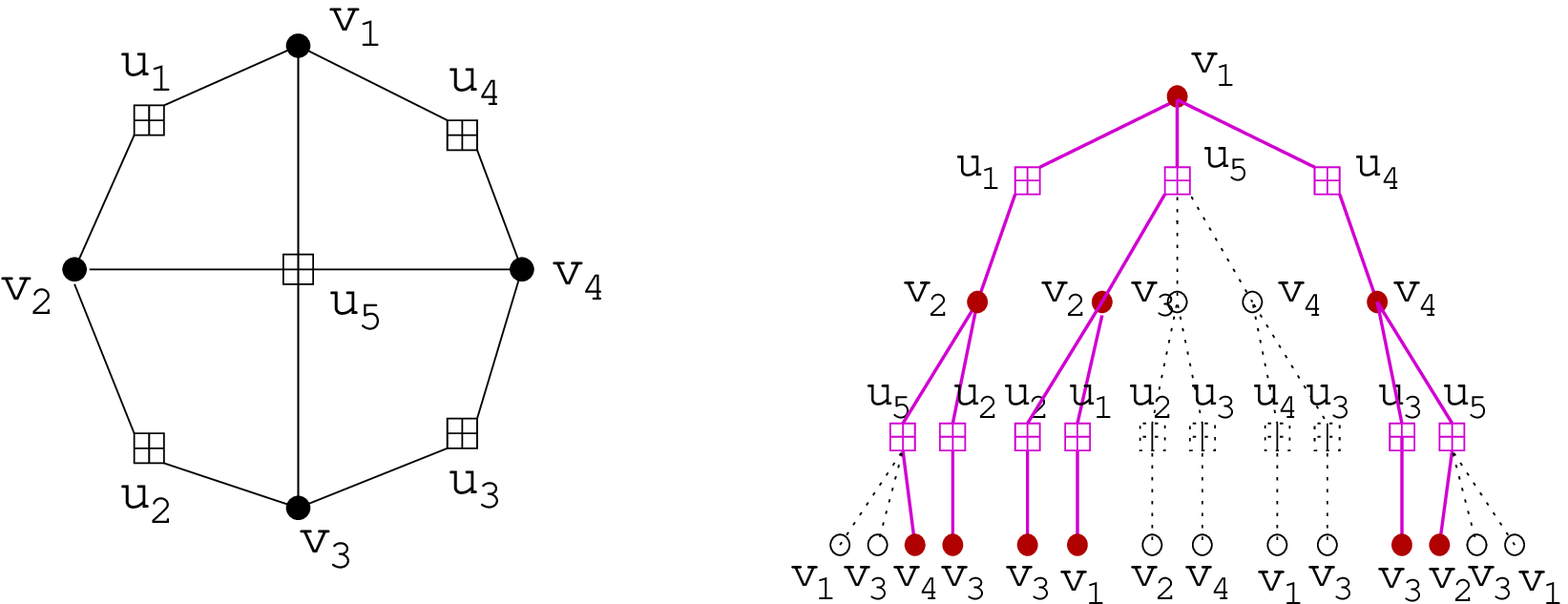}}}
\caption{Modified Tanner graph and CT for the [4,1,4] repetition
code.} \label{rep4_1_4new2}
\end{figure}

Suppose we add a redundant check node to the graph, then we obtain a new LDPC constraint
graph, shown in Figure~\ref{rep4_1_4new1}, for the same code. Even on this graph, the only
lift realizable pseudocodewords are $(0,0,0,0)$ and $(k,k,k,k)$, for some
positive integer $k$. Therefore the polytope of \cite{pascal} contains $(0,0,0,0)$ and
$(1,1,1,1)$ as the vertex points and has no bad pseudocodewords (as in Definition~\ref{bad_pscw_defn}).
However, on the computation tree, there are several valid assignments that
do not have an equivalent representation in the graph-covers-polytope. The assignment where
all nodes on the computation tree are assigned the same value, say $1$, (as highlighted in Figure~\ref{rep4_1_4new1})
 corresponds to a codeword in the code. For this assignment on the computation tree, the local configuration
at check $u_1$ is (1,1) corresponding to $(v_1,v_2)$, at check $u_2$ it is $(1,1)$ corresponding to
$(v_2,v_3)$, at check $u_3$ it is $(1,1)$ corresponding to $(v_3,v_4)$, at check $u_4$ it is
$(1,1)$ corresponding to $(v_1,v_4)$, and at check $u_5$ it is $(1,1,1,1)$ corresponding to
$(v_1,v_2,v_3,v_4)$. Thus, the pseudocodeword vector $(1,1,1,1)$ corresponding to
$(v_1,v_2,v_3,v_4)$ is consistent locally with all the local configurations at the individual
check nodes.

However, an assignment where some nodes are assigned different
values compared to the rest (as highlighted in
Figure~\ref{rep4_1_4new2}) corresponds to a nc pseudocodeword on the
Tanner graph. For the assignment shown in Figure~\ref{rep4_1_4new2},
the local configuration at check $u_1$ is $(1,1)$, corresponding to
$(v_1,v_2)$, as there are two check nodes $u_1$ in the computation
tree with $(1,1)$ as the local codeword at each of them. Similarly,
the local configuration at check $u_2$ is $(2/3,2/3)$, corresponding
to $(v_2,v_3)$, as there are three $u_2$ nodes on the computation
tree, two of which have $(1,1)$ as the local codeword and the third
which has $(0,0)$ as the local codeword. Similarly, the local
configuration at check $u_3$ is $(1/3,1/3)$ corresponding to
$(v_3,v_4)$, the local configuration at check $u_4$ is $(1/2,1/2)$
corresponding to $(v_1,v_4)$, and the local configuration at check
$u_5$ is $(1/3,1,0,2/3)$ corresponding to $(v_1,v_2,v_3,v_4)$. Thus,
there is no pseudocodeword vector that is consistent locally with
all the above local configurations at the individual check nodes.

Clearly, as the computation tree grows with the number of decoding iterations,
the number of nc-pseudocodewords in the graph grows exponentially with the depth of the tree.
Thus, even in the simple case of the repetition code, the graph-covers-polytope of \cite{pascal}
fails to capture all min-sum-iterative-decoding-pseudocodewords of a Tanner graph.

\begin{figure}[h]
\centering{\resizebox{4.5in}{3.5in}{\includegraphics{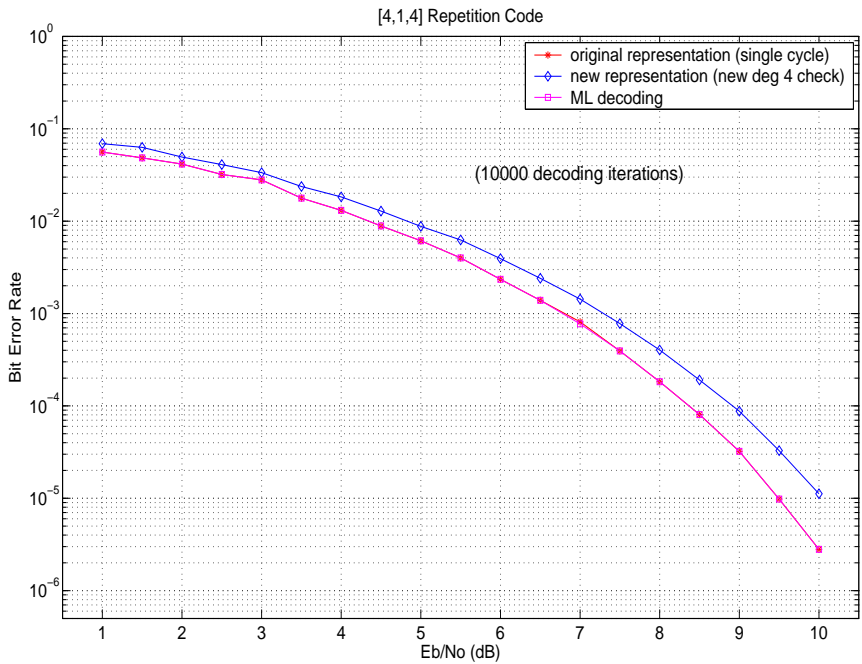}}}
\caption{Performance of different representations of [4,1,4] repetition code over the BIAWGNC.}
\label{rep4_1_4}
\end{figure}

Figure~\ref{rep4_1_4} shows the performance of MS iterative decoding
on the constraint graphs of Figures \ref{rep4_1_4orig} and
\ref{rep4_1_4new1} when simulated over the binary input additive
white Gaussian noise channel (BIAWGNC) with signal to noise ratio
$E_b/N_o$. The ML performance of the code is also shown as
reference. With a maximum of $10^4$ decoding iterations, the
performance obtained by the iterative decoder on the single cycle
constraint graph of Figure~\ref{rep4_1_4orig} is the same as the
optimal ML performance (the two curves are one on top of the other),
thereby confirming that the graph has no nc-pseudocodewords. The
iterative decoding performance deteriorates when a new degree four
check node is introduced as in Figure~\ref{rep4_1_4new1}. (A
significant fraction of detected errors, i.e., errors due to the
decoder not being able to converge to any valid codeword within
$10^4$ iterations, were obtained upon simulation of this new graph.)
\vspace{0.15in}

This example illustrates that the polytope $P^{GC}(G)$ does not
capture the entire set of MS pseudocodewords on the computation
tree. In general, we state the following results: \vspace{0.15in}
\begin{claim}
A bipartite graph $G$ representing an LDPC code $\mathcal{C}$
contains no irreducible nc-pseudocodewords on the computation tree
$C(G)$ of any depth if and only if either (i) $G$ is a tree, or
(ii) $G$ contains only degree two check nodes.
\label{comptree_claim}
\end{claim}
\vspace{0.15in}

\begin{claim}
A bipartite graph $G$ representing an LDPC code $\mathcal{C}$
contains either exactly one or zero irreducible lift-realizable nc-pseudocodewords
 if either (i) $G$ is a tree, or (ii) there is at least one path
between any two variable nodes in $G$ that traverses only via check nodes
having degree two.
\label{graphcovers_claim}
\end{claim}
\vspace{0.15in} \indent Note that condition (ii) in
Claim~\ref{graphcovers_claim} states that if there is at least one
path between every pair of variable nodes that has only degree two
check nodes, then $G$ contains at most one irreducible
lift-realizable nc-pseudocodeword. However, condition (ii) in
Claim~\ref{comptree_claim} requires that {\em every } path between
every pair of variable nodes has only degree two check nodes.

For the rest of the paper, unless otherwise mentioned, we will restrict
our analysis of pseudocodewords to the set of lift-realizable
pseudocodewords as they have an elegant mathematical description in terms
of the polytope $P^{GC}(G)$ that makes the analysis tractable.

\section{Bounds on minimal pseudocodeword weights}

In this section, we derive lower bounds on the pseudocodeword weight
for the  BSC and AWGN channel, following
Definition~\ref{pscw_wt_defn}. The support size of a pseudocodeword
${\bf p}$ has been shown to upper bound its weight on the BSC/AWGN
channel \cite{forney}. Hence, from
Lemma~\ref{supportpscw_sset_lemma}, it follows that
$w^{BSC/AWGN}_{\min} \le s_{\min}$.  We establish the following
lower bounds for the minimum pseudocodeword weight: \vspace{0.15in}

\begin{theorem} Let $G$ be a regular bipartite graph with girth $g$ and
smallest left degree $d$. Then the minimal pseudocodeword weight is lower
bounded by \vspace{-0.05in}
{ \[ w^{BSC/AWGN}_{\min} \ge \left\{ \begin{array}{cc}
            1+d+d(d-1)+\cdots+d(d-1)^{\frac{g-6}{4}},& \frac{g}{2} \mbox{ odd }\\
            1+d+\cdots+d(d-1)^{\frac{g-8}{4}} +(d-1)^{\frac{g-4}{4}}, & \frac{g}{2}
\mbox{ even }
\end{array} \right. . \]}
\label{tree_bound_basic}
\vspace{-0.0in}
\end{theorem}
\vspace{0.15in} Note that this lower bound holds analogously for the
minimum distance $d_{\min}$ of $G$ \cite{tanner}, and also for the
size of the smallest stopping set, $s_{\min}$, in a graph with girth
$g$ and smallest left
degree $d$ \cite{orlitsky}.\\

For generalized LDPC codes, wherein the right nodes in $G$ of degree
$k$ represent constraints of a $[k,k', \epsilon k]$
sub-code\footnote{Note that $\epsilon k$ and $\epsilon$ are the
minimum distance and the relative minimum distance, respectively of
the sub-code.}, the above result is extended as: \vspace{0.15in}
\begin{theorem} Let $G$ be a $k$-right-regular bipartite graph with girth $g$
and smallest left degree $d$ and let the right nodes represent constraints
of a $[k,k',\epsilon k]$ subcode, and let $x=(\epsilon k -1)$.
Then: \vspace{-0.05in}
{ \[  w^{BSC/AWGN}_{\min} \ge \left\{
\begin{array}{cc}
            1+dx+d(d-1)x^2+\cdots+d(d-1)^{\frac{g-6}{4}}x^{\frac{g-2}{4}},& \frac{g}{2}
\mbox{ odd }\\
            1+dx+\cdots+d(d-1)^{\frac{g-8}{4}}x^{\frac{g-4}{4}}
+(d-1)^{\frac{g-4}{4}}x^{\frac{g}{4}}, & \frac{g}{2} \mbox{ even }
\end{array} \right. . \] }
\label{tree_bound_gen}
\vspace{-0.0in}
\end{theorem}
\vspace{0.15in}

\begin{definition}{\rm A stopping set for a generalized LDPC code
    using $[k,k', \epsilon k]$ sub-code constraints
    may be defined as a set of
variable nodes $S$ whose neighbors are each connected at least $\epsilon
k$ times to $S$ in $G$.}
\label{gen_sset_defn}
\end{definition}
\vspace{0.15in}

This definition makes sense since an optimal decoder on an erasure
channel can recover at most $\epsilon k -1$ erasures in a linear
code of length $k$ and minimum distance $\epsilon k$. Thus if all
constraint nodes are connected to a set $S$, of variable nodes, at
least $\epsilon k$ times, and if all the bits in $S$ are erased,
then the iterative decoder will not be able to recover any erasure
bit in $S$. Note that Definition~\ref{gen_sset_defn} assumes there
are no idle components in the subcode, i.e. components that are zero
in all the codewords of the subcode. For the above definition of a
stopping set in a generalized Tanner graph, the lower bound holds
for $s_{\min}$ also. That is, the minimum stopping set size
$s_{\min}$ in a $k$-right-regular bipartite graph $G$ with girth $g$
and smallest left degree $d$, wherein the right nodes represent
constraints of a $[k,k',\epsilon k]$ subcode with no idle
components, is lower bounded as: \vspace{-0.05in} { \[  s_{\min} \ge
\left\{
\begin{array}{cc}
            1+dx+d(d-1)x^2+\cdots+d(d-1)^{\frac{g-6}{4}}x^{\frac{g-2}{4}},& \frac{g}{2}
\mbox{ odd }\\
            1+dx+\cdots+d(d-1)^{\frac{g-8}{4}}x^{\frac{g-4}{4}}
+(d-1)^{\frac{g-4}{4}}x^{\frac{g}{4}}, & \frac{g}{2} \mbox{ even }
\end{array} \right. , \] where $x = (\epsilon k-1)$. }

\indent The {\em max-fractional weight} of a vector ${\bf
x}=[x_1,\ldots,x_n]$ is defined as $w_{\mathrm{max-frac}}({\bf
x})=\frac{\sum_{i=1}^{n}x_i}{\max_i x_i}$. The max-fractional weight of
pseudocodewords in LP decoding (see \cite{feldman}) characterizes
the performance of the LP decoder, similar to the role of the pseudocodeword weight
 in MS decoding. It is worth noting that for any pseudocodeword ${\bf p}$, the
pseudocodeword weight of ${\bf p}$ on the BSC and AWGN channel
relates to the max-fractional weight of ${\bf p}$ as follows:
\vspace{0.15in}

\begin{lemma} For any pseudocodeword ${\bf p}$, $w^{BSC/AWGN}({\bf p}) \ge
w_{\mathrm{max-frac}}({\bf p})$. \label{feldman_lemma} \end{lemma}
\vspace{0.15in}

It follows that $w^{BSC/AWGN}_{\min} \ge d_{frac}^{\max}$, the
max-fractional distance which is the minimum max-fractional weight
over all ${\bf p}$.  Consequently, the bounds established in
\cite{feldman} for $d_{frac}^{\max}$ are also lower bounds for
$w_{\min}$. One such bound is given by the following theorem.
\vspace{0.15in}

\begin{theorem} (Feldman {\rm \cite{feldman}}) Let $deg_l^-$ (respectively, $deg_r^-$) denote
the smallest left degree (respectively, right degree) in a bipartite graph
$G$. Let $G$ be a factor graph with $deg_l^- \ge 3, deg_r^- \ge 2,$ and
girth $g$, with $g >4$. Then
\[d_{frac}^{\max} \ge (deg_l^- -1)^{\lceil \frac{g}{4}\rceil - 1}. \]
\label{feldman_theorem}
\vspace{-0.0in}
\end{theorem}
\vspace{0.15in}

\begin{corollary} Let $G$ be a factor graph with $deg_l^- \ge 3, deg_r^-
\ge 2,$ and girth $g$, with $g >4$. Then \[w^{BSC/AWGN}_{\min} \ge
(deg_l^- -1)^{\lceil \frac{g}{4}\rceil - 1}. \]
\label{corollary_feldman}
\end{corollary}
\vspace{0.15in}

Note that Corollary~\ref{corollary_feldman}, which is essentially the
result obtained in Theorem~\ref{tree_bound_basic}, makes sense due to the
equivalence between the LP polytope and GC polytope (see Section 2).

Recall that any pseudocodeword can be expressed as a sum of
irreducible pseudocodewords, and further, it has been shown in
\cite{pascal} that the weight of any pseudocodeword is lower bounded
by the smallest weight of its constituent pseudocodewords.
Therefore, given a graph $G$, it is useful to find the smallest lift
degree needed to realize all irreducible lift-realizable
pseudocodewords (and hence, also all minimum weight
pseudocodewords).

One parameter of interest is the maximum component $t$ which can occur in
any irreducible lift-realizable pseudocodeword of a given graph $G$, i.e., if a
pseudocodeword ${\bf p}$
has a component larger than $t$, then ${\bf p}$ is reducible.
\vspace{0.15in}

\begin{definition}{\rm Let $G$ be a Tanner graph. Then the maximum component value
an irreducible pseudocodeword of G can have is called the {\em $t$-value} of $G$,
and will be denoted by $t$.}
\label{t_value_defn2}
\end{definition}
\vspace{0.15in}

We first show that for any finite bipartite graph, the following
holds: \vspace{0.15in}

\begin{theorem}
Every finite bipartite graph $G$ representing a finite length LDPC code
has a finite $t$. \label{existence_t_lemma}
\end{theorem}
\vspace{0.15in}

\begin{theorem}
Let $G$ be an LDPC constraint graph with largest right degree $ d_r^+$ and $t$-value $t$ . That is, any irreducible
lift-realizable pseudocodeword ${\bf p} = ( p_1,\ldots,p_n)$ of
$G$ has $0 \le p_i \le t$, for $i =1,\dots,n$. Then the smallest lift degree
$m_{\min}$ needed to realize all irreducible pseudocodewords of $G$ satisfies
\[ m_{\min} \le\max_{u_j} \frac{\sum_{v_i\in N(u_j)}p_i}{2}\le \frac{t d_r^+}{2},\]
where the maximum is over all check nodes $u_j$ in the graph and $N(u_j)$ denotes
the variable node neighbors of $u_j$.
\label{min_lift_lemma}
\end{theorem}
\vspace{0.15in}

If such a $t$ is known, then Theorem~\ref{min_lift_lemma} may be
used to obtain the smallest lift degree needed to realize all
irreducible lift-realizable pseudocodewords. This has great
practical implications, for  an upper bound on the lift degree
needed to obtain a pseudocodeword of minimum weight would
significantly lower the complexity of determining the minimum
pseudocodeword weight $w_{\min}$. \vspace{0.15in}

\begin{corollary}If ${\bf p}$ is any lift-realizable pseudocodeword and $b$ is the
maximum component, then the smallest lift degree needed to realize ${\bf
p}$ is at most $\frac{bd_r^+}{2}$.
\end{corollary}
\vspace{0.15in}

\begin{example}{\rm Some graphs with known $t$-values are:
 $t\le2$ for cycle codes  \cite{wiberg, horn},
 $t=1$ for LDPC codes whose Tanner graphs are trees, and
 $t\le2$ for LDPC graphs having a single cycle, and
 $t\le s$ for tail-biting trellis codes represented on
Tanner-Wiberg-Loeliger graphs \cite{wiberg} with state-space sizes $s_1, \ldots, s_m$ and $s =
\mbox{max}\{s_1,\ldots,s_m\}$.}
\end{example}
\vspace{0.15in}

We now bound the weight of a pseudocodeword ${\bf p}$ based on its
maximal component value $t$ and its support size
$|\operatorname{supp}({\bf p})|$. \vspace{0.15in}

\begin{lemma} Suppose in an LDPC constraint graph $G$ every irreducible
lift-realizable pseudocodeword ${\bf p}=(p_1,p_2,\dots,p_n)$ with support
set $V$ has components $0\le p_i \le t$, for $1\le i\le n$, then:
$($a$)$ $w^{AWGN}({\bf p}) \ge \frac{2t^2}{(1+t^2)(t-1)+2t} |V|$, and
$($b$)$ $w^{BSC}({\bf p}) \ge \frac{1}{t}|V|$. \label{wmin_t_smin_lemma}
\end{lemma}
\vspace{0.15in} For many graphs, the $t$-value may be small and this
makes the above lower bound large. Since the support of any
pseudocodeword is a stopping set
(Lemma~\ref{supportpscw_sset_lemma}), $w_{\min}$ can be lower
bounded in terms of $s_{\min}$ and $t$. Thus, stopping sets are also
important in the BSC and the AWGN channel.

Further, we can bound the weight of good and bad pseudocodewords
(see Definitions~\ref{good_pscw_defn}, \ref{bad_pscw_defn})
separately, as shown below: \vspace{0.15in}
\begin{theorem} For an $[n,k,d_{\min}]$ code represented by an LDPC constraint
graph $G$: $($a$)$ if ${\bf p}$ is a {\rm good} pseudocodeword
of $G$, then $w^{BSC/AWGN}({\bf p})
\ge w_{\mathrm{max-frac}}({\bf p})\ge d_{\min}$, and $($b$)$ if ${\bf p}$ is a
{\rm bad} pseudocodeword {\rm \cite{horn}} of $G$, then $w^{BSC/AWGN}({\bf
p}) \ge w_{\mathrm{max-frac}}({\bf p}) \ge \frac{ s_{\min}}{t}$, where $t$ is
as in the previous lemma.
\label{good_bad_pscw_wt_thm}
\end{theorem}
\vspace{0.15in}

\section{Examples}
In this section we present three different examples of Tanner graphs
which give rise to different types of pseudocodewords and examine
their performance on the binary input additive white Gaussian noise
channel (BIAWGNC), with signal to noise ratio $E_b/N_o$, with MS,
SP, and ML decoding. The MS and SP iterative decoding is performed
for 50 decoding iterations on the respective LDPC constraint graphs.
\vspace{0.15in}

\begin{example}{\rm Figure~\ref{example1_fig} shows a graph that has no
pseudocodeword with weight less than $d_{\min}$ on the BSC and
AWGN channel.  For this code (or more
precisely, LDPC constraint graph), the minimum distance, the minimum
stopping set size, and the minimum pseudocodeword weight on the AWGN
channel, are all equal to 4, i.e., $d_{\min}=s_{\min}=w_{\min}=4$, and the
$t$-value (see Definition~\ref{t_value_defn2}) is $2$. An irreducible
nc-pseudocodeword with a component of value 2 may be observed by
assigning value 1 to the nodes in the outer and inner rings and
assigning value 2 to exactly one node in the middle ring, and
zeros elsewhere.

\indent Figure~\ref{example1_sim} shows the
performance of this code on a BIAWGNC with MS,
SP, and ML decoding. It is evident that all three algorithms perform almost
identically. Thus, this LDPC code does not have low weight
(relative to the minimum distance) bad pseudocodewords, implying that the
performance of the MS decoder, under i.i.d. Gaussian noise, will be
close to the optimal ML performance.}
\label{ex1}
\end{example}
\vspace{0.15in}

{\begin{center}
\begin{figure}
            \centering{
            \resizebox{1.5in}{1.5in}{\includegraphics{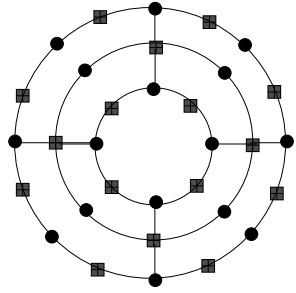}}}
          \caption{A graph with all pseudocodewords having weight at least $d_{\min}$.}
\label{example1_fig}
\end{figure}

\begin{figure}
            \centering{
\resizebox{3.2in}{2.5in}{\includegraphics{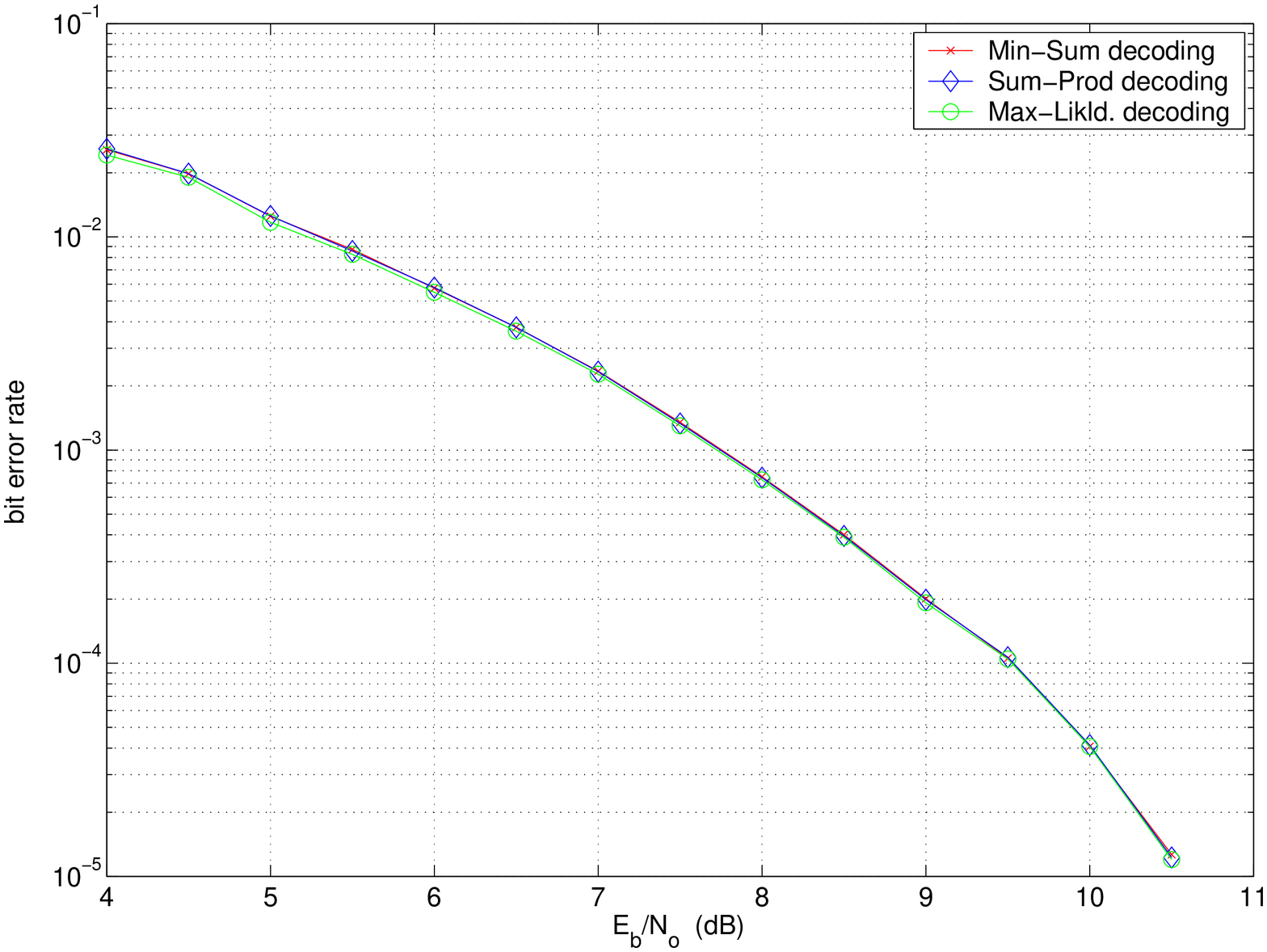}}}
          \caption{Performance of Example~\ref{ex1} - LDPC code: MS, SP, ML
decoding over the BIAWGNC.}
\label{example1_sim}
\end{figure}
\end{center}
} \vspace{0.15in}

\begin{example}{\rm  Figure~\ref{example2_fig} shows a graph that has both good
and bad pseudocodewords. Consider ${\bf p}=(1,0,1,1,1,1,3,0,0,1,1,1,1,0)$.
Letting
${\bf w} =(1,0,0,0,0,0,-1,0,0,0,0,0,0,0)$, we obtain ${\bf p}{\bf w}^T = -2$ and
${\bf c}{\bf w}^T \ge {\bf 0}$ for all codewords ${\bf c}$. Therefore,
${\bf p}$ is a bad pseudocodeword for min-sum iterative decoding. In
particular, this pseudocodeword has a weight of $w^{BSC/AWGN}({\bf p}) =
8$ on both the BSC and the AWGN channel. This LDPC graph results in an
LDPC code of minimum distance $d_{\min} = 8$, whereas the minimum stopping
set size and minimum pseudocodeword weight (AWGN channel) of the graph are
3, i.e., $s_{\min}=w_{\min}=3$, and the $t$-value is $8$. \\
\indent Figure~\ref{example2_sim} shows the performance of this code
on a BIAWGNC with MS, SP, and ML decoding. It is evident in the
figure that the MS and the SP decoders are inferior in performance
in comparison to the optimal ML decoder.  Since the minimal
pseudocodeword weight $w_{\min}$ is significantly smaller than the
minimum distance of the code $d_{\min}$, the performance of the MS
iterative decoder at high signal to noise ratios (SNRs) is dominated
by low-weight bad pseudocodewords.} \label{ex2}
\end{example}
\vspace{0.15in}

{\begin{center}
\begin{figure}
          \centering
                   {
            \resizebox{2in}{1.5in}{\includegraphics{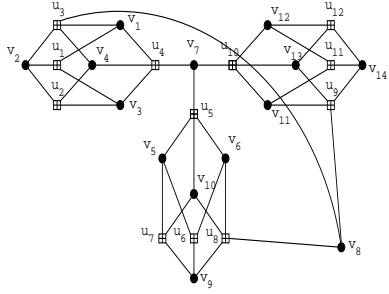}}}
          \caption{A graph with good and bad pseudocodewords.}
\label{example2_fig}
\end{figure}
\begin{figure}
                     \centering{
\resizebox{3.2in}{2.5in}{\includegraphics{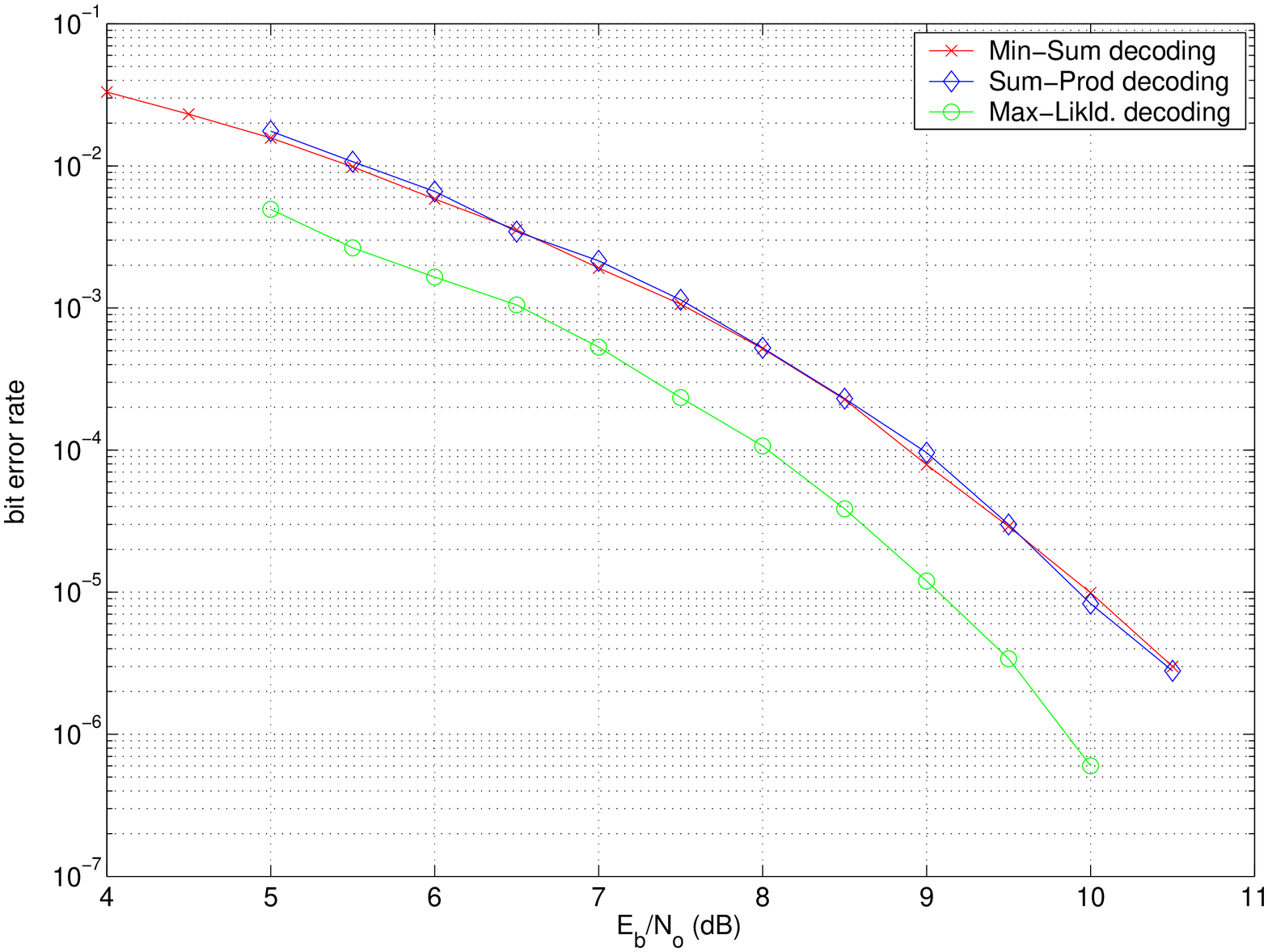}}}
          \caption{Performance of Example~\ref{ex2} - LDPC code: MS, SP, ML
decoding over the BIAWGNC.}
\label{example2_sim}
            \end{figure}
\end{center}
}

\vspace{0.15in}

\begin{figure}
\centering{\resizebox{2in}{1in}{\includegraphics{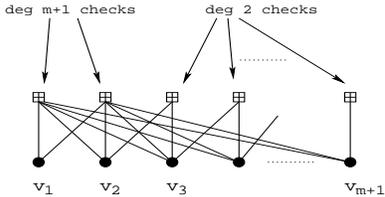}}}
\caption{A graph with only bad nc-pseudocodewords.}
\label{example3_fig}
\end{figure}

\begin{example}{\rm Figure~\ref{example3_fig} shows a graph on $m+1$
variable nodes, where the set of all variable nodes except $v_1$ form a minimal
stopping set of size $m$, i.e., $s_{\min}=m$. When $m$ is even, the only
irreducible pseudocodewords are of the form $(k,1,1,\ldots,1)$, where
$0\le k \le m$ and $k$ is even, and the only nonzero codeword is $(0,1,1,\ldots,1)$. When $m$ is odd, the
irreducible pseudocodewords have the form $(k,1,1,\ldots,1)$, where $1 \le k \le m$,
and $k$ is odd, or $(0,2,2,\ldots,2)$, and the only nonzero codeword is $(1,1,\ldots,1)$. In general, any
pseudocodeword of this graph is a linear combination of these irreducible pseudocodewords.
When $k$ is not 0 or 1, then these are irreducible nc-pseudocodewords; the weight vector ${\bf w} = (w_1,\ldots,w_{m+1})$, where $w_1 = -1, w_2 = +1,$ and $w_3=\cdots=w_{m+1}=0$, shows that these pseudocodewords are bad.
 When $m$ is even or odd, any reducible pseudocodeword of this graph
that includes at least one
irreducible nc-pseudocodeword in its sum, is also bad (according to
Definition~\ref{bad_pscw_defn}). We also observe that for both the BSC and AWGN channel, all of the irreducible pseudocodewords
have weight at most $d_{\min} = m$ or $m+1$, depending on whether $m$ is even or odd. The minimum
pseudocodeword weight is $w_{\min}^{AWGN} = 4m/(m+1)$, and the LDPC
constraint graph has a $t$-value of $m$.\\
\indent Figures~\ref{example3a_sim} and \ref{example3b_sim} show the performance
of the code for odd and even $m$, respectively, on a BIAWGNC with MS, SP, and
ML decoding. The performance difference between the MS
(respectively, the SP) decoder and the optimal ML decoder is more
pronounced for odd $m$. (In the case of even $m$, $(0,2,2\dots,2)$ is not
a bad pseudocodeword, since it is twice a codeword, unlike in the case for odd $m$; thus, one can argue
that, relatively, there are a fewer number of bad pseudocodewords when
$m$ is even.)  Since the graph has low weight bad pseudocodewords, in
comparison to the minimum distance, the performance of the MS decoder
in the high SNR regime is clearly inferior to that of the ML decoder.} \label{ex3}
\end{example}
\vspace{0.15in}

{\begin{center}
\begin{figure}
            \centering
                   {
            \resizebox{3.2in}{2.5in}{\includegraphics{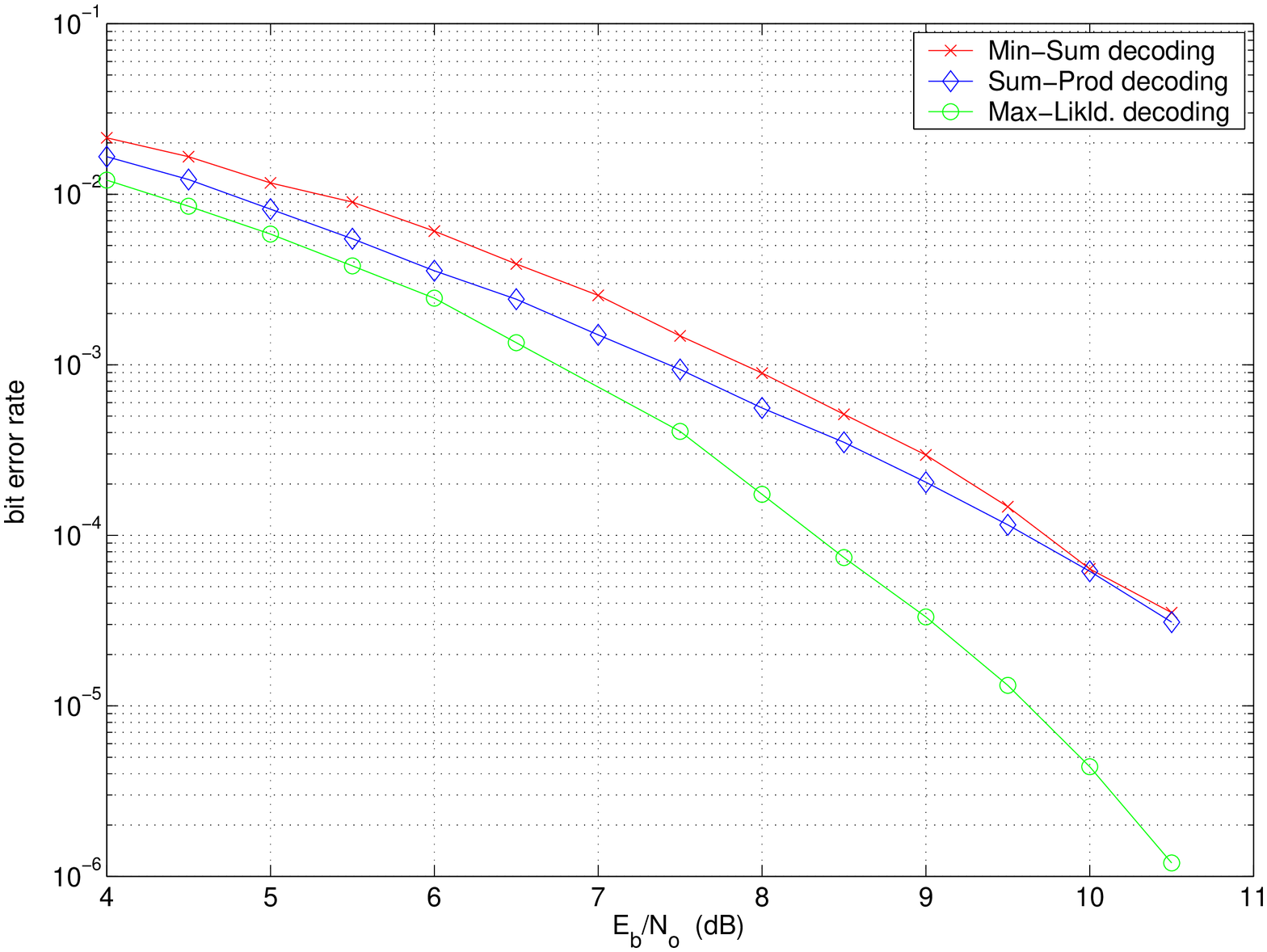}}}
          \caption{Performance of Example~\ref{ex3} - LDPC code for $m=11$: MS, SP, ML decoding over the BIAWGNC.}
                \label{example3a_sim}
            \centering{
\resizebox{3.2in}{2.5in}{\includegraphics{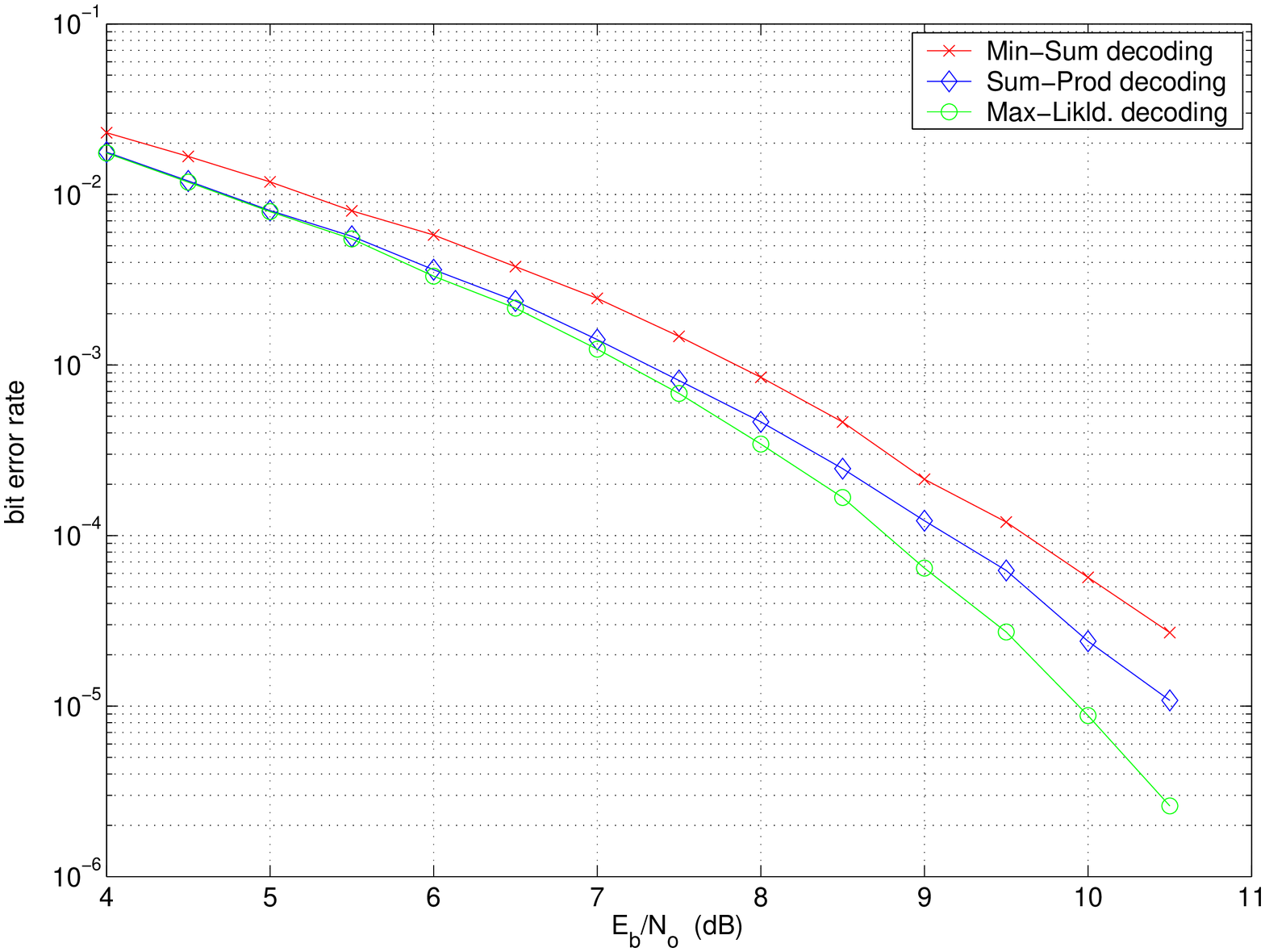}}}
          \caption{Performance of Example~\ref{ex3} - LDPC code for $m=10$: MS, SP, ML decoding over the BIAWGNC.}
\label{example3b_sim}
\end{figure}
\end{center}
}

\begin{figure}
\centering{\resizebox{4in}{3.2in}{\includegraphics{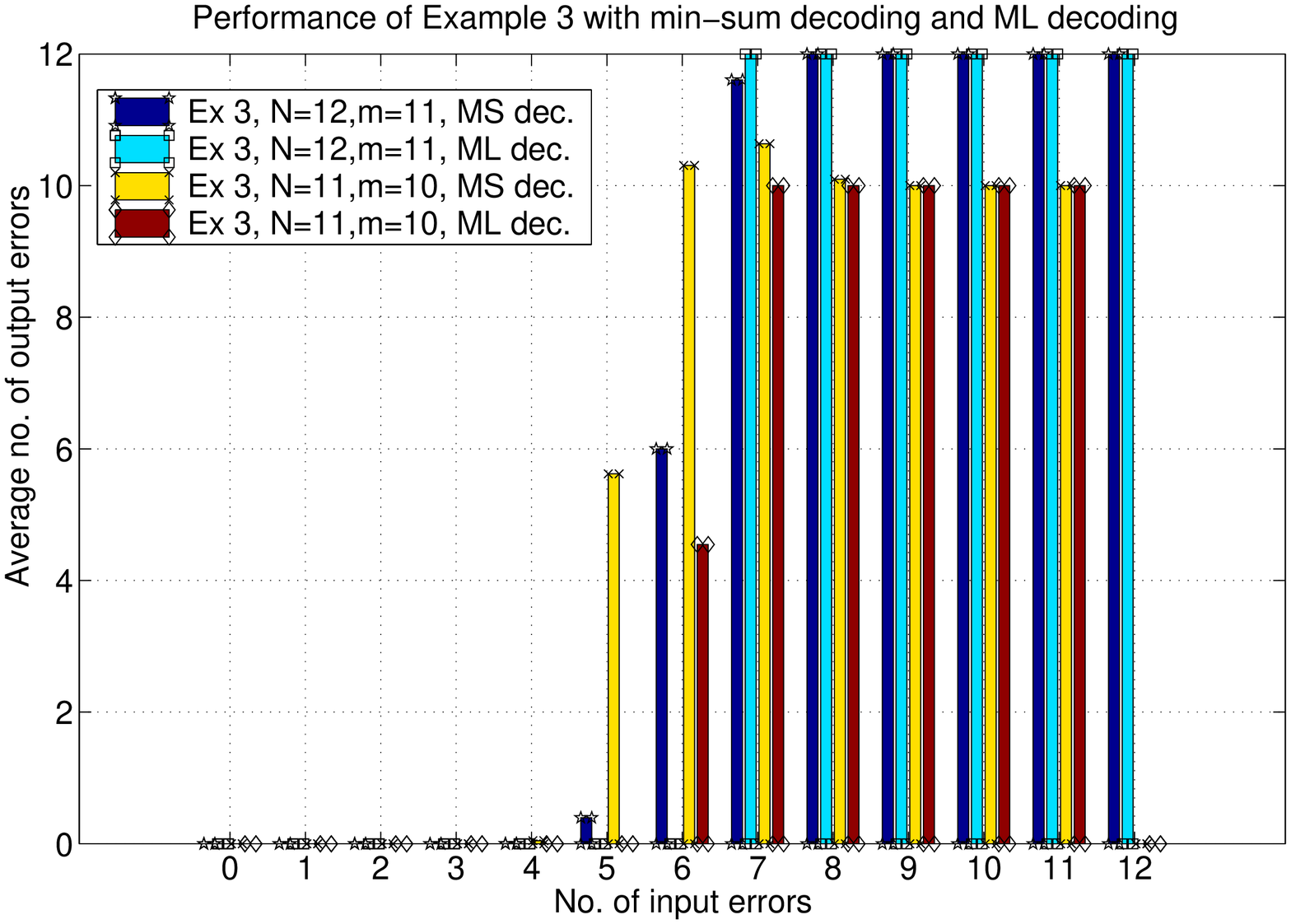}}}
\caption{Performance of Example~\ref{ex3}-LDPC code for $m=11,m=12$ with MS
and ML decoding over the BSC.}
\label{example3_bsc}
\end{figure}

Figure~\ref{example3_bsc} shows the performance of Example 3 for $m=10$
and $m=11$ over the BSC channel with MS iterative decoding. Since there
are only $2^{m+1}$ different error patterns possible for the BSC channel,
the performance of MS decoding for each error pattern was determined and
the average number of output errors were computed. The figure shows that
all four-bit or less error patterns were corrected by the MS and the ML
decoders. However, the average number of output bit errors for five-bit
error patterns with MS decoding was around 0.48 for the $m=11$ code and
was around 5.55 for the $m=10$ code, while the ML decoder corrected all
five-bit error patterns for both the codes. The average number of output
bit errors for six-bit error patterns with MS decoding was 6 for the
$m=11$ code and 10.5 for the $m=10$ code, whereas the ML decoder corrected
all six bit errors for the $m=11$ code and yielded an average number of
output bit errors of 4.545 for the $m=10$ code. The figure also shows that
MS decoding is closer to ML decoding for the $m=10$ code than for the
$m=11$ code.

This section has demonstrated three particular LDPC constraint graphs
having different types of pseudocodewords, leading to different
performances with iterative decoding in comparison to optimal decoding. In
particular, we observe that the presence of low weight irreducible
nc-pseudocodewords, with weight relatively smaller than the minimum distance of the code,
can adversely affect the performance of iterative decoding.

\section{Structure of pseudocodewords}

This section examines the structure of lift-realizable
pseudocodewords and identifies some sufficient conditions for
certain pseudocodewords to  potentially cause the min-sum iterative
decoder to fail to converge to a codeword. Some of these conditions
relate to subgraphs of the base Tanner graph. We recall that we are
only considering the set of lift-realizable pseudocodewords and that
by Definition~\ref{pscw_defn}, the pseudocodewords have non-negative
integer components, and hence are unscaled.

\vspace{0.15in}
\begin{lemma}
Let ${\bf p} = (p_1,p_2,\dots,p_n)$ be a pseudocodeword in the graph $G$
that represents the LDPC code $\mathcal{C}$. Then the vector ${\bf x} =
{\bf p} \mbox{ mod } 2$, obtained by reducing the entries in ${\bf p}$,
modulo 2, corresponds to a codeword in $\mathcal{C}$.
\label{pmod2_lemma}
\end{lemma}\vspace{-0.0in}
\vspace{0.15in}

The following implications follow from the above lemma:
\bi
\item If a pseudocodeword ${\bf p}$ has at least one odd component, then
it has at least $d_{\min}$ odd components.
\item If a pseudocodeword ${\bf p}$ has a support size $|\operatorname{supp}({\bf p})|<
d_{\min}$, then it has no odd components.
\item If a pseudocodeword ${\bf p}$ does not contain the support
  of any non-zero codeword in
its support, then ${\bf p}$ has no odd components.
\ei

\vspace{0.15in}
\begin{lemma} A pseudocodeword ${\bf p} = (p_1,\dots,p_n)$ can be written
as ${\bf p} = {\bf c}^{(1)}+{\bf c}^{(2)} +\dots + {\bf c}^{(k)} + {\bf
r}$, where ${\bf c}^{(1)}, \dots, {\bf c}^{(k)}$, are $k$ (not necessarily
distinct) codewords and ${\bf r}$ is some residual vector, not
containing the support of any nonzero codeword in its support, that remains after subtracting the codeword vectors ${\bf
c}^{(1)}, \dots, {\bf c}^{(k)}$ from ${\bf p}$. Either ${\bf r}$ is the
all-zeros vector, or ${\bf r}$ is a vector comprising of $0$ or even
entries only.
\label{pscwresidue_even_lemma}
\end{lemma}
\vspace{0.15in}

This lemma describes a particular composition of a pseudocodeword ${\bf
p}$. Note that the above result does not claim that ${\bf p}$ is reducible
even though the {\em vector} ${\bf p}$ can be written as a sum of codeword
vectors ${\bf c}^{(1)},\dots,{\bf c}^{(k)}$, and ${\bf r}$. Since ${\bf
r}$ need not be a pseudocodeword, it is not necessary that ${\bf p}$ be
reducible {\em structurally} as a sum of codewords and/or pseudocodewords
(as in Definition~\ref{pscw_irred_defn}). It is also worth noting that the decomposition of a
pseudocodeword, even that of an irreducible pseudocodeword, is not unique.

\vspace{0.15in}
\begin{example} For representation B of the $[7,4,3]$ Hamming code as shown in Figure~\ref{rep_Hmg} in Section 6,
label the vertices clockwise from the top as $v_1,v_2, v_3,v_4,v_5,v_6$, and $v_7$.
The vector ${\bf p}=(p_1,\ldots,p_n) = (1,2,1,1,1,0,2)$ is an irreducible pseudocodeword and
may be decomposed as ${\bf p} = (1,0,1,0,0,0,1)+(0,0,0,1,1,0,1)+(0,2,0,0,0,0,0)$ and
also as ${\bf p} = (1,0,1,1,1,0,0) + (0,2,0,0,0,0,2)$. In each of these decompositions,
each vector in the sum is a codeword except for the last vector which is the residual vector ${\bf r}$.
\end{example}
\vspace{0.15in}

\begin{theorem} Let ${\bf p} = (p_1,\dots,p_n)$ be a pseudocodeword. If there is a decomposition of
${\bf p}$ as in Lemma~\ref{pscwresidue_even_lemma} such that ${\bf r} = {\bf 0}$,
then ${\bf p}$ is a good pseudocodeword as in Definition~\ref{good_pscw_defn}.
\label{rzero_good_thm}
\end{theorem}
\vspace{0.15in}

\begin{theorem} The following are sufficient conditions for a pseudocodeword ${\bf p} = (p_1,\dots,p_n)$ to be bad,
as in Definition~\ref{bad_pscw_defn}:
\begin{enumerate}
\item $w^{BSC/AWGN}({\bf p}) < d_{\min}$.
\item $|\operatorname{supp}({\bf p})|< d_{\min}$.
\item If ${\bf p}$ is an irreducible nc-pseudocodeword and $|\operatorname{supp}({\bf p})|\ge \ell + 1$, where $\ell$ is the number of distinct codewords whose support is contained in $\operatorname{supp}({\bf p})$.
\end{enumerate}
\label{bad_conditions_thm}
\end{theorem}
\vspace{0.15in}

Intuitively, it makes sense for good pseudocodewords, i.e., those
pseudocodewords that are not problematic for iterative decoding, to have a
weight larger than the minimum distance of the code, $d_{\min}$. However,
we note that bad pseudocodewords can also have weight larger than
$d_{\min}$.

\vspace{0.15in}
\begin{definition}   A stopping set $S$ {\em has property $\Theta$} if  $S$ contains at least one pair of variable nodes $u$ and $v$ that are not connected by any path that traverses only via degree two check nodes in the subgraph $G_{|_S}$ of $G$ induced by $S$ in $G$.
\end{definition}
\vspace{0.15in}

\begin{figure}
\centering{\resizebox{2.8in}{0.45in}{\includegraphics{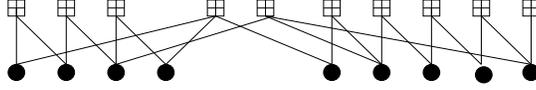}}}
\caption{A minimal stopping set with property $\Theta$.}
\label{min_sset_propX}
\end{figure}

\begin{example} In Figure~\ref{example2_fig} in Section 4, the set
  $\{v_1,v_2,v_4\}$ is a minimal stopping set and does not have
  property $\Theta$, whereas the set
  $\{v_1,v_3,v_4,v_5,v_6,v_7,v_{10},v_{11},v_{12},v_{13} \}$ is
  not minimal but has property $\Theta$. The graph in
  Figure~\ref{min_sset_propX} is a minimal stopping set
  that has property $\Theta$. The graph in Example~\ref{ex1} has  no minimal stopping sets with property $\Theta$, and all
stopping sets have size at least the minimum distance $d_{\min}$.
\end{example}
\vspace{0.15in}

\vspace{0.15in}
\begin{lemma} Let $S$ be a stopping set in $G$. Let $t_S$ denote the
  largest component an irreducible pseudocodeword with support
  $S$ may have in $G$. If $S$ is a minimal stopping set and does not have property $\Theta$, then a pseudocodeword with
support $S$ has maximal component 1 or 2. That is, $t_S =1$ or
$2$.
\label{partial_t_thm}
\end{lemma}
\vspace{0.15in}

Subgraphs of the LDPC constraint graph may also give rise to bad
pseudocodewords, as indicated below. \vspace{0.15in}

\begin{definition} A variable node $v$ in an LDPC constraint
  graph $G$ is said to be {\em problematic} if there is a
  stopping set $S$ containing $v$ that is not minimal but
  nevertheless has  no proper stopping set $S' \subsetneq S$  for
  which $v \in S'$.
\label{problematic_node_def}
\end{definition}
\vspace{0.15in}

Observe that all graphs in the examples of Section 4 have problematic nodes and conditions 1 and 3
in Theorem~\ref{bad_conditions_thm} are met in  Examples~\ref{ex2} and ~\ref{ex3}. The problematic nodes are
the variable nodes in the inner ring in Example~\ref{ex1}, the nodes  $v_7, v_8$ in Example~\ref{ex2},
and $v_1$ in Example~\ref{ex3}. Note that if a graph $G$ has a problematic node, then
$G$ necessarily contains a stopping set with property $\Theta$.

The following result classifies bad nc-pseudocodewords,
with respect to the AWGN channel, using the graph structure of
the underlying pseudocodeword supports, which, by
Lemma~\ref{supportpscw_sset_lemma}, are stopping sets
in the LDPC constraint graph.

\vspace{0.15in}
\begin{theorem}
Let $G$ be an LDPC constraint graph representing an LDPC code
$\mathcal{C}$, and let $S$ be a stopping set in $G$. Then, the following hold:
\begin{enumerate}
\item If there is no non-zero codeword in $\mathcal{C}$ whose support is contained in $S$, then all nc-pseudocodewords of $G$, having support
 equal to $S$, are bad as in Definition~\ref{bad_pscw_defn}. Moreover, there exists a bad pseudocodeword in $G$
  with support equal to $S$.
\item  If there is at least one codeword ${\bf c}$ whose support
  is contained in $S$, then we have the following cases:\begin{enumerate}
    \item[(a)] if $S$ is minimal,
      \begin{enumerate} \item[(i)] there exists a nc-pseudocodeword ${\bf p}$ with support equal to $S$ iff $S$
      has property $\Theta$.
      \item[(ii)] all nc-pseudocodewords with support
        equal to $S$ are bad.
        \end{enumerate}
     \item[(b)] if $S$ is not minimal,
       \begin{enumerate} \item[(i)] and $S$ contains a problematic
         node $v$ such that  $v\notin S'$ for any proper stopping set\footnote{A proper stopping set $S'$ of $S$ is a non-empty stopping set that is a
strict subset of $S$.}
         $S'\subsetneq S$, then there exists a bad pseudocodeword ${\bf
           p}$ with support $S$. Moreover, any irreducible
         nc-pseudocodeword ${\bf p}$ with support $S$ is bad.
\item[(ii)] and $S$ does not contain any problematic nodes, then
 every variable node in $S$ is contained in a minimal stopping
 set within $S$. Moreover,
  there exists a bad nc-pseudocodeword with support $S$
  iff either one of these minimal stopping sets is not the support of
  any non-zero codeword in $\mathcal{C}$ or one of these minimal
  stopping sets has property $\Theta$.
\end{enumerate}
\end{enumerate}
\end{enumerate}
\label{main_theorem}
\end{theorem}
\vspace{0.15in}

\begin{figure}
\centering{\resizebox{2.2in}{0.6in}{\includegraphics{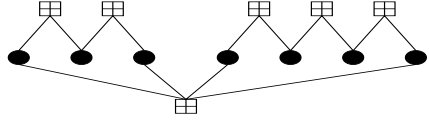}}}
\caption{A non-minimal stopping set as in case 2(b)(ii) of Theorem~\ref{main_theorem}.}
\label{nonmin_sset_fig}
\end{figure}

The graph in Figure~\ref{nonmin_sset_fig} is an example of case
2(b)(ii)  in Theorem~\ref{main_theorem}. Note that the stopping
set in the figure is a disjoint union of two codeword supports
and therefore, there are no irreducible nc-pseudocodewords.

\begin{figure}
\centering{\resizebox{2.2in}{0.8in}{\includegraphics{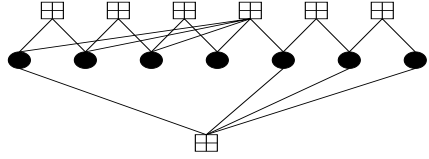}}}
\caption{A minimal stopping set as in case 2(a) of Theorem~\ref{main_theorem}.}
\label{min_sset_noprob}
\end{figure}

The graph in Figure~\ref{min_sset_noprob} is an example of case
2(a). The graph has property $\Theta$ and therefore has
nc-pseudocodewords, all of which are bad.

\subsection{Remarks on the weight vector and channels}
In \cite{frey}, Frey et. al show that the max-product iterative
decoder (equivalently, the MS iterative decoder) will always
converge to an irreducible pseudocodeword (as in
Definition~\ref{pscw_irred_defn}) on the AWGN channel.
However, their result does not explicitly show that for a given
irreducible pseudocodeword ${\bf p}$, there is a weight vector
${\bf w}$ such that the cost ${\bf p}{\bf w}^T$ is the smallest among
all possible pseudocodewords. In the previous subsection, we have
given sufficient conditions under which such a weight vector can
explicitly be found for certain irreducible pseudocodewords. We
believe, however, that finding such a weight vector ${\bf w}$ for any
irreducible pseudocodeword ${\bf p}$ may not always be possible.
In particular, we state the following definitions and results.

\vspace{0.15in}
\begin{definition}
A {\em truncated AWGN} channel, parameterized by $L$ and denoted by $TAWGN(L)$, is an AWGN
channel whose output log-likelihood ratios corresponding to the
received values from the channel are truncated, or limited, to the interval $[-L,L]$.
\label{awgn_t_l}
\end{definition}
\vspace{0.15in} \indent In light of \cite{feldman2,
isit05_pascal_feldman}, we believe that there are fewer problematic
pseudocodewords on the BSC than on the truncated AWGN channel or the
AWGN channel.

\vspace{0.15in}
\begin{definition}
For an LDPC constraint graph $G$ that defines an LDPC code
$\mathcal{C}$, let $P^B_{AWGN}(G)$ be the set of lift-realizable
pseudocodewords of $G$ where for each pseudocodeword ${\bf p}$ in the set,
there exists a weight vector ${\bf w}$ such that the cost ${\bf p}{\bf
  w}^T$ on the AWGN channel is the smallest among all possible
lift-realizable pseudocodewords in $G$.
\label{pb_awgn}
\end{definition}
\vspace{0.15in}

Let $P^B_{BSC}(G)$ and $P^B_{TAWGN(L)}(G)$ be defined analogously
for the BSC and the truncated AWGN channel, respectively. Then, we
have the following result: \vspace{0.15in}
\begin{theorem}
For an LDPC constraint graph $G$, and $L\ge 1$, we have
\[P^B_{BSC}(G)\subseteq P^B_{TAWGN(L)}(G)\subseteq P^B_{AWGN}(G). \]
\label{bad_pscw_theorem}
\end{theorem}
\vspace{0.15in}

The above result says that there may be fewer problematic
irreducible pseudocodewords for the BSC than over the
TAWGN(L) channel and the AWGN channel. In other words, the above result implies that
MS iterative decoding may be more accurate for the BSC than
over the AWGN channel. Thus, quantizing
or truncating the received information from the
channel to a smaller interval before performing MS iterative
decoding may be beneficial. (Note that while the above result considers all possible weight vectors that can occur for a given channel,
it does not take into account the probability distribution of weight vectors for the different channels, which is essential
when comparing the performance of MS decoding across different channels.) Since the set of lift-realizable
pseudocodewords for MS iterative decoding is the set of
pseudocodewords for linear-programming (LP) decoding (see Section 2), the same
analogy carries over to LP decoding as well. Indeed, at high
enough signal to noise ratios,  the above observation has been shown true for
the case of LP decoding in \cite{feldman2}
and more recently in \cite{isit05_pascal_feldman}.

\section{Graph Representations and Weight Distribution}

In this section, we examine different representations of individual LDPC codes and
analyze the weight distribution of lift-realizable pseudocodewords in each representation and how it
affects the performance of the MS iterative decoder. We use the classical
$[7,4,3]$ and $[15,11,3]$ Hamming codes as examples.

{\begin{center}
\begin{figure}
\centering{
     \begin{minipage}[b]{0.30\linewidth} 
            \centering
                   {
            \resizebox{1.2in}{0.9in}{\includegraphics{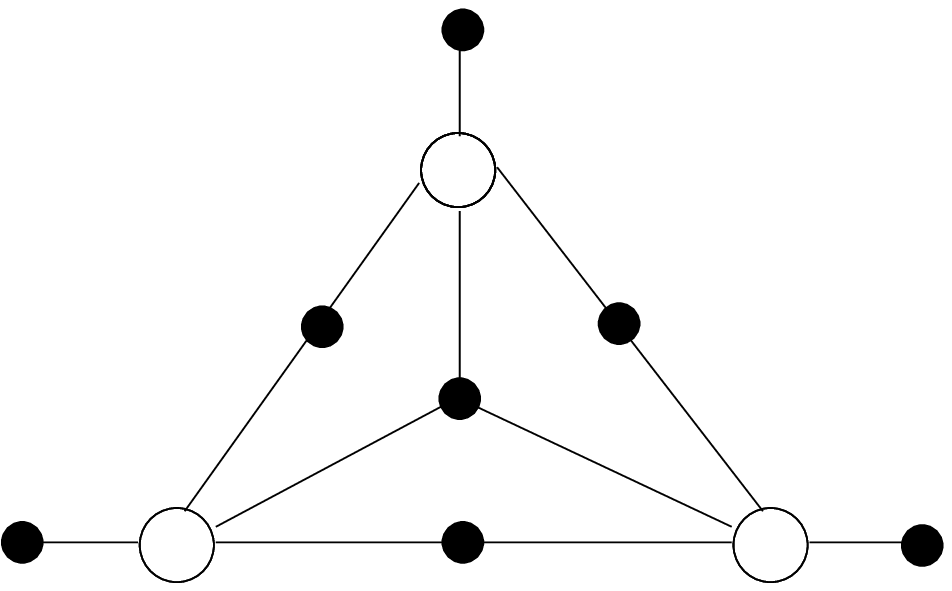}}}
          \centerline{Representation A}\label{rep_A}
         \end{minipage}
            \hspace{0.1in}
            \begin{minipage}[b]{0.30\linewidth}
            \centering{
\resizebox{1.2in}{0.9in}{\includegraphics{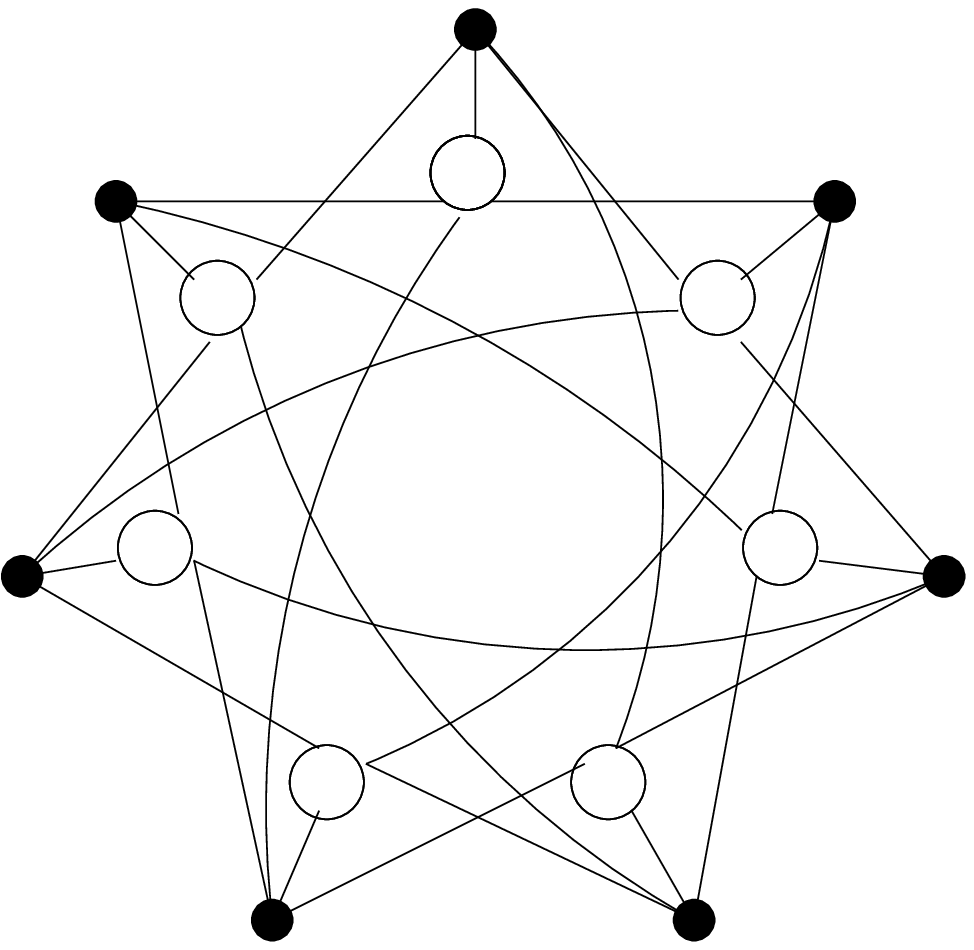}}}
          \centerline{Representation B}\label{rep_B}
            \end{minipage}
            \hspace{0.1in}
            \begin{minipage}[b]{0.30\linewidth}
            \centering{
              \resizebox{1.2in}{0.9in}{\includegraphics{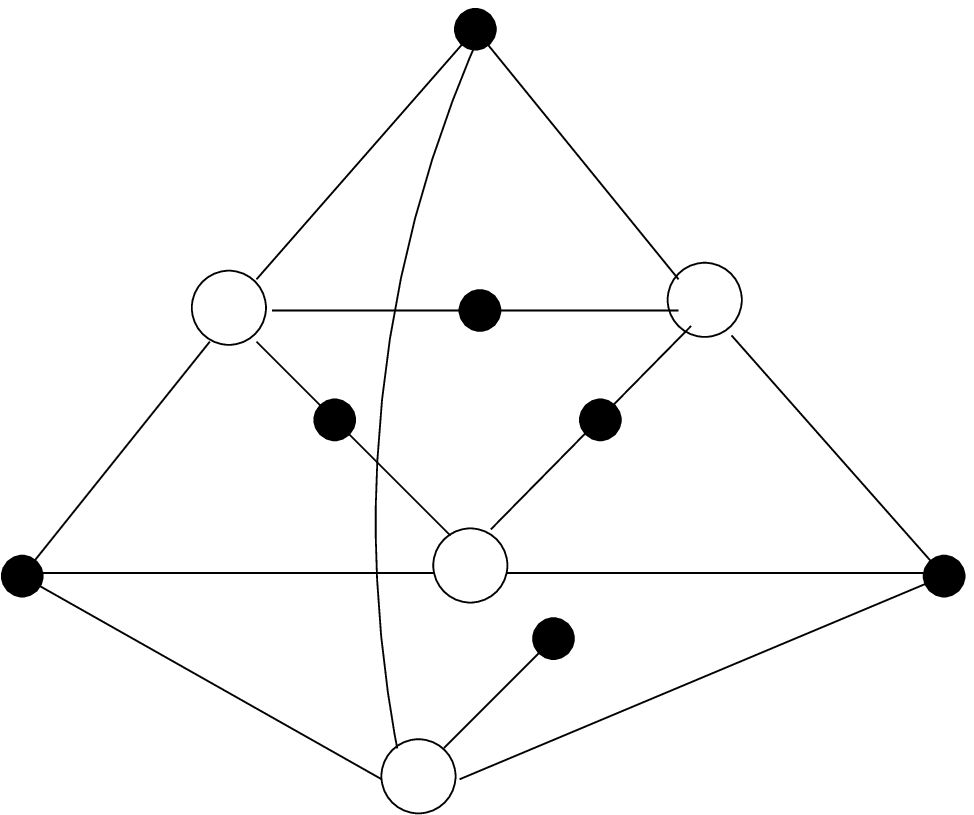}}}
          \centerline{Representation C}\label{rep_C}
            \end{minipage}

}
\caption{Three different representations of the [7,4,3] Hamming code.}
\label{rep_Hmg}
\end{figure}
\end{center}
}
Figure~\ref{rep_Hmg} shows three different graph representations of the
$[7,4,3]$ Hamming code. We will call the representations $A$, $B$, and $C$, and
moreover, for convenience, also refer to the graphs in the three respective
representations as $A$, $B$, and $C$. The graph $A$ is based on the systematic
parity check matrix representation of the $[7,4,3]$ Hamming code and hence,
contains three
degree one variable nodes, whereas the graph $B$ has no degree one nodes and
is more structured (it results in a circulant parity check matrix) and contains
4 redundant check equations compared to $A$, which has none, and $C$, which
has one. In particular, $A$ and $C$ are subgraphs of $B$, with the same set of
variable nodes. Thus, the set of lift-realizable pseudocodewords of $B$ is contained in the
set of lift-realizable pseudocodewords of $A$ and $C$, individually. Hence, $B$ has fewer
number of lift-realizable pseudocodewords than $A$ or $C$. In particular, we state the
following result:

\vspace{0.15in}
\begin{theorem}
The number of lift-realizable pseudocodewords in an LDPC graph $G$ can only reduce
with the addition of redundant check nodes to $G$.
\label{redundancy_theorem}
\end{theorem}
\vspace{0.15in}

The proof is obvious since with the introduction of new check nodes
in the graph, some previously valid pseudocodewords may not satisfy the
new set of inequality constraints imposed by the new check nodes. (Recall that at a
check node $c$ having variable node
neighbors $v_{i_1},\dots,v_{i_k}$, a pseudocodeword
${\bf p}=(p_1,\dots,p_n)$, must satisfy the following inequalities
$p_{i_j}\le\sum_{h\ne j, h=1,\dots,k}p_{i_h}, \mbox{ for }j=1,\dots,k$
(see equation (\ref{pscw_ineq})).)
However, the set of valid codewords in the graph remains the same, since we are
introducing only redundant (or, linearly dependent) check nodes.
Thus, a graph with more check nodes can only have fewer number of lift-realizable pseudocodewords
and possibly a better pseudocodeword-weight distribution.

If we add all possible redundant check nodes to the graph, which, we note, is an
exponential number in the number of linearly
dependent rows of the parity check matrix
of the code, then the resulting graph would have the smallest number of
lift-realizable pseudocodewords among all possible representations of the code.
If this graph does not have any
bad nc-pseudocodewords (both lift-realizable ones and those arising on the computation tree)
then the performance obtained with iterative decoding
is the same as the optimal ML performance.

\vspace{0.15in}
\begin{remark}{\rm Theorem~\ref{redundancy_theorem} considers only the set of
lift-realizable pseudocodewords of a Tanner graph. On adding
redundant check nodes to a Tanner graph, the shape of the
computation tree is altered and thus, it is possible that some new
pseudocodewords arise in the altered computation tree, which can
possibly have an adverse effect on iterative decoding. The $[4,1,4]$
repetition code example from Section 2.C illustrates this. Iterative
decoding is optimal on the single cycle representation of this code.
However, on adding a degree four redundant check node, the iterative
decoding performance deteriorates due to the introduction of bad
pseudocodewords to the altered computation tree. (See
Figure~\ref{rep4_1_4}.) (The set of lift-realizable pseudocodewords
however remains the same for the new graph with redundant check
nodes as for the original graph.)}
\end{remark}
\vspace{0.15in}

\begin{figure}
\centering{\resizebox{4.5in}{3.5in}{\includegraphics{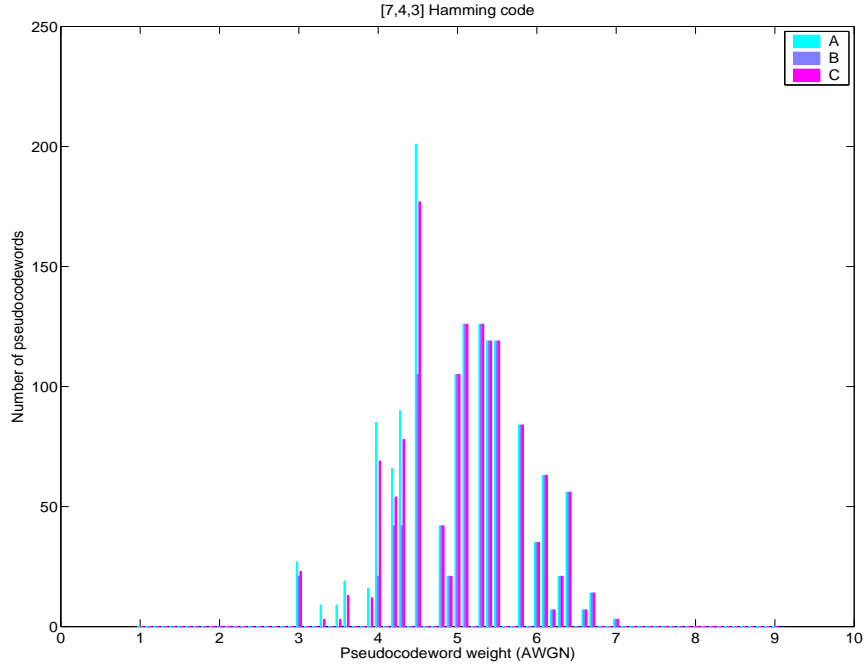}}}
\caption{Pseudocodeword-weight (AWGN) distribution of representations
A,B,C of the [7,4,3] Hamming code.}
\label{dist_awgn_hmg}
\end{figure}

Returning to the Hamming code example, graph $B$ can be obtained by adding edges to either $A$ or
$C$, and thus,
$B$ has more cycles than $A$ or $C$.
The distribution of the weights of the irreducible lift-realizable pseudocodewords for the three
graphs $A$, $B$, and $C$
is shown\footnote{The plots considered all pseudocodewords
in the three graphs that had a maximum component value of at most
3. Hence, for each codeword ${\bf c}$, ${2\bf c}$ and ${3\bf c}$
are also counted in the histogram, and each has weight at least
$d_{\min}$. However, each irreducible nc-pseudocodeword ${\bf p}$
is counted only once, as ${\bf p}$ contains at least one entry
greater than 1, and any nonzero multiple of ${\bf p}$ would have
a component greater than 3. The $t$-value (see Section 5) is 3 for the graphs $A$, $B$, and $C$ of the $[7,4,3]$ Hamming
code.} in Figure~\ref{dist_awgn_hmg}. (The distribution considers all irreducible pseudocodewords in the graph,
since irreducible pseudocodewords may potentially prevent the MS decoder to
converge to any valid codeword \cite{frey}.)
Although, all three graphs
have a pseudocodeword of weight three\footnote{Note that this pseudocodeword is a
valid codeword
in the graph and is thus a {\em good} pseudocodeword for iterative
decoding.}, Figure~\ref{dist_awgn_hmg} shows that
$B$ has most of its lift-realizable pseudocodewords of high weight, whereas $C$, and more
particularly, $A$,  have more
low-weight lift-realizable pseudocodewords. The corresponding weight distributions over
the BEC and the BSC channels are shown in Figure~\ref{bec_bsc_7}.
$B$ has a better weight distribution than $A$ and $C$ over these channels
as  well.

\vspace{0.15in}

{\begin{center}
\begin{figure}
\centering{
     \begin{minipage}[b]{0.46\linewidth} 
            \centering
                   {              \resizebox{2.2in}{1.8in}{\includegraphics{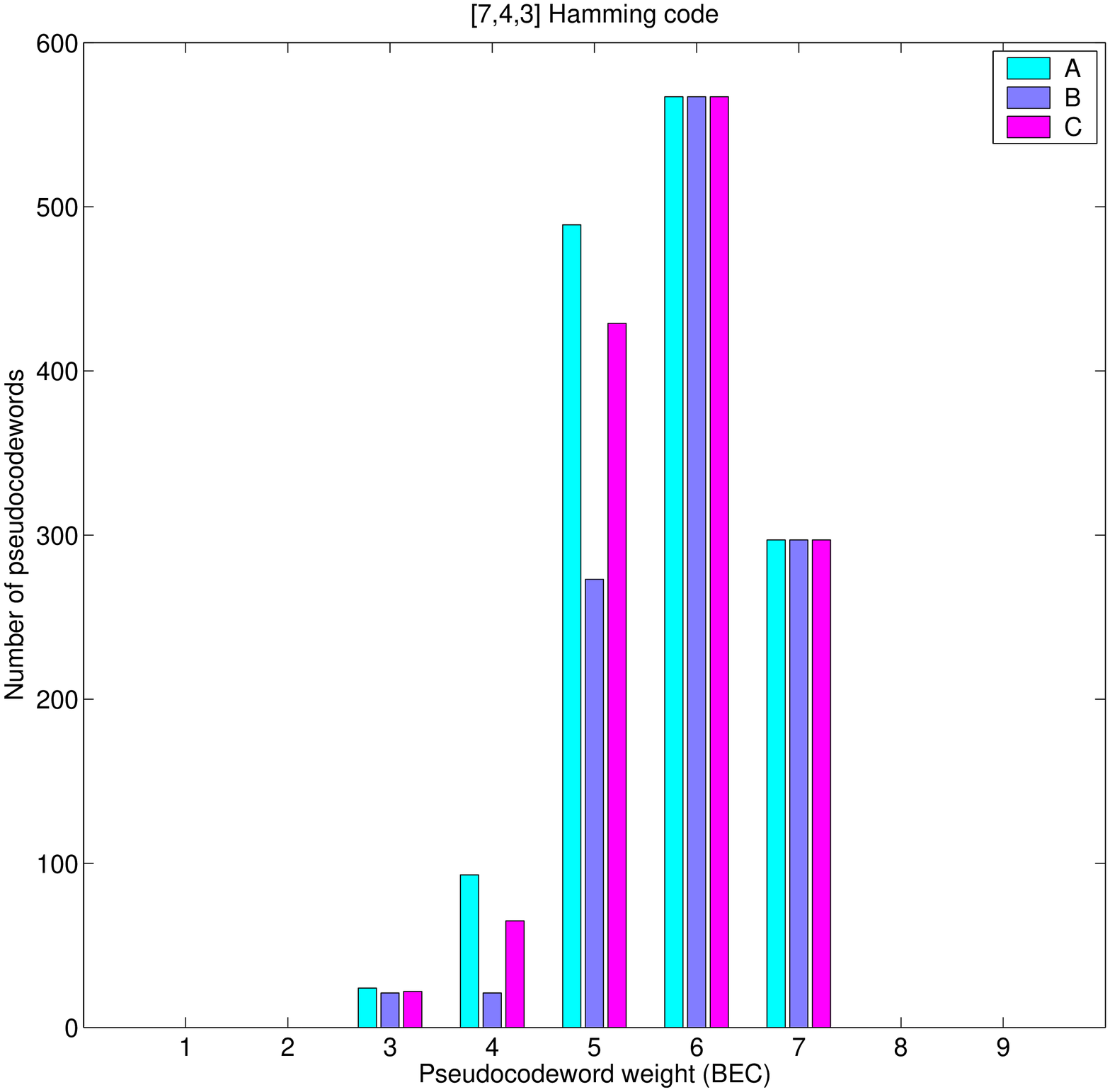}}}
          \newline\centerline{BEC}\label{bec_wt_7}
         \end{minipage}
            \hspace{0.2in}
            \begin{minipage}[b]{0.46\linewidth}
            \centering{
\resizebox{2.2in}{1.8in}{\includegraphics{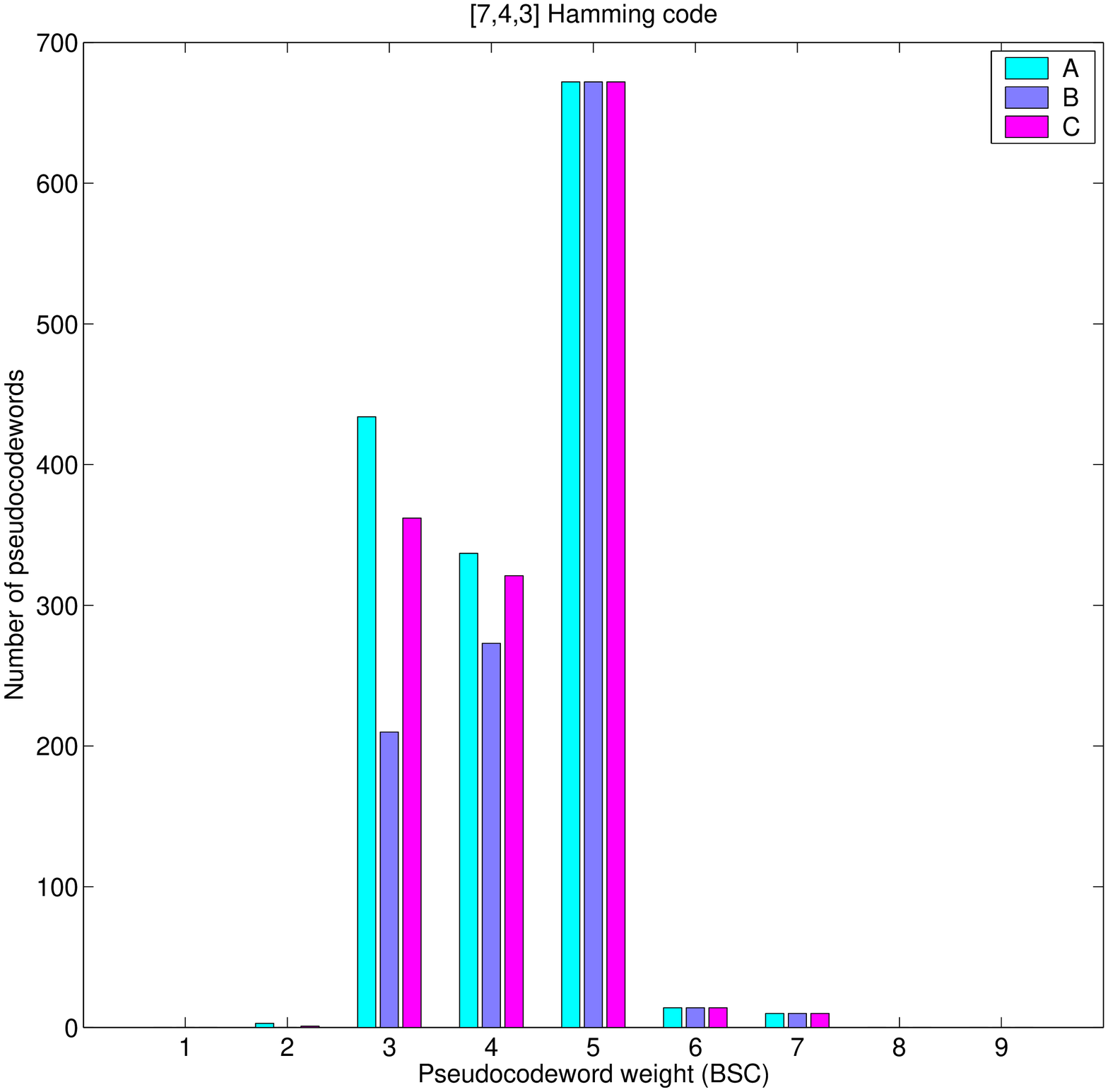}}}
          \newline\centerline{BSC}\label{bsc_wt_7}
            \end{minipage}
}
\caption{Pseudocodeword-weight distribution (BEC and BSC channels) of
representations $A$,
$B$, $C$ of the $[7,4,3]$ Hamming code.}
\label{bec_bsc_7}
\end{figure}
\end{center}
}

The performance of MS iterative decoding of $A$, $B$,
and $C$ on the BIAWGNC with signal to noise ratio  $E_b/N_o$
is shown in Figures~\ref{hmg_A}, \ref{hmg_B},
and \ref{hmg_C}, respectively. (The maximum number of decoding iterations was
fixed at 100.) The performance plots show
both the bit error rate and the frame error rate, and further, they also
distinguish between undetected decoding errors, that are caused
due to the decoder converging to an incorrect but valid codeword, and
detected errors, that are caused due to the decoder failing to converge to any valid
codeword within the maximum specified number of decoding iterations, 100 in this
case. The detected errors can be attributed to the decoder trying to converge to
an nc-pseudocodeword rather than to any valid codeword.

Representation $A$ has a significant detected error rate, whereas
representation $B$ shows no presence of detected errors at all. All errors
in decoding $B$ were due to the decoder converging to a wrong codeword. (We note
that an optimal ML decoder would yield a performance closest to that of
the iterative decoder on representation $B$.) This is interesting since
the graph $B$ is obtained by adding 4 redundant check nodes to the graph
$A$. The addition of these 4 redundant check nodes to the graph removes
most of the low-weight nc-pseudocodewords that were present in $A$. (We
note here that representation $B$ includes all possible redundant
parity-check equations there are for the [7,4,3] Hamming code.)
Representation $C$ has fewer number of pseudocodewords compared to
$A$. However, the set of irreducible pseudocodewords of $C$ is not a subset
of the set of irreducible pseudocodewords of $A$. The performance of iterative decoding on
representation $C$ indicates a small fraction of detected errors.

{\begin{center}
\begin{figure}[h]
\centering{
     \begin{minipage}[b]{0.48\linewidth} 
            \centering
                   {
            \resizebox{3.2in}{2.5in}{\includegraphics{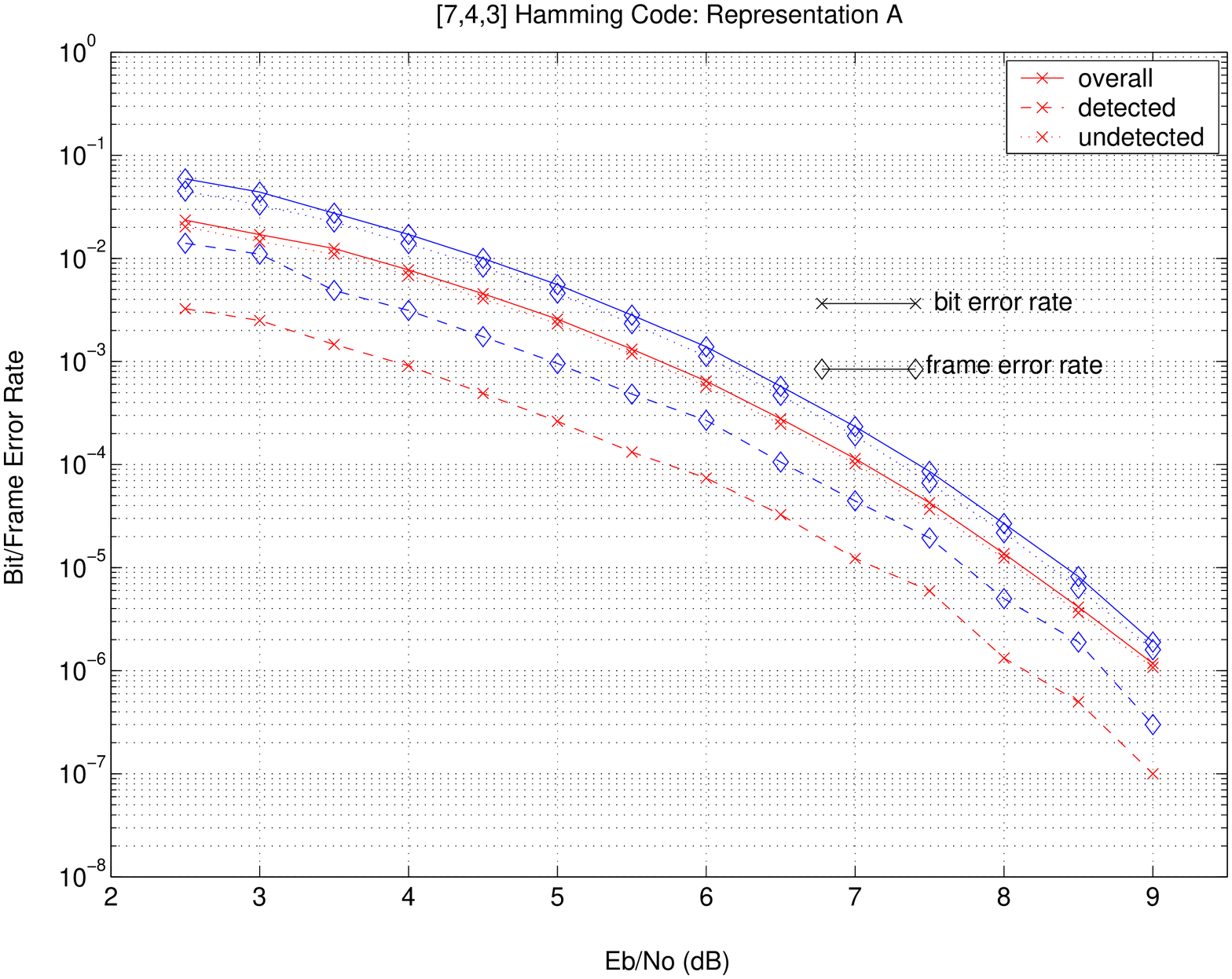}}}
          \caption{Representation A.}\label{hmg_A}
         \end{minipage}
            \hspace{0.1in}
            \begin{minipage}[b]{0.48\linewidth}
            \centering{
\resizebox{3.2in}{2.5in}{\includegraphics{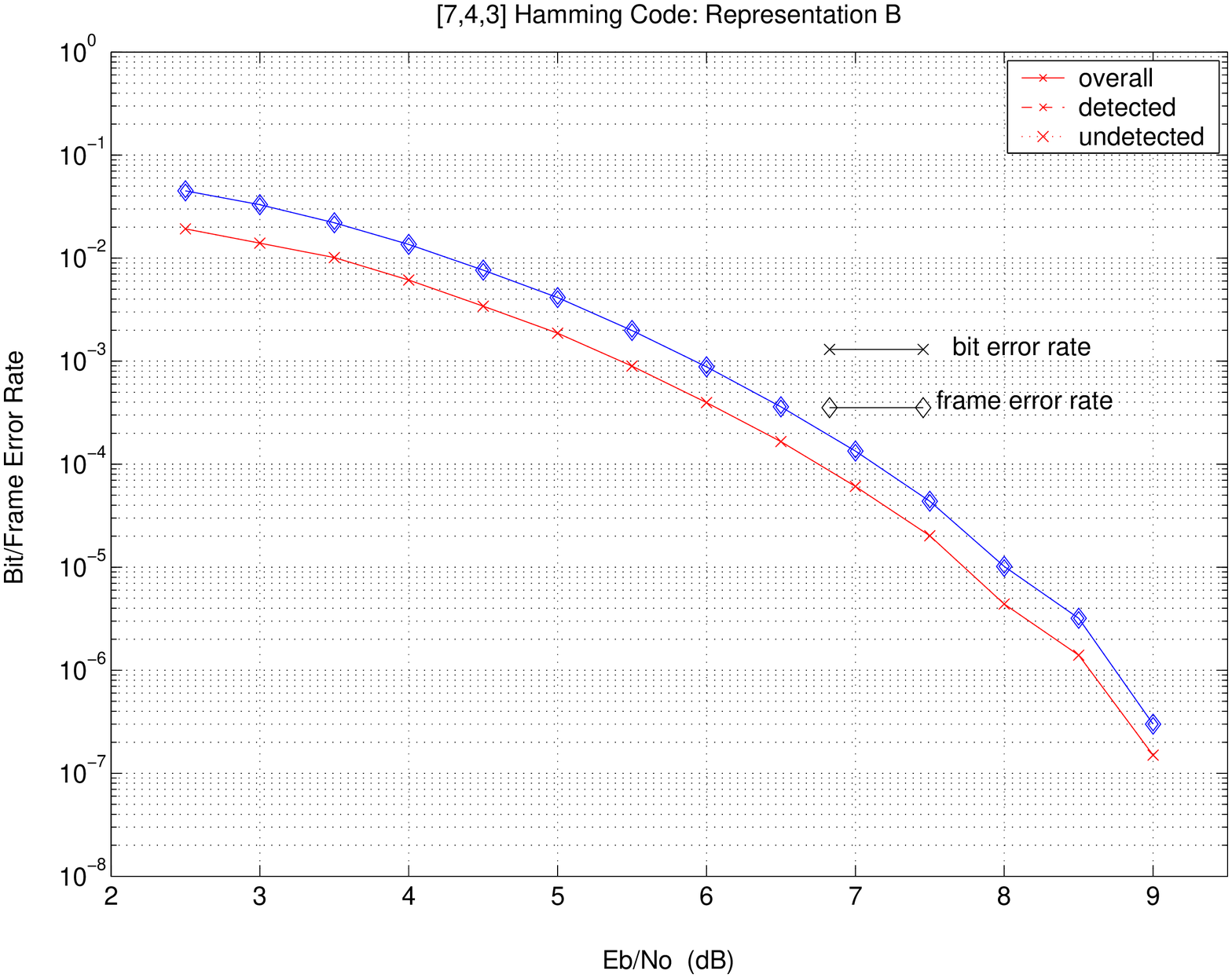}}}
          \caption{Representation B.}\label{hmg_B}
            \end{minipage}
}
\centering{
     \begin{minipage}[b]{0.48\linewidth} 
            \centering
                   {
            \resizebox{3.2in}{2.5in}{\includegraphics{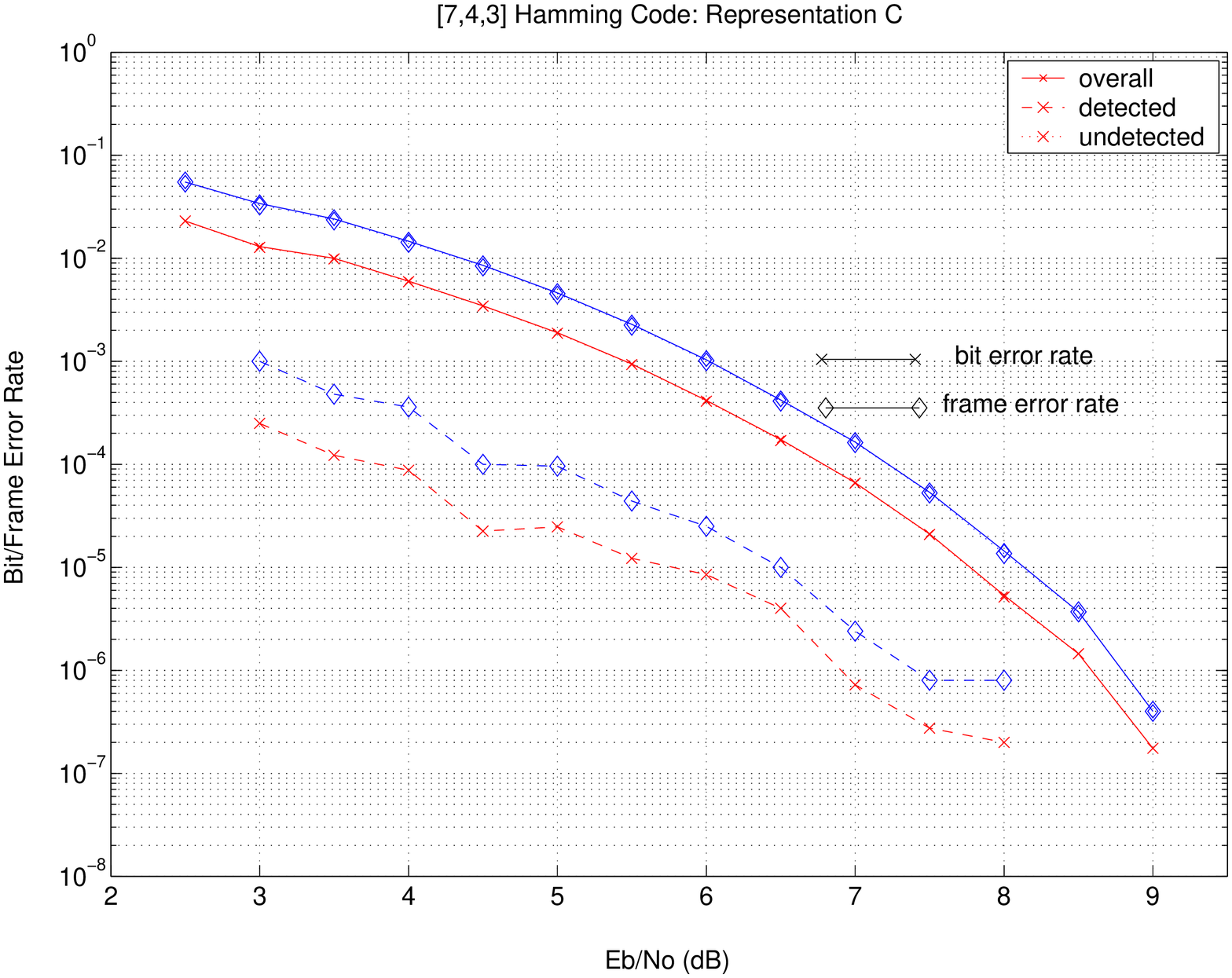}}}
          \caption{Representation C.}\label{hmg_C}
         \end{minipage}
            \hspace{0.1in}
            \begin{minipage}[b]{0.48\linewidth}
            \centering{
\resizebox{3.2in}{2.5in}{\includegraphics{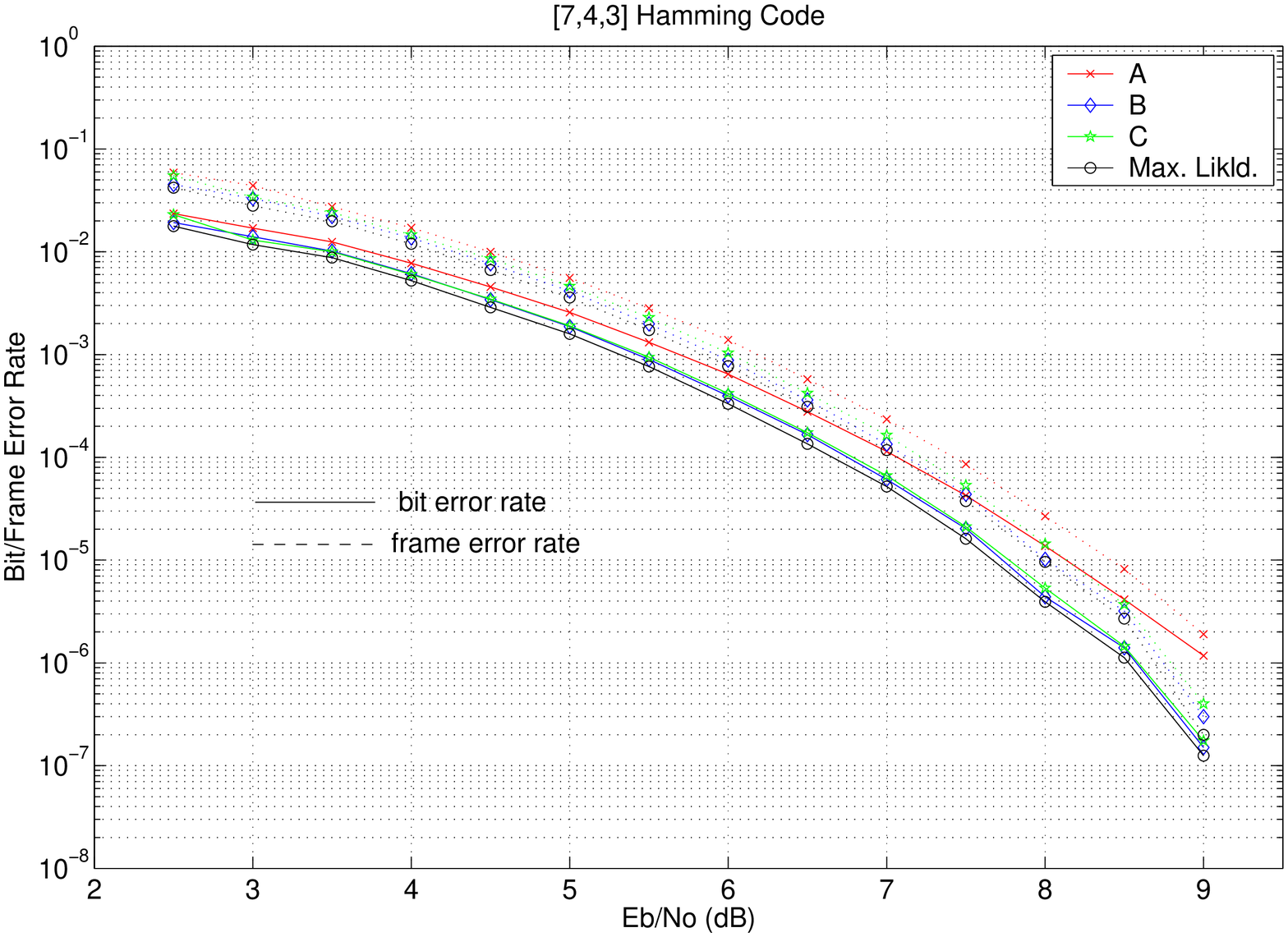}}}
          \caption{Comparison between representations.}\label{hmg_comp}
            \end{minipage}
}
\centerline{Performance of the [7,4,3] Hamming code with min-sum iterative decoding over the BIAWGNC.}
\end{figure}
\end{center}
}
Figure~\ref{hmg_comp} compares the performance of min-sum decoding on the
three representations. Clearly, $B$, having the best pseudocodeword weight
distribution among the three representations, yields the best performance
with MS decoding, with performance almost matching that of the optimal ML
decoder.

\begin{figure}
\centering{\resizebox{4in}{3.2in}{\includegraphics{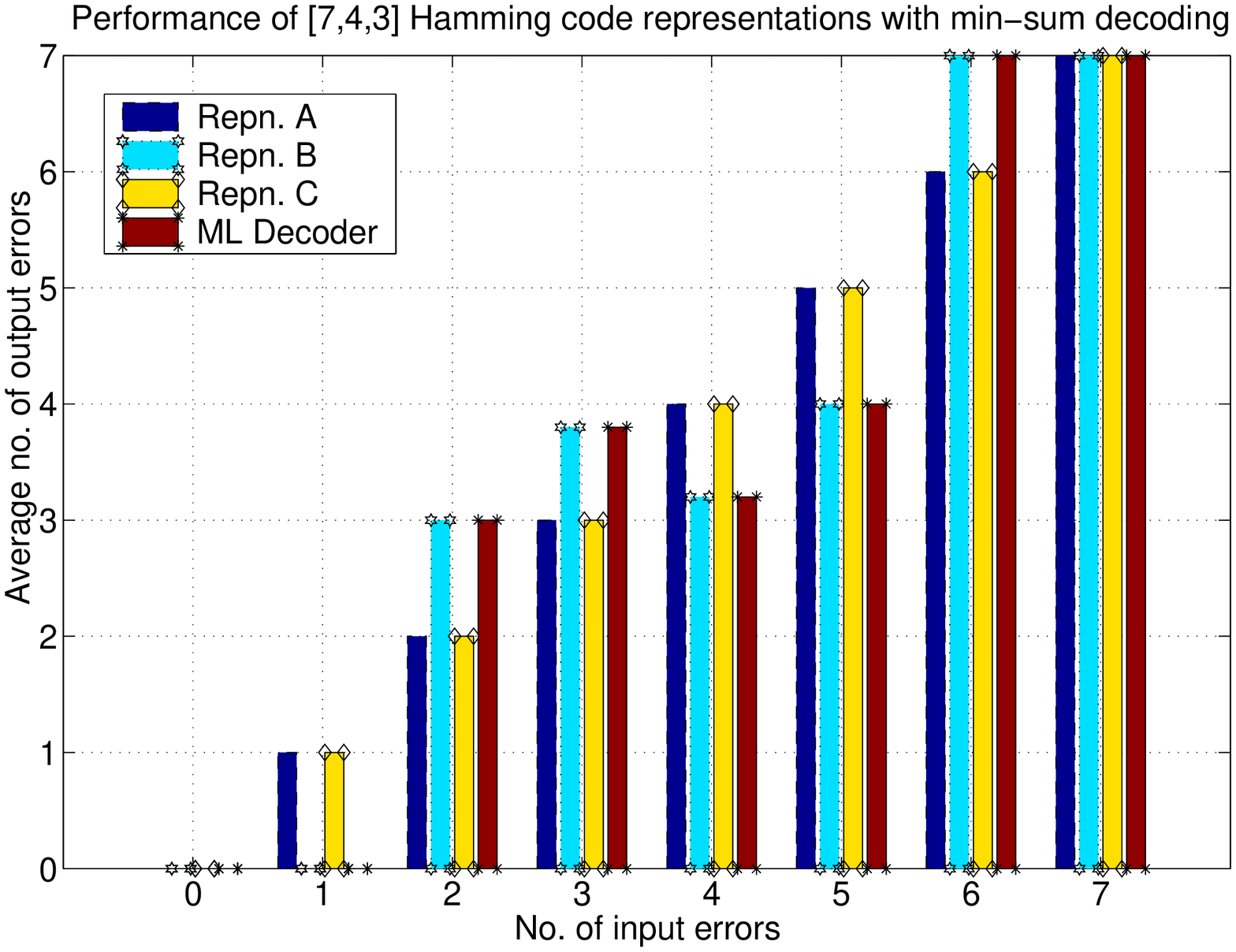}}}
\caption{Performance of the [7,4,3] Hamming code representations with MS
decoding over the BSC.}
\label{hmg_743_bsc}
\end{figure}

Figure~\ref{hmg_743_bsc} shows the performance of the three different
representations over the BSC channel with MS iterative decoding. Since
there are only $2^7=128$ different error patterns, the performance of MS
decoding for each error pattern was determined and the average number of
output errors were computed. Representations A and C failed to correct any
non-zero error pattern whereas representation B corrected all one-bit
error patterns. The performance of MS decoding using representation B was
identical to the performance of the ML decoder and the MS decoder always
converged to the ML codeword with representation B. This goes to show that
representation B is in fact the optimal representation for the BSC
channel.

Similarly, we also analyzed three different representations of the
$[15,11,3]$ Hamming code. Representation $A$ has its parity
check matrix in the standard systematic form and thus, the corresponding
Tanner graph has 4 variable nodes of degree one.  Representation $B$
includes all possible redundant parity check equations of representation
$A$ and has the best pseudocodeword-weight distribution.
Representation $C$ includes up to order-two redundant parity check
equations from the parity check matrix of representation $A$, meaning, the
parity check matrix of representation $C$ contained all linear
combinations of every pair of rows in the parity check matrix of
representation $A$. Thus, its (lift-realizable) pseudocodeword-weight distribution is
superior to that of $A$ but inferior to that of $B$. (See Figure \ref{hmg_comp_15}.)

The analogous performance of MS iterative decoding of representations
$A$, $B$, and $C$ of the $[15,11,3]$ Hamming code on a BIAWGNC with signal
to noise ratio $E_b/N_o$ is shown in Figures~\ref{hmg_A_15},
\ref{hmg_B_15}, and \ref{hmg_C_15}, respectively. (The maximum number of
decoding iterations was fixed at 100.) We observe similar trends in the performance curves as in the previous example.
$A$ shows a
prominent detected error rate, whereas $B$ and $C$ show no presence of
detected errors at all. The results suggest that merely adding order two
redundant check nodes to the graph of $A$ is sufficient to remove most of
the low-weight pseudocodewords.
{\begin{center}
\begin{figure}[h]
\centering{
     \begin{minipage}[b]{0.48\linewidth} 
            \centering
                   {
            \resizebox{3.2in}{2.5in}{\includegraphics{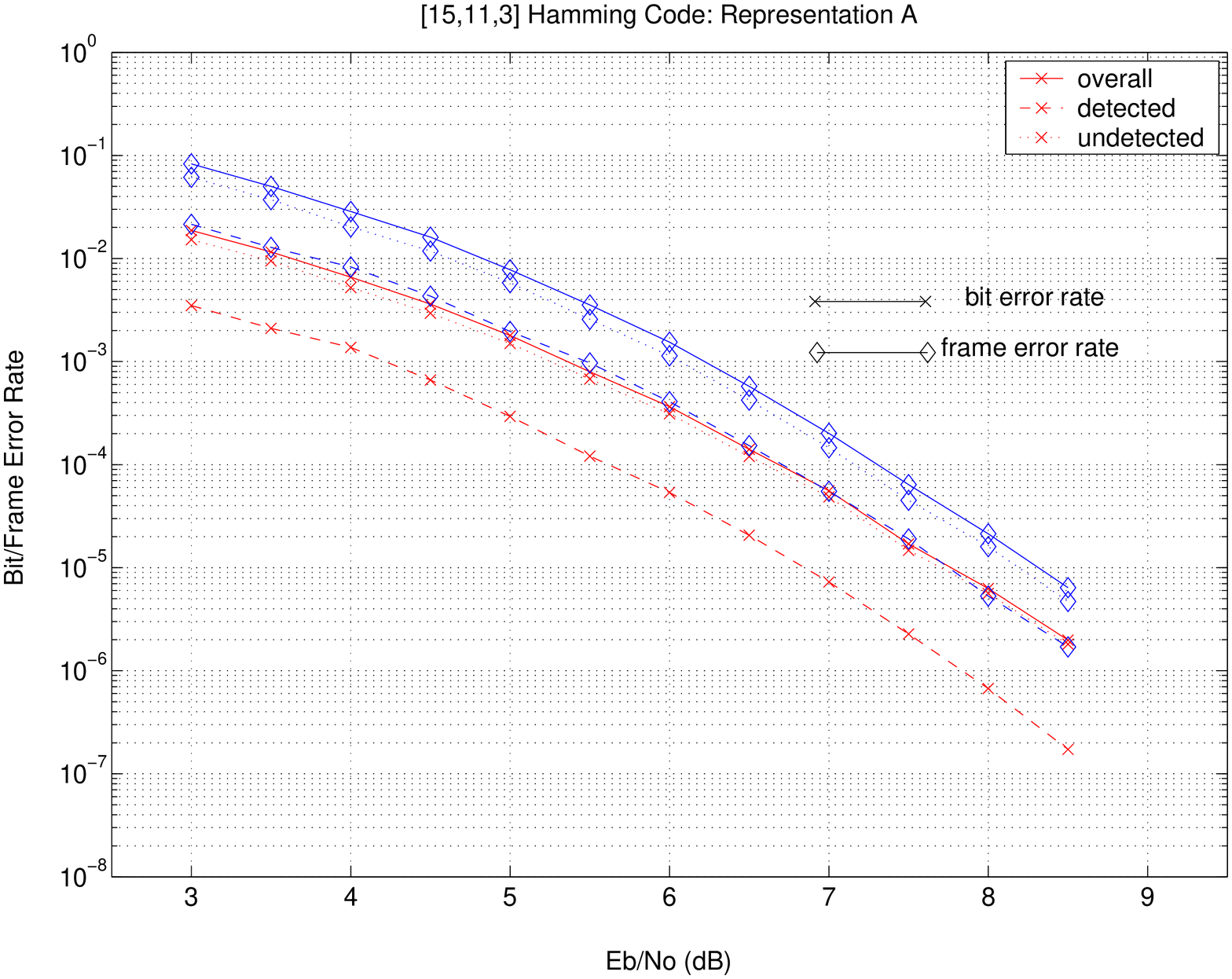}}}
          \caption{Representation A.}\label{hmg_A_15}
         \end{minipage}
            \hspace{0.1in}
            \begin{minipage}[b]{0.48\linewidth}

\centering{\resizebox{3.2in}{2.5in}{\includegraphics{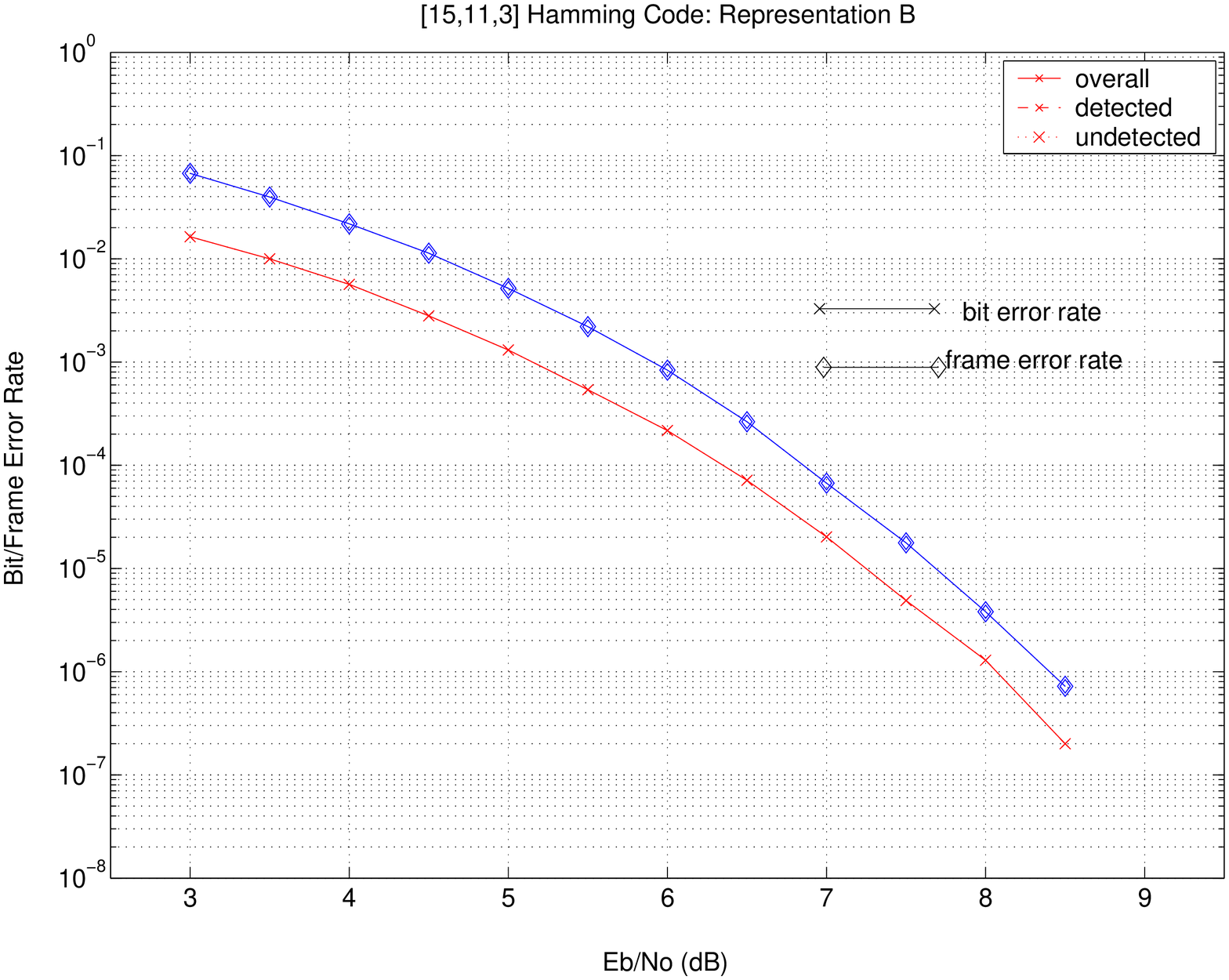}}}
          \caption{Representation B.}\label{hmg_B_15}
            \end{minipage}
}
\vspace{0.15in}
\centering{
     \begin{minipage}[b]{0.48\linewidth} 
            \centering
                   {
            \resizebox{3.2in}{2.5in}{\includegraphics{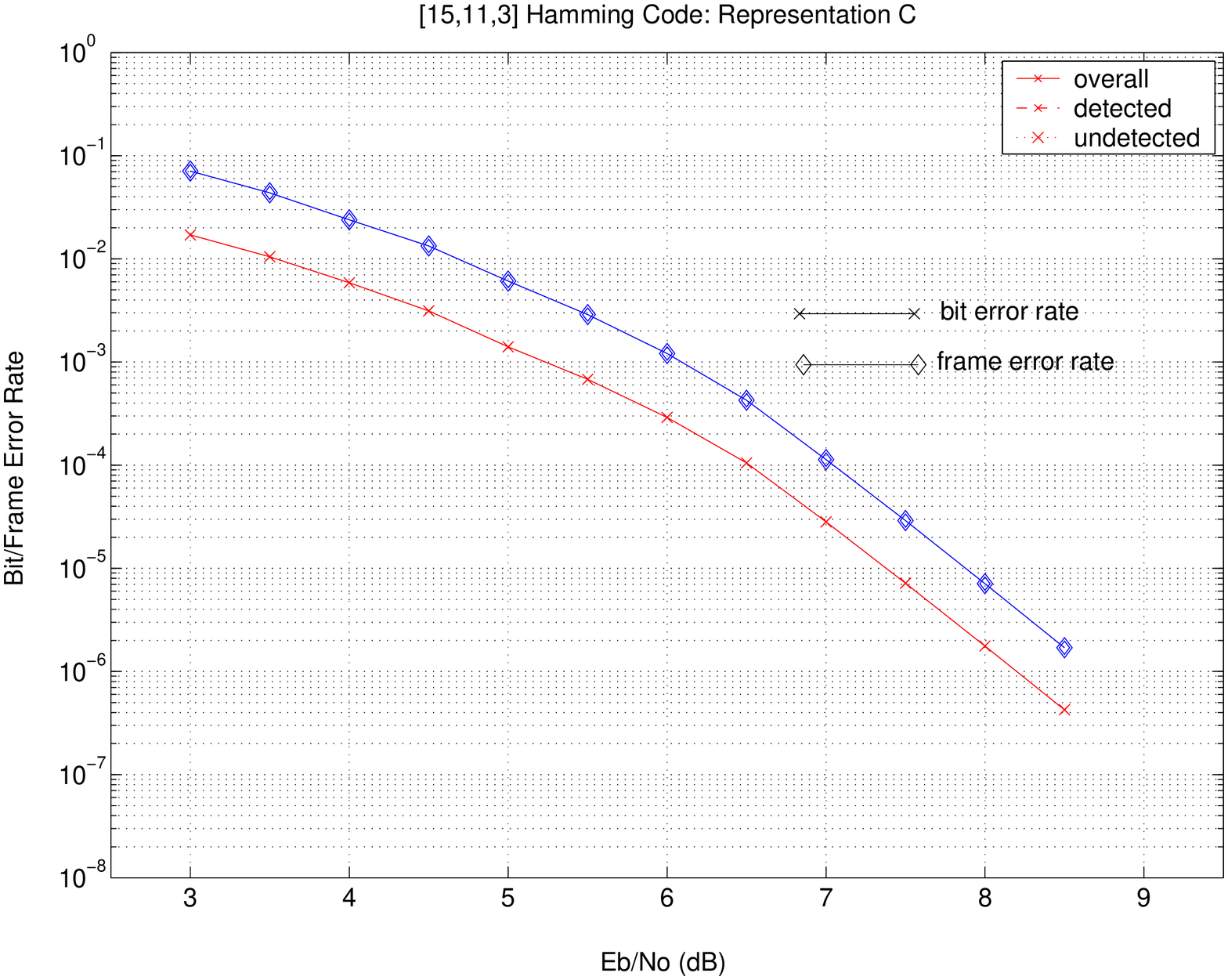}}}
          \caption{Representation C.}\label{hmg_C_15}
         \end{minipage}
            \hspace{0.1in}
            \begin{minipage}[b]{0.48\linewidth}
            \centering{
\resizebox{3.2in}{2.5in}{\includegraphics{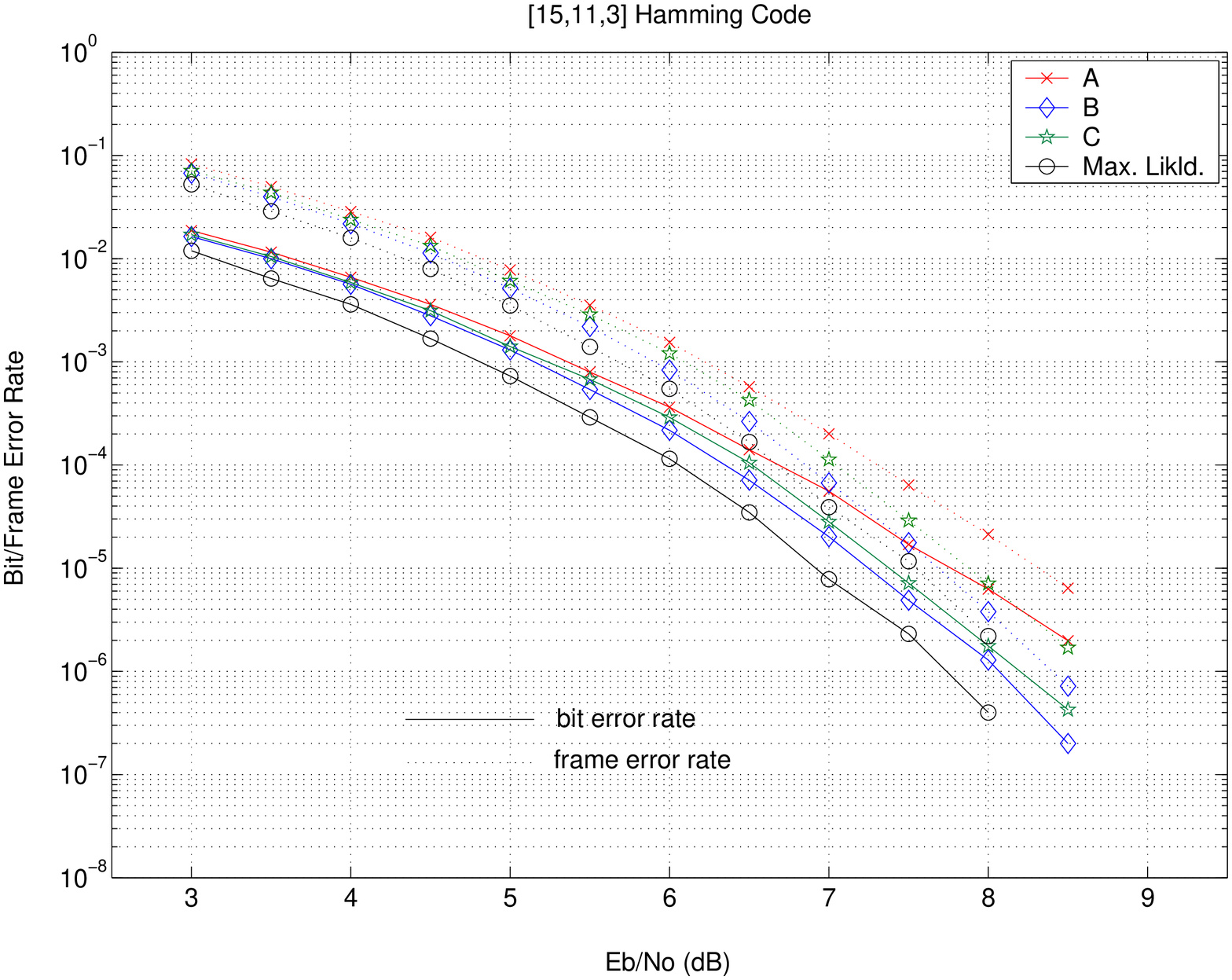}}}
          \caption{Comparison between representations.}\label{hmg_comp_15}
            \end{minipage}
}
\centerline{Performance of the [15,11,3] Hamming code with min-sum iterative decoding over the
BIAWGNC.}
\end{figure}
\end{center}
}

Inferring from the empirical results of this section,  we comment that
LDPC codes that have structure and redundant check nodes, for example,
the class of LDPC codes obtained from finite geometries \cite{lin},
are likely to have fewer number of low-weight pseudocodewords
in comparison to other randomly constructed LDPC graphs of
comparable parameters. Despite the presence of
a large number of short cycles (i.e., 4-cycles and 6-cycles), the class of LDPC
codes in \cite{lin} perform very well with iterative decoding. It is worth investigating
how the set of pseudocodewords among existing LDPC constructions can be
improved, either by adding redundancy or modifying the Tanner graphs,
so that the number of (bad) pseudocodewords, both lift-realizable ones
as well as those occurring on the computation tree, is lowered.

\section{Conclusions}
This paper analyzed pseudocodewords of Tanner graphs, with the
 focus on the structure, bounds on the
minimum pseudocodeword weight, and iterative decoding performance.
It would be worthwhile to relate the results in Section 6 to the
stopping redundancy, as introduced in \cite{stopping_redn}.  Since
this paper primarily dealt with lift-realizable pseudocodewords, the
results presented are also applicable in the analysis of LP
decoding. We hope the insights gained from this paper will aid in
the design of LDPC codes with good minimum pseudocodeword weights.

\section*{Acknowledgments}
We thank Joachim Rosenthal, Pascal Vontobel, and the two reviewers for their careful reading of the
paper and their insightful comments which have considerably improved this paper. We also thank Reviewer 1
for providing an elegant proof of Theorem \ref{existence_t_lemma}.
\nocite{*}
\bibliographystyle{ieeetr}

\appendix

{\em Proof of Lemma~\ref{supportpscw_sset_lemma}:}
Consider the subgraph $G_{|S}$ of $G$ induced by the set of
vertices $S =
\operatorname{supp}({\bf p})$.
Observe that
every cloud of check nodes in the
corresponding cover of $G_{|S}$ is connected to either none or at least two of
the variable clouds in the support of ${\bf p}$. If this were not the
case, then there would be a cloud of check nodes in the cover with at
least one check node in that cloud connected to exactly one variable node
of bit value one, thus, leaving the check node constraint unsatisfied.
Therefore, the corresponding variable nodes in the base graph $G$ satisfy
the conditions of a stopping set.$\hfill \QED$

\vspace{0.15in}

{\em Proof of Claim~\ref{comptree_claim}:} Let $G$ represent the
constraint graph of an LDPC code $\mathcal{C}$. Suppose $G$ is a
tree, then clearly, any pseudocodeword of $G$ can be expressed as a
linear combination of codewords of $G$. Hence, suppose $G$ is not a
tree, and suppose all check nodes in $G$ are of degree two. Then the
computation tree contains only check nodes of degree two and hence,
for a valid assignment on the computation
 tree, the value of any child variable node $v_1$ on the computation tree that stems from a parent check node $u$ is
the same as the value of the variable node $v_2$ which is the parent node of $u$. Thus, the only local codeword
configurations at each check node is the all-ones configurations when the root node of the tree is assigned the value
one. Hence, the only valid solutions on the computation tree
correspond to the all ones vector and the all zeros vector -- which are valid
codewords in $\mathcal{C}$.\\
Conversely, suppose $G$ is not a tree and suppose there is a check
node $u$ of degree $k$ in $G$. Let $v_1,v_2,\dots,v_k$ be the
variable node neighbors of $u$. Enumerate the computation tree
rooted at $v_1$ for a sufficient depth such that the node $u$
appears several times in the tree and also as a node in the final
check node layer of the tree. Then there are several possible valid
assignments in the computation tree, where the values assigned to
the leaf nodes that stem from $u$ yield a solution that is not a
valid codeword in $G$. Thus, $G$ contains irreducible
nc-pseudocodewords on its computation tree.$\hfill \QED$
\vspace{0.15in}

{\em Proof of Claim~\ref{graphcovers_claim}:} Let $G$ represent the
constraint graph of an LDPC code $\mathcal{C}$. Suppose $G$ is a
tree, then clearly, any pseudocodeword of $G$ can be expressed as a
linear combination of codewords of $G$. Hence, suppose $G$ is not a
tree, and between every pair of variable nodes in $G$ there is a
path that contains only degree two check nodes in $G$. Then $G$
contains only lift-realizable pseudocodewords of the form
$(k,k,\dots,k)$, where $k$ is a positive integer. Hence, the only
irreducible pseudocodewords in $G$ are the all-zeros vector
$(0,0,\dots,0)$ and either the all-ones vector $(1,1,\dots,1)$ or
the all-twos vector $(2,2,\dots,2)$ (if the all-ones vector is not a
codeword in $G$). $\hfill \QED$ \vspace{0.15in}

{\em Proof of Theorem~\ref{tree_bound_basic}:}
{\begin{center}
\begin{figure}[h]
\centering{
            \begin{minipage}[b]{0.35\linewidth}
            \centering
                 {
             \resizebox{1in}{1in}{\includegraphics{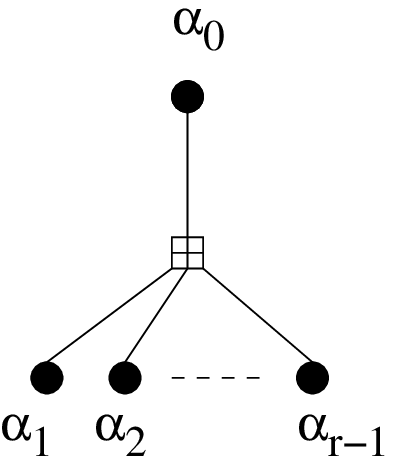}\vspace{1in}}
             \centerline{Single parity check code.}\label{spc}\\
\vspace{0.2in}
             $\alpha_i\le \sum_{j\ne i}\alpha_j$\vspace{0.9in}}
         \end{minipage}
            \hspace{0.1in}
      \begin{minipage}[b]{0.6\linewidth}\vspace{0in}
    \centering{
            \resizebox{1.8in}{1.8in}{\includegraphics{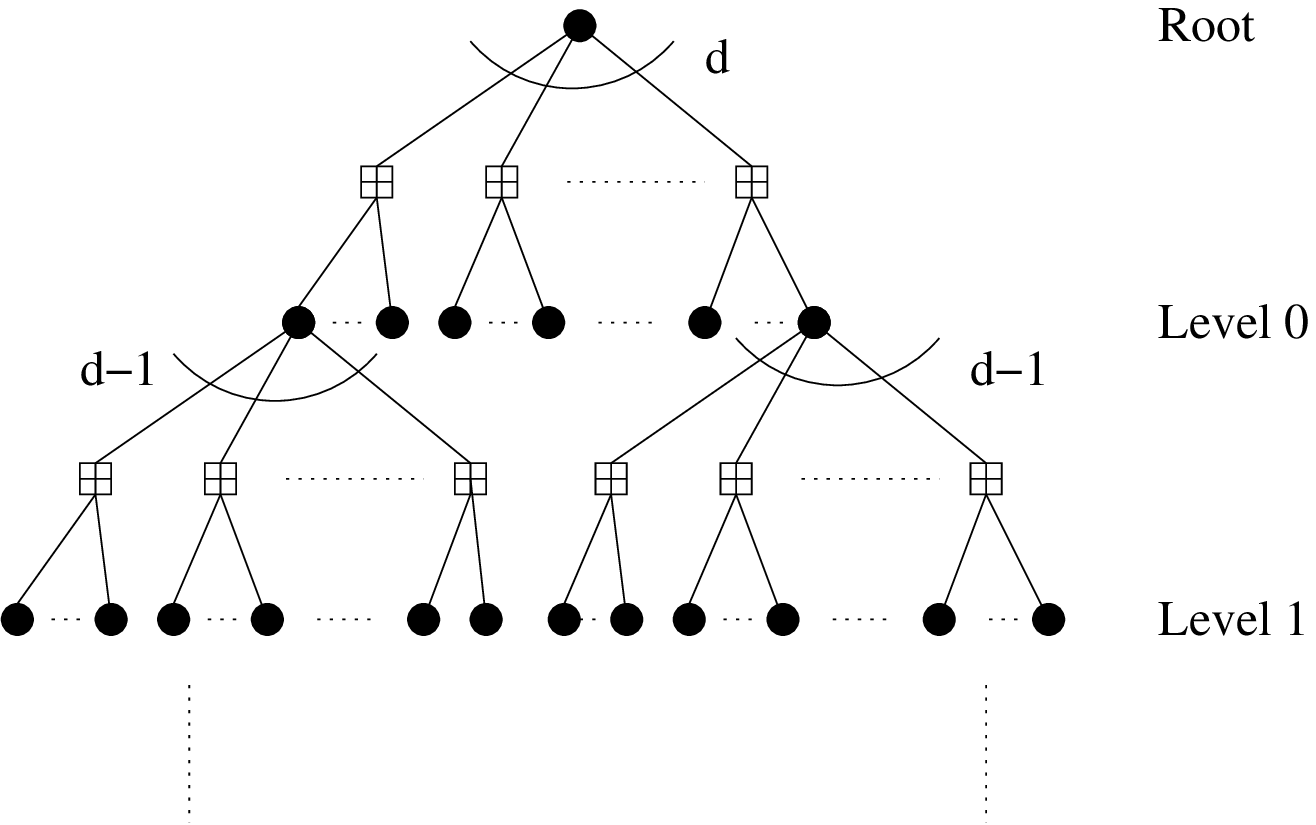}}}
            \centerline{Local tree structure for a $d$-left regular
graph.}
        \label{tree}\\
$d\alpha_1 \le \sum_{j\in
          L_0}\alpha_j$,\vspace{-0.05in}\\
 $d(d-1)\alpha_1\le \sum_{j\in L_1}\alpha_j$\vspace{-0.05in}\\
   $:$\vspace{-0.05in}\\
           \end{minipage}}
\caption{A single constraint node and a $d$-left regular (bipartite) tree enumerated from an arbitrary variable node.}
\end{figure}
\end{center}
}
\vspace{-0.25in}

\underline{Case:} $\frac{g}{2}$ odd. At a single constraint node, the
following inequality holds (see equation \ref{pscw_ineq}): \[ \alpha_i \le \sum_{j\ne i} \alpha_j \]
Applying this inequality in the LDPC constraint graph enumerated as a tree with the
root node corresponding to the dominant pseudocodeword component
$\alpha_1$, we have \[d\alpha_1 \le \sum_{j\in L_0}\alpha_j,\] where $L_0$
corresponds to variable nodes in the first level (level 0) of the tree. Similarly,
we have
 \[ d(d-1)\alpha_1\le \sum_{j\in L_1}\alpha_j,\] and so on, until,
\[d(d-1)^{\frac{g-6}{4}}\alpha_1\le \sum_{j\in L_{\frac{g-6}{4}}}\alpha_j\]
Since the LDPC graph has girth $g$, the variable nodes up to level
$L_{\frac{g-6}{4}}$\footnote{Note that $L_i$ refers to the level for which
the exponent of the $(d-1)$ term is $i$.} are all distinct. The above
inequalities yield:
    \[{\left[ 1+d+d(d-1)+\cdots+d(d-1)^{\frac{g-6}{4}}\right] \alpha_1 \le \sum_{i\in
\{1\}\cup L_0\cup\dots L_{\frac{g-6}{4}}} \alpha_i}{\le
\sum_{all}\alpha_i}\] { Without loss of generality, let us
assume, $\alpha_1\ge \alpha_2\ge \dots\ge \alpha_{e}$
to be the $e$ dominant components in ${\bf p}$. That is, $\alpha_1+\alpha_2+\cdots+\alpha_e\ge \frac{\sum_{i}\alpha_i}{2}$.  Since, each is at most
$\alpha_1$}, we have ${{\sum_{i=1}^{e} \alpha_i \le e \alpha_1.}}$
This implies that \[{ e\alpha_1\ge \sum_{i=1}^{e}\alpha_i \ge
\sum_{i}\frac{\alpha_i}{2}}{\ge
    \frac{\left[ 1+d+d(d-1)+\cdots+d(d-1)^{\frac{g-6}{4}}\right] \alpha_1}{2} }\]
     \[{\Rightarrow e \ge
    \frac{\left[ 1+d+d(d-1)+\cdots+d(d-1)^{\frac{g-6}{4}}\right] }{2}}.\]
Since ${w}_{BSC} = 2e$, the result follows.
(The case when $\frac{g}{2}$ is even is treated similarly.)

{\flushleft \em \underline{AWGN case:}}
Let $x =  \frac{\left[ 1+d+d(d-1)+\cdots+d(d-1)^{\frac{g-6}{4}}\right] }{2} $.  Since,
\[
\sum_{i=1}^{e}\alpha_i\ge
    \frac{\left[ 1+d+d(d-1)+\cdots+d(d-1)^{\frac{g-6}{4}}\right] \alpha_1}{2} ,\] we can
write ${ \sum_{i=1}^{e}\alpha_i = (x+y)\alpha_1 },$  where $y$ is some
non-negative quantity. {Suppose $\alpha_1+ \cdots +\alpha_{e}=
  \alpha_{e+1} + \cdots \alpha_{n}$.}

Then, since we have $\sum_{i=1}^{n}\alpha_i^2 \le 2\sum_{i=1}^{e}\alpha_i^2
\le 2\alpha_1(\sum_{i=1}^{e}\alpha_i)=2(x+y)\alpha_1^2$.,
we get,

\[ {w =\frac{(\sum_{i=1}^{n}\alpha_i)^2}{\sum_{i=1}^{n}\alpha_i^2}}
{\ge
\frac{(2\sum_{i=1}^{e}\alpha_i)^2}{2\sum_{i=1}^{e}\alpha_i^2}
}\]

\[{  w \ge \frac{4(x+y)^2\alpha_1^2}{2(x+y)\alpha_1^2}=2(x+y)}{\ge
2x.}\]
{(The case $\alpha_1+ \cdots +\alpha_{e}> \alpha_{e+1} + \cdots \alpha_{n}$ is treated similarly.)} $\hfill \QED$

\vspace{0.15in}

{\em Proof of Theorem~\ref{tree_bound_gen}:}
As in the proof of Theorem~\ref{tree_bound_basic}, where we note that for a single
constraint with neighbors having pseudocodeword
components $\alpha_1,\dots,\alpha_k$,
we  have the following relation (for $\alpha_i, i=1,\dots,k$):
\[ (\epsilon k)\alpha_i \le \sum_{j=1}^{k}\alpha_j. \]
The result follows by applying this inequality at every constraint node as
in the proof of Theorem~\ref{tree_bound_basic}. $\hfill \QED$

\vspace{0.15in}

{\em Proof of Lemma~\ref{feldman_lemma}:}
Let ${\bf p} = (\alpha_1,\dots,
\alpha_n)$ be a
pseudocodeword of $G$, and without loss of generality, let
$\alpha_1 \ge \alpha_2 \ge \ldots \ge \alpha_n$.
To establish the inequality for the AWGN channel, we need to
show
\[ \frac{(\sum_{i = 1}^{n} \alpha_i)^2}{\sum_{i = 1}^{n} \alpha_i^2} \ge
\frac{\sum_{i = 1}^{n}
\alpha_i}{\max_{i} \alpha_i} = \frac{\sum_{i=1}^{n}\alpha_i}{\alpha_1}. \]

Since $\sum_{i} \alpha_i^2 \le
\alpha_1^2+\alpha_1\alpha_2+\cdots+\alpha_1\alpha_n = \alpha_1(\sum_i \alpha_i)$,
this implies \[\frac{(\sum_{i = 1}^{n} \alpha_i)^2}{\sum_{i = 1}^{n} \alpha_i^2} \ge
\frac{(\sum_{i = 1}^{n} \alpha_i)^2}{\alpha_1\sum_{i = 1}^{n} \alpha_i} \ge
\frac{\sum_{i=1}^{n}\alpha_i}{\alpha_1}. \]
Hence, $w^{AWGN}({\bf p})
\ge w_{\mathrm{max-frac}}({\bf p})$.\\
 \indent To establish the bound for the BSC, let $e$ be the
smallest number such that $\sum_{i=1}^{e}
\alpha_i
\ge
\frac{\sum_{i=1}^{n}\alpha_i}{2}$. First suppose $\sum_{i=1}^{e} \alpha_i =
\sum_{i=e+1}^{n}\alpha_i$. Then
$w^{BSC}({\bf p}) = 2e$. Moreover,  $w_{\mathrm{max-frac}}({\bf p}) =
\frac{\sum_{i=1}^{n} \alpha_i}{\alpha_1} =
\frac{2\sum_{i=1}^{e} \alpha_i}{\alpha_1}$. Each $\alpha_i \le \alpha_1 \Rightarrow
w_{\mathrm{max-frac}}({\bf p}) \le
\frac{2e\alpha_1}{\alpha_1} = 2e = w^{BSC}({\bf p})$.
Now suppose  $\sum_{i=1}^{e} \alpha_i > \sum_{i=e+1}^{n}\alpha_i$.
Then, for some $\delta >0$, we have
$\sum_{i=1}^{e} \alpha_i = \sum_{i=e+1}^{n}\alpha_i + \delta$.
We have $w_{\mathrm{max-frac}}({\bf p}) =
\frac{\sum_{i=1}^{n} \alpha_i}{\alpha_1} = \frac{\sum_{i=1}^{e} \alpha_i+
\sum_{i=e+1}^n \alpha_i}{\alpha_1}$. Note
that $\sum_{i=1}^e\alpha_i + \sum_{i=e+1}^{n}\alpha_i < 2\sum_{i=1}^e \alpha_i <
(2e)\alpha_1$.
Thus, $w_{\mathrm{max-frac}}({\bf p}) < \frac{2e\alpha_1}{\alpha_1} = 2e$.  $\hfill \QED$

\vspace{0.15in}

{\em Proof of Corollary~\ref{corollary_feldman}:}
Follows from Lemma~\ref{feldman_lemma} and Theorem~\ref{feldman_theorem}. $\hfill \QED$

\vspace{0.15in}

{\em Proof of Theorem~\ref{existence_t_lemma}:}
In the polytope representation introduced by Koetter and Vontobel in \cite{pascal},
irreducible pseudocodewords correspond to the edges of the fundamental cone, which can be
described by $\{ {\bf x}\ |\ x_i \ge 0, i = 1,\ldots,n \}$ and $\{ {\bf x}\ |\ x_i \leq \sum_{i' \in N(j)\backslash \{ i \} } x_{i'} \}$
for all check nodes $j$ and all variable nodes $i$ in
$N(j)$. Consider the polytope that is the intersection of the fundamental cone with the hyperplane
$x_1 + \cdots + x_n = 1$. The
vertices of this polytope have a one-to-one correspondence with the edges  of the fundamental cone.
Let $v$ be a fixed vertex of this
new polytope. Then $v$ satisfies at least $n - 1$ of the above inequalities with equality.
Together with the hyperplane equality, $v$ meets at least $n$ inequalities with equality.
The resulting system of linear equations contains only integers as coefficients. By
Cramer's rule, the vertex $v$ must have rational coordinates. Taking the least common multiple
of all denominators of all coordinates of all vertices gives an upper bound on $t$.
Therefore, $t$ is finite. $\hfill \QED$

\vspace{0.15in}

{\em Proof of Theorem~\ref{min_lift_lemma}:}
Let $m$ be the minimum degree lift needed to realize the given
pseudocodeword ${\bf p}$. Then, in a degree $m$ lift graph $\hat{G}$ that realizes
${\bf p}$, the maximum number of active check nodes in any check cloud is at most
$m$. A check cloud $u$ is connected to $\sum_{i\in N(u)}p_i$ active variable nodes from the variable clouds adjoining check
cloud $u$.
(Note that $N(u)$ represents all the variable clouds adjoining $u$.)
Since every active check node in any check cloud has at least two (an even number)
active variable nodes connected to it, we have that $2m \le max_{u}\sum_{i\in N(u)}p_i$. This quantity can be upper-bounded
by $t d_r^{+}$ since $p_i \le t$, for all $i$, and $|N(u)|\le d_r^{+}$, for all $u$. $\hfill \QED$

\vspace{0.15in}

{\em Proof of Lemma~\ref{wmin_t_smin_lemma}:}
(a) \underline{AWGN case:} Let $n_k$ be the number of $p_i$'s
that are equal to $k$, for $k=1,\dots,t$. The pseudocodeword weight
is then equal to:
{\[ w^{AWGN}({\bf p})=
\frac{(n_1+2n_2+\dots+tn_t)^2}{(n_1+2^2n_2+\dots+t^2n_t)}.\]}
Now, we have to find a number $r$ such that
$w^{AWGN}({\bf p}) \ge r |\operatorname{supp}({\bf p})|$.
Note however, that $|\operatorname{supp}({\bf p})| = n_1+n_2+\cdots+n_t$.
This implies that for an appropriate choice of $r$, we have
{ \[\frac{(n_1+2n_2+\cdots+tn_t)^2}{(n_1+2^2n_2+\cdots+t^2n_t)}\ge r(n_1+\cdots n_t)
\mbox{\hspace{0.1in} or \hspace{0.1in}}
  \sum_{i=1}^t i^2 n_i^2 \ge
\sum_{i=1}^t\sum_{j=i+1}^t
\frac{(i^2+j^2)r - 2ij}{1-r} n_i n_j\hspace{0.1in} (*)\]} Note that $r<1$ in the above.
Clearly, if we set $r$ to be the minimum over all $1\le i<j\le n$ such that
$(i^2+j^2)r \le  2ij$, then it can be verified that this choice of $r$ will
ensure that  $(*)$ is true. This implies $r =\frac{2t}{1+t^2}$
(for $i=1, j=t$).

However,  observe that left-hand-side (LHS) in $(*)$ can be written as the following LHS:
{\[ \frac{1}{t-1} \sum_{i=1}^t\sum_{j=i+1}^t (i^2n_i^2+j^2n_j^2) \ge
\sum_{i=1}^t\sum_{j=i+1}^t \frac{(i^2+j^2)r - 2ij}{1-r} n_i n_j \]}
Now, using the inequality $a^2+b^2 \ge 2ab$, $r$ can be taken as the
minimum over all $1\le i<j\le t$
such that {\scriptsize $\frac{1}{t-1} (i^2n_i^2+j^2n_j^2)  \ge
\frac{r(i^2+j^2) - 2ij}{1-r}n_in_j$}. The smallest value of $r$ for which this inequality
holds for all $1\le i<j\le t$ is given by $r=\frac{2t^2}{(1+t^2)(t-1) +2t}$, thereby proving the lemma
in the AWGN case.

\underline{BSC case:}

Since $w^{BSC}({\bf p})\ge w_{\mathrm{max-frac}}({\bf p})$, we have from Lemma~\ref{feldman_lemma}
that
$w^{BSC}({\bf p})\ge \frac{p_1+\cdots + p_n}{\max p_i} \ge \frac{|\operatorname{supp}({\bf
p})|}{t}$. (Note that the $p_i$'s are non-negative integers.)
Therefore, $w^{BSC}({\bf p})\ge  \frac{1}{t}|\operatorname{supp}({\bf p})|$.$\hfill \QED$

\vspace{0.15in}

{\em Proof of Theorem~\ref{good_bad_pscw_wt_thm}:}
Let ${\bf p}$ be a
good pseudocodeword. This means that if
for any weight vector
${\bf w}$ we have ${\bf c}{\bf w}^T\ge 0$ for all ${\bf 0}\ne {\bf c}\in C$, then,
${\bf p}{\bf w}^T \ge 0$. Let us now consider the BSC and the AWGN cases
separately.

\underline{BSC case:} Suppose at most $\frac{d_{\min}}{2}$ errors
occur in channel. Then, the corresponding weight vector ${\bf w}$ will have
$\frac{d_{\min}}{2}$ or fewer components equal to $-1$ and the
remaining components equal to $+1$.
This implies that the cost of any ${\bf 0}\ne {\bf c}\in C$ (i.e., ${\bf c}{\bf w}^T$)
is at least $0$ since there
are at least $d_{\min}$ $1$'s in support of any  ${\bf 0}\ne {\bf c}\in
C$. Since ${\bf p}$ is a good pseudocodeword, it must also have
positive cost, i.e. ${\bf p}{\bf w}^T\ge 0$. Let us assume that the $-1$'s occur in the dominant
$\frac{d_{\min}}{2}$ positions of ${\bf p}$, and without loss of generality,
assume $p_1 \ge p_2 \ge \ldots \ge  p_n$. (Therefore, ${\bf
w}=(-1,-1,\dots,-1,+1,+1,\dots,+1)$.) Positive cost of
${\bf p}$ implies  $p_1+\cdots+p_{d_{\min}/2} \le p_{(d_{\min}/2) +1}+\cdots +p_n$.
So we have $e \ge \frac{d_{\min}}{2}$, where $e$ is as defined in the
pseudocodeword weight of $p$ for the BSC. The result follows.

\underline{AWGN case:} Without loss of generality, let $p_1$ be dominant component of ${\bf p}$.
Set the weight vector ${\bf w} = (1-d_{\min}, 1,\ldots,1)$.
Then it can be verified
that ${\bf c}{\bf w}^T \ge {\bf 0}$ for any ${\bf 0}\ne {\bf c}\in C$.
 Since ${\bf p}$ is a good pseudocodeword, this implies $ {\bf p}$ also must
have positive cost.
 Cost of ${\bf p}$ is $(1-d_{\min})p_1+p_2+\cdots+ p_n  \ge 0 \Rightarrow d_{\min} \le
\frac{p_1+\cdots+p_n}{p_1}$. Note that the right-hand-side (RHS) is $w_{\mathrm{max-frac}}({\bf p})$; hence, the result follows from Lemma~\ref{feldman_lemma}.

Now let us consider ${\bf p}$ to be a bad pseudocodeword. From
Lemma~\ref{supportpscw_sset_lemma},
we have $|\operatorname{supp}({\bf p})| \ge s_{\min}$. Therefore,
$w_{\mathrm{max-frac}}({\bf p}) \ge
\frac{s_{\min}}{t}$ (since $p_1=t$ is the maximum component of ${\bf p}$),
and hence, the result follows by
Lemmas~\ref{feldman_lemma} and
\ref{wmin_t_smin_lemma}. \hspace{0in} $\hfill\QED$

\vspace{0.15in}

{\em Proof of Lemma~\ref{pmod2_lemma}:}
Consider a graph $H$ having a single check node which is connected to
variable nodes $v_1.\dots,v_k$. Suppose ${\bf b} = (b_1,\dots,b_k)$ is a
pseudocodeword in $H$, then ${\bf b}$ corresponds to a codeword in a lift
$\hat{H}$ of $H$. Every check node in $\hat{H}$ is connected to an even
number of variable nodes that are assigned value $1$, and further, each
variable node is connected to exactly one check node in the check cloud.
Since the number of variable nodes that are assigned value 1 is equal to
the sum of the $b_i$'s, we have $\sum_i b_i \equiv 0 \mbox { mod } 2$.

Let $\hat{G}$ be the corresponding lift of $G$ wherein ${\bf p}$ forms a
valid codeword. Then each check node in $\hat{G}$ is connected to an even
number of variable nodes that are assigned value $1$. From the above
observation, if nodes $v_{i_1},\dots,v_{i_k}$ participate in the check
node $u_i$ in $G$, then $p_{i_1}+\cdots+p_{i_k} \equiv 0 \mbox{ mod } 2$.
Let $x_i = p_i \mbox { mod } 2$, for $i = 1,\dots, n$ ($n$ being the
number of variable nodes, i.e., the block length of $\mathcal{C}$, in
$G$).  Then, at every check node $u_i$, we have $x_{i_1}+\cdots +
x_{i_k} \equiv 0 \mbox { mod } 2$. Since ${\bf x} = (x_1,\dots,x_n) = {\bf p}
\mbox{ mod } 2$ is a binary vector satisfying all checks, it is a codeword
in $\mathcal{C}$. $\hfill \QED$

\vspace{0.15in}

{\em Proof of Lemma~\ref{pscwresidue_even_lemma}:}
Suppose ${\bf c}\in \mathcal{C}$ is in the support of ${\bf p}$, then form ${\bf p'}={\bf p}-{\bf c}$. If ${\bf p'}$ contains a codeword in its support,
then repeat the above step on ${\bf p'}$. Subtracting codewords from the
pseudocodeword vector in this manner will lead to a decomposition of the
vector ${\bf p}$ as stated.  Observe that the residual vector ${\bf r}$
contains no codeword in its support.

From Lemma~\ref{pmod2_lemma}, ${\bf x} = {\bf p} \mbox{ mod } 2$ is a
codeword in $\mathcal{C}$. Since ${\bf p} = {\bf c}^{(1)}+\cdots {\bf c}^{(k)} +
{\bf r}$, we have ${\bf x} = ({\bf c}^{(1)} + \cdots + {\bf c}^{(k)}) \mbox{ mod }
2 + {\bf r} \mbox{ mod } 2$. But since ${\bf x} \in \mathcal{C}$, this
implies ${\bf r} \mbox{ mod } 2 \in \mathcal{C}$. However, since ${\bf r}$
contains no codeword in its support, ${\bf r} \mbox{ mod } 2 $ must be the
all-zero codeword. Thus, ${\bf r}$ contains only even (possibly 0)
components. \hspace{0in} $\hfill \QED$

\vspace{0.15in}

{\em Proof of Theorem~\ref{rzero_good_thm}:}
Let ${\bf p}$ be a pseudocodeword of a code $\mathcal{C}$, and
suppose ${\bf p}$ may be decomposed as ${\bf p} = {\bf c}^{(1)} +
{\bf c}^{(2)} +\cdots +{\bf c}^{(k)}$, where $\{{\bf c}^{(i)}\}_{i=1}^k$ is a
set of not necessarily distinct codewords. Suppose ${\bf p}$ is
bad. Then there is a weight vector ${\bf w}$ such that ${\bf
  pw}^T < 0$ but for all codewords ${\bf c} \in \mathcal{C}$,
${\bf cw}^T \ge 0$. Having ${\bf pw}^T < 0$ implies that  ${\bf
  c}^{(1)}{\bf w}^T+ {\bf c}^{(2)}{\bf w}^T +\cdots +{\bf
  c}^{(k)}{\bf w}^T = -x$, for some
positive real value $x$. So there is at least one $i$ for which
${\bf c}^{(i)}{\bf w}^T < 0$, which is a contradiction. Therefore, ${\bf p}$ is a good pseudocodeword. \hspace{0in}
$\hfill\QED$

\vspace{0.15in}

{\em Proof of Theorem~\ref{bad_conditions_thm}:}
Let $M$ be an arbitrarily large finite positive integer.
\begin{enumerate}
\item If $w^{BSC/AWGN}({\bf p}) < d_{\min}$, then ${\bf p}$ is a bad pseudocodeword by Theorem~\ref{good_bad_pscw_wt_thm}.
\item If $|\operatorname{supp}({\bf p})| < d_{\min}$, then there is no codeword in the support of ${\bf p}$, by Lemma~\ref{pmod2_lemma}. Let  ${\bf w} = (w_1,\ldots,w_n)$, be a weight vector where for $i = 1,2,\ldots,n$,
\[w_i = \left\{ \begin{array}{cc}
-1 & \mbox{ if } v_i \in \operatorname{supp}({\bf p}),\\
M& \mbox{ if } v_i \notin \operatorname{supp}({\bf p})
\end{array} \right .\]
Then ${\bf pw}^T < 0$ and for all codewords ${\bf c} \in \mathcal{C}$, ${\bf cw}^T \ge 0$.
\item Suppose ${\bf p}$ is a irreducible nc-pseudocodeword. Without
  loss of generality, assume ${\bf
    p}=(p_1,p_2,\dots,p_s,0,0,\dots,0)$, i.e., the first $s$ positions of ${\bf p}$
  are non-zero and the rest are zero. Suppose ${\bf p}$ contains $\ell$
  distinct codewords ${\bf c}^{(1)}$,${\bf c}^{(2)}$, $\dots$,
  ${\bf c}^{(\ell)}$ in its support. Then if $s\ge \ell+1$, we
  define a weight wector ${\bf w}=(w_1,w_2,\dots, w_n)$ as
  follows. Let $w_i=M$ for $i\notin \operatorname{supp}({\bf p})$. Solve
  for $w_1, w_2,\dots, w_s$ from the following system of linear
  equations:
{
\[{\bf c}^{(1)}{\bf w}^T=+1,\]
\[\vspace{-0.1in}\vdots\]
\[\vspace{-0.1in}{\bf c}^{(\ell)}{\bf w}^T=+1,\]
\[\vspace{-0.1in}{\bf p}{\bf w}^T=p_1w_1+p_2w_2+\cdots+p_sw_s=-2,\]
}

The above system of equations involves $s$ unknowns $w_1$, $w_2$,
$\dots$, $w_s$ and there are $\ell+1$ equations. Hence, as long
as $s\ge \ell+1$, we have a solution for the $w_i$'s. (Note that for the case $s=\ell+1$, when
the $(\ell+1)\times (\ell+1)$ matrix containing the first $s$ components of the codewords ${\bf c}^{(1)}, \dots,{\bf c}^{(\ell)}$ and the pseudocodeword ${\bf p}$ as its rows,
has a non-zero determinant, there is exactly one solution for the $w_i$'s and when this determinant is zero, there is either
zero or more than one solution.)

Thus, there
exists a weight vector ${\bf w}=(w_1,\dots,w_s,M,\dots,M)$ such that ${\bf p}{\bf w}^T<0$ and ${\bf  c}{\bf w}^T\ge 0$ for all codewords ${\bf c}$ in the code. This
proves that ${\bf p}$ is a bad pseudocodeword.
\end{enumerate}
$\hfill\QED$

\vspace{0.15in}

{\em Proof of Theorem~\ref{main_theorem}:}
Let $M$ be an arbitrarily large finite positive integer.
\begin{enumerate}
\item Let $S$ be a stopping set. {\bf Suppose there are no non-zero codewords whose support is  contained in $S$}.  The pseudocodeword ${\bf p}$ with
component value $2$ in the positions of $S$, and $0$ elsewhere is
then a bad pseudocodeword on the AWGN channel, which may be seen by the weight vector
${\bf w}_{\mathrm{a}} = (w_1,\ldots,w_n)$, where for $i = 1,2,\ldots,n$,
\[ w_i = \left\{ \begin{array}{cc}
-1 & \mbox{ if } i \in S,\\
M& \mbox{ if } i \notin S
\end{array} \right .\]
In addition, since all nonzero components have the same value,
the weight of ${\bf p}$ on the BSC and AWGN channel is $|S|$.

Suppose now that ${\bf p}$ is a nc-pseudocodeword with support
$S$. Then the weight vector ${\bf w}_{\mathrm{a}}$ again shows that ${\bf p}$ is bad, i.e., ${\bf p}{\bf w}_{\mathrm{a}}^T<0$ and ${\bf c'}{\bf w}_{\mathrm{a}}^T\ge 0$ for all ${\bf c'}\in \mathcal{C}$.

\item {\bf Suppose there is at least one non-zero codeword ${\bf
    c}$ whose support is in $S$. }\\
(a) {\bf Assume $S$ is a minimal stopping set.} Then this
    means that ${\bf c}$ is the only non-zero codeword whose
    support is in $S$ and $\operatorname{supp}({\bf c})=S$.\\
\indent (i) Suppose $S$ has property $\Theta$, then we can divide the variable nodes in $S$ into disjoint equivalence classes such that the
nodes belonging to each class are connected pairwise by a path
traversing only via degree two check nodes in $G_{|_S}$. Consider
the pseudocodeword ${\bf p}$ having component value $3$ in the
positions corresponding to all nodes of one equivalence class, component value $1$ for the remaining positions of $S$, and component value $0$ elsewhere.
(It can be quickly verified that this is indeed a pseudocodeword by considering the subgraph induced by $S$. In this subgraph,
variable nodes from different equivalence classes will
be connected together at check nodes of degree greater than two. Since there is a non-zero codeword with support in $S$, all such check nodes have
even degree. This implies that the pseudocodeword inequalities of the form $p_i\le \sum_{j\ne i} p_j$ (see equation \ref{pscw_ineq}) at each check node is satisfied, implying
that the chosen ${\bf p}$ is a pseudocodeword.)
Let ${\bf r} = {\bf p} - {\bf c}$, and let $\hat{i}$ denote the index of the first non-zero component of ${\bf r}$ and $i^*$ denote the index of the first non-zero component in $\operatorname{supp}({\bf p})-\operatorname{supp}({\bf r})$. The weight vector ${\bf w}_{\mathrm{b}} = (w_1,\ldots,w_n)$, where for $i = 1,2,\ldots,n$,
\[w_i = \left\{ \begin{array}{cc}
-1 & \mbox{ if } i = \hat{i},\\
+1 & i = i^*\\
M& \mbox{ if } i \notin S\\
0 & \mbox{ otherwise }
\end{array} \right .\]
 ensures that
${\bf p}$ is bad as in Definition~\ref{bad_pscw_defn}, and it is easy to show that the weight of ${\bf p}$ on the AWGN channel is strictly
less than $|S|$.\\

Conversely, suppose $S$ does not have property $\Theta$. Then
every pair of variable nodes in $S$ is connected by a path in
$G_{|_S}$ that contains only degree two check nodes. This means
that any pseudocodeword ${\bf p}$ with support $S$ must have all
its components in $S$ of the same value. Therefore, the only
pseudocodewords with support $S$ that arise have the form ${\bf p} = k{\bf c}$,
for some positive integer $k$. (By Theorem~\ref{rzero_good_thm},
these are good pseudocodewords.) Hence, there exists no bad
pseudocodewords with support $S$.\\

\indent (ii)  Let ${\bf p}$ be a nc-pseudocodeword with
  support $S$. If $S$ contains a codeword ${\bf c}$ in its
  support, then since $S$ is minimal $\operatorname{supp}({\bf c})=S$. Let $k$ denote the number of times ${\bf c}$ occurs in
  the decomposition (as in Lemma~\ref{pscwresidue_even_lemma}) of ${\bf p}$. That is,  ${\bf p} = k{\bf   c}+{\bf r}$. Note that ${\bf r}$ is non-zero
  since ${\bf p}$ is a nc-pseudocodeword. Let $\hat{i}$
  denote an index of the maximal component of ${\bf r}$, and let
  $i^*$ denote the index of the first nonzero component in
  $\operatorname{supp}({\bf p})-\operatorname{supp}({\bf r})$. The weight vector ${\bf w}_{\mathrm{b}}$, defined above,  again ensures that ${\bf p}$ is bad.\\

(b) {\bf Suppose $S$ is not a minimal stopping set and there is at least one
  non-zero codeword ${\bf c}$ whose support is in $S$.}\\

\indent (i) Suppose $S$ contains a problematic node $v$. By
  definition, assume that $S$ is the only stopping set
  among all stopping sets in $S$ that contains $v$.
  Define a set $S_v$ as \[S_v := \{u\in S| \mbox{ there is a path
    from $u$ to $v$ containing only degree two check nodes in }
  G_{|_S}\}\]

Then, the pseudocodeword ${\bf p}$ that has component value $4$
on all nodes in $S_v$, component value $2$ on all nodes in
$S-S_v$ and component value $0$ everywhere else is a valid
pseudocodeword. (It can be quickly verified that this is indeed a pseudocodeword by
considering all the check nodes in the graph induced by $S$ in $G$.
Any check node that is in the path from some $u\in S_v$ to $v$ is either
of degree two, in which case the inequality in equation (\ref{pscw_ineq}) is satisfied,
or is of degree greater than two, in which case the check node is possibly connected to a variable node in $S-S_v$ and to at least two variable nodes in $S_v$.
The choice of the components guarantee that the pseudocodeword inequality in equation (\ref{pscw_ineq}) is still satisfied at this check node. For any other check node in $G_{|_S}$,
the degree is at least two, and it is easy to verify that the inequality in equation (\ref{pscw_ineq}) is satisfied.)
Let $i^*$ be the index of the variable node $v$ in
$G$, and let $i'$ be the index of some variable node in
$S-S_v$. Then, the weight vector ${\bf w}_{\mathrm{c}} = (w_1,\ldots,w_n)$, where for $i = 1,2,\ldots,n$,
\[w_i = \left\{ \begin{array}{cc}
-1 & \mbox{ if } i = i^*,\\
+1 & i = i'\\
M& \mbox{ if } i \notin S\\
0 & \mbox{ otherwise }
\end{array} \right .\]
 ensures that ${\bf p}{\bf w}_{\mathrm{c}}^T<0$ and ${\bf c'}{\bf w}_{\mathrm{c}}^T\ge 0$
 for all non-zero codewords ${\bf c'}$ in $\mathcal{C}$. Hence,
 ${\bf p}$ is a bad pseudocodeword with support $S$. This shows
 the existence of a bad pseudocodeword on $S$.

Suppose now that ${\bf p'}$ is some irreducible nc-pseudocodeword
with support $S$.  If there is a non-zero codeword ${\bf c}$ such
that ${\bf c}$ has support in $S$ and contains $v$ in its
support, then since $v$ is a problematic node in $S$, $v$ cannot
lie in a smaller stopping set in $S$. This means that support of
${\bf c}$ is equal to $S$. We will show that ${\bf p'}$ is a bad
pseudocodeword by constructing a suitable weight vector ${\bf w}_{\mathrm{d}}$.  Let $i'$ be the index of some variable node in
$S-S_v$. Then we can define a weight vector ${\bf w}_{\mathrm{d}} = (w_1,\ldots,w_n)$, where for $i = 1,2,\ldots,n$,
\[w_i = \left\{ \begin{array}{cc}
-1 & \mbox{ if } i = i^*,\\
+1 & i = i'\\
M& \mbox{ if } i \notin S\\
0 & \mbox{ otherwise }
\end{array} \right .\]
(Note that since ${\bf p'}$ is a irreducible
nc-pseudocodeword and contains $v$ in its support, $p'_{i^*}>p'_{i'}$.)
This weight vector ensures that ${\bf p'}{\bf w}_{\mathrm{d}}^T<0$ and ${\bf
  c'}{\bf w}_{\mathrm{d}}^T\ge 0$ for all non-zero codewords ${\bf c'}\in
\mathcal{C}$. Thus, ${\bf p'}$ is a bad pseudocodeword.

If there is no non-zero codeword ${\bf c}$ such that ${\bf c}$
has support in $S$ and also contains $v$ in its support. Then,
the weight vector ${\bf w}_{\mathrm{e}}=(w_1,\ldots,w_n)$, where for $i = 1,2,\ldots,n$,
\[w_i = \left\{ \begin{array}{cc}
-1 & \mbox{ if } i = i^*,\\
M& \mbox{ if } i \notin S\\
0 & \mbox{ otherwise }
\end{array} \right .\]
ensures that ${\bf p'}{\bf w}_{\mathrm{e}}^T<0$ and ${\bf
  c'}{\bf w}_{\mathrm{e}}^T\ge 0$ for all non-zero codewords ${\bf c'}\in
\mathcal{C}$. Thus, ${\bf p'}$ is a bad pseudocodeword. This proves that any irreducible nc-pseudocodeword
with support $S$ is bad.\\

\indent (ii) Suppose $S$ is not a minimal stopping set and suppose
  $S$ does not contain any problematic nodes. Then, any node in
  $S$ belongs to a smaller stopping set within $S$. We claim that
  each node within $S$ belongs to a minimal stopping
  set within $S$. Otherwise, a node belongs to a proper non-minimal
  stopping set $S'\subsetneq S$ and is not contained in any smaller stopping set
  within $S'$ -- thereby, implying that
  the node is a problematic node.  Therefore, all nodes in $S$
  are contained in minimal stopping sets within
  $S$.

To prove the last part of the theorem, suppose one of these
 minimal stopping sets, say $S_j$, is not the support of any non-zero
 codeword in $\mathcal{C}$. Then,
 there exists a bad nc-pseudocodeword ${\bf p}=(p_1,\dots,p_n)$
 with support $S$, where $p_i= x$, for an appropriately chosen
 positive even integer $x$ that is at least 4, for $i\in S_j$ and $p_i=2$ for
 $i\in S-S_j$, and $p_i=0$ for $i\notin S$. Let $i^*$ be the
 index of a variable node $v^*$ in $S_j$. If there are distinct codewords
 ${\bf c}^{(1)}$, ${\bf c}^{(2)}$, $\dots$, ${\bf c}^{(t)}$  whose supports contain $v^*$  and whose supports
are contained in $S$, then let $i_1$, $i_2$, $\dots$, $i_{t'}$
be the indices of variable nodes in the supports of these
codewords outside of $S_j$. (Note that we choose the smallest
number $t'\le t$ of indices such
that each codeword contains one of the variable nodes
$v_{i_1},\dots, v_{i_{t'}}$ in its support.)
 The following weight vector
 ${\bf w}=(w_1,\dots,w_n)$ ensures that ${\bf p}$ is a bad pseudocodeword.
\[w_i = \left\{ \begin{array}{cc}
-1 & \mbox{ if } i =i^*\\
+1 & \mbox{ if } i=i_1, i_2,\dots, i_{t'}\\
M& \mbox{ if } i \notin S\\
0 & \mbox{ otherwise }
\end{array} \right .\]
(Note that $x$ can be chosen so that ${\bf p}{\bf w}^T <0$ and it
is clear that ${\bf c}{\bf w}^T \ge 0$ for all codewords in the code.)

Now suppose $S_j$ contains the support of a codeword ${\bf c'}$ and
has property $\Theta$, then we can construct a bad pseudocodeword
on $S_j$ using the previous argument and that in part 2(a)(i) above (since $S_j$ is minimal) and allow the remaining
components of ${\bf p}$ in $S-S_j$ to have a component value of
2. It is easy to verify that such a pseudocodeword is bad.

Conversely, suppose every minimal stopping set $S_j$ within $S$ does
not have property $\Theta$ and contains the support of some non-zero codeword
${\bf c}^{(j)}$ within it. Then, this means that $S_j=\operatorname{supp}({\bf
  c}^{(j)})$ and that between every pair of nodes within $S_j$ there is
a path that contains only degree two check nodes in
$G_{|_{S_j}}$. Then, for any pseudocodeword ${\bf p}$ with
support $S$, there is a decomposition of ${\bf p}$, as in
Lemma~\ref{pscwresidue_even_lemma}, such that ${\bf p}$ can be expressed as a linear
combination of codewords ${\bf c}^{(j)}$'s. Thus, by Theorem~\ref{rzero_good_thm}, there are no bad
pseudocodewords with support $S$.
\end{enumerate} $\hfill \QED$

\vspace{0.15in}

{\em Proof of Lemma~\ref{partial_t_thm}:}
 Suppose $S$ is minimal and does not have property $\Theta$. Then each pair of nodes in $S$ are
connected by a path via degree two check nodes, and hence all components of a pseudocodeword ${\bf p}$
on $S$ are equal. If $S$ is the support of a codeword ${\bf c}$, then any pseudocodeword is a multiple
of ${\bf c}$, and so $t_S = 1$.  If $S$ is not the support of a codeword, then the all-two's vector is
the only irreducible pseudocodeword, and $t_S = 2$. $\hfill \QED$

\vspace{0.15in}

{\em Proof of Theorem~\ref{bad_pscw_theorem}:}
Let the LDPC code $\mathcal{C}$ represented by the LDPC
constraint graph $G$ have blocklength $n$.
Note that the weight vector ${\bf w}=(w_1,w_2,\dots,w_n)$ has
only $+1$ or $-1$ components on the BSC, whereas it has
every component $w_i$ in the interval $[-L,+L]$ on the
truncated AWGN channel $TAWGN(L)$, and has every component $w_i$
in the interval $(-\infty,+\infty)$ on the AWGN channel. That is,
\[P^B_{BSC}(G)=\{{\bf p}| \exists {\bf w}\in \{+1,-1\}^n\
s.t.\ {\bf p}{\bf w}^T <0,  {\bf c}{\bf w}^T \ge 0, \forall 0\ne
{\bf c}\in \mathcal{C}\}. \]
\[P^B_{TAWGN(L)}(G)=\{{\bf p}| \exists {\bf w}\in [+L,-L]^n\
s.t.\ {\bf p}{\bf w}^T <0,  {\bf c}{\bf w}^T \ge 0, \forall 0\ne
{\bf c}\in \mathcal{C}\}. \]
\[P^B_{AWGN}(G)=\{{\bf p}| \exists {\bf w}\in (-\infty,+\infty)^n\
s.t.\ {\bf p}{\bf w}^T <0,  {\bf c}{\bf w}^T \ge 0, \forall 0\ne
{\bf c}\in \mathcal{C}\}. \] It is clear from the above that, for
$L\ge 1$, $P^B_{BSC}(G)\subseteq P^B_{TAWGN(L)}(G) \subseteq
P^B_{AWGN}(G)$.  $\hfill \QED$ \vspace{0.15in}

\vspace{0.15in}

{\bf Christine Kelley} received the B.S. in Mathematics from the
University of Puget Sound, WA, in 1999, and the M.S. and Ph.D. in
Mathematics from the University of Notre Dame, IN, in 2003 and 2006,
respectively.  She participated in the Budapest Semesters in
Mathematics program in Fall 1998, worked at Los Alamos National
Laboratory in 1999-2000, and studied at Cambridge University in
England in 2000-2001. She also did a year of her graduate study at
the University of Zurich, Switzerland.

Since receiving her Ph.D., she has been a postdoctoral fellow at the
Fields Institute in Toronto, Canada, and a VIGRE Arnold Ross
Assistant Professor at The Ohio State University in Columbus, OH. As
of August 2007, she is an assistant professor at the University of
Nebraska-Lincoln, though she will be on leave and remain at the Ohio
State University until the summer of 2008. Her research interests
include codes on graphs, algebraic coding theory, discrete
mathematics, cryptography, and information theory.

\vspace{0.15in}

{\bf Deepak Sridhara} received the B.Tech degree in electrical engineering from
Indian Institute of Technology, Madras, in 1998, the M.S. and Ph.D.
degrees in electrical engineering from University of Notre Dame, Notre
Dame, IN, in 2000 and 2003, respectively.

Between June and December 2003, he was as a postdoctoral Research
Associate in the Department of Mathematics at University of Notre
Dame. In the year 2004, he was a Research Associate in the
Department of Mathematics at the Indian Institute of Science,
Bangalore. Between January 2005 and April 2006, he was a
post-doctoral Research Associate in the Institute of Mathematics at
the Univeristy of Zurich, Switzerland. He is currently a Research
Staff Member at Seagate Technology research center in Pittsburgh,
USA. His research interests include coding theory, codes on
graphs and iterative techniques, and information theory. \\

\end{document}